\documentclass[
 reprint,
 amsmath,
 amssymb,
 aps,
 nofootinbib
]{revtex4-2}

\usepackage{graphicx}
\usepackage{dcolumn}
\usepackage{bm}
\usepackage{hyperref}
\usepackage{amsmath}
\usepackage{subcaption}
\usepackage{amssymb,amsfonts}
\usepackage{empheq}
\usepackage{xcolor}
\usepackage{mathrsfs}
\usepackage{esint}
\usepackage{soul}

\begin{document}

\preprint{APS/123-QED}

\title{Gravitationally Induced Polarization-Angle Rotation in Pulsar–Black Hole Binaries\\}

\author{Jyotijwal Debnath}
\email{jyotijwal.debnath@inaf.it}
\author{Amodio Carleo}
\author{Delphine Perrodin}
\author{Pratik Tarafdar}
\affiliation{INAF - Osservatorio Astronomico di Cagliari, Via della Scienza 5, 09047 Selargius (CA), Italy\\}



\date{\today}

\begin{abstract}
Radio pulsars emit highly polarized radiation whose polarization properties are commonly described by the Rotating Vector Model (RVM). In this work, we investigate the effects of gravitational light bending and parallel transport on the polarization profile of a pulsar in a hypothetical binary system with a black hole (BH) companion. We combine the RVM with numerical integration of null geodesics to calculate the polarization direction along the photon trajectory and model the orbital-phase-dependent variation of the observed polarization position angle (PPA). We consider both non-spinning and spinning black hole companions and include the evolution of the orbital geometry due to relativistic spin-orbit coupling. For a non-spinning black hole, the change in the observed polarization direction arises from the geometric shift associated with gravitational light bending. For a spinning black hole, frame dragging introduces an additional geometric shift, while parallel transport produces an additional polarization rotation measured relative to our chosen reference construction. We investigate the dependence of these polarization changes on the binary parameters and the orientation of the pulsar spin. We find that gravitational light bending can significantly distort the characteristic S-shaped PPA profile, particularly near superior conjunction, and can reverse its overall sense of variation for the supermassive black hole system considered here. Relativistic spin-orbit evolution also affects the orbital-phase dependence of the polarization profile, particularly in the supermassive black hole system. These results show that relativistic photon propagation can modify both the shape and the phase dependence of pulsar polarization profiles and may provide a useful probe of the geometry of pulsar--BH binaries when such systems are discovered.
\end{abstract}

\maketitle

\section{Introduction}
\label{introduction}

Radio pulsars are rapidly rotating neutron stars that emit highly polarized radio radiation \citep{Manchester1970}. The polarization position angle (PPA) across the pulse profile provides information about the geometry of the pulsar emission region and the orientations of its magnetic and rotation axes \citep{fcl11}. The observed PPA is commonly described by the Rotating Vector Model (RVM), which predicts a characteristic S-shaped variation of the polarization angle across the pulse \citep{rc69,Rankin83}.

The exceptional timing precision of pulsars, particularly millisecond pulsars, makes them powerful probes of a wide range of astrophysical phenomena \citep{pan21}. Their stable rotational periods enable stringent tests of general relativity \citep{Stairs2003, ksm21}, studies of the interstellar medium \citep{Krishnakumar:2021}, and searches for low-frequency gravitational waves \citep{agazie24, Maggiore:2007ulw}. Pulsar-BH binaries are of particular interest because their timing and propagation signals can provide information about both the pulsar and the gravitational field of the black hole companion \citep{Seymour:2018}. Intermediate-mass black holes (IMBHs) are also expected to occur in globular clusters, although their existence has not yet been firmly established observationally \citep{Chen:2025, FCA2005}. Pulsar-BH systems in such environments are therefore of particular interest for searches for these otherwise difficult-to-detect black holes. The increasing sensitivity of modern radio telescopes can improve the prospects for finding such systems by enabling pulsar surveys to be carried out with shorter integration times \citep{Liu:2014}.

When a pulsar is part of a binary system, the propagation of its radiation can be significantly affected by the gravitational field of the companion \citep{dk1995, RL06, hkc22, km02}. Depending on the nature and geometry of the binary, the observed pulsar signal can be affected by several propagation effects. These include the Rømer delay associated with the orbital motion of the pulsar, the Shapiro delay produced by the curvature of spacetime, and the geometric propagation delay \citep{Lorimer2004}. In addition, gravitational bending of the light ray introduces a bending delay that depends on the geometry of the pulsar-companion system \citep{dk1995,RL06}. These relativistic propagation effects have been studied extensively in the context of pulsar-BH binaries and can be investigated through precise pulsar timing \citep{DellaMonica:2025}.

The effect of gravitational light bending becomes particularly important when the pulsar is observed close to superior conjunction, where the line of sight passes close to the companion \citep{DBB23}. Light bending changes both the direction and the path of the light ray between the pulsar and the observer. Previous studies have investigated the latitudinal and longitudinal displacements of the observed signal and the corresponding bending delay in pulsar-BH systems \citep{DBB23,DB25}. The change in the direction of the light ray also affects the observed polarization of the pulsar radiation. Since the polarization direction predicted by the RVM depends on the emission direction, gravitational light bending produces an apparent geometric shift in the observed polarization angle.

In addition to this geometric effect, the polarization vector is parallel transported along the null geodesic followed by the radiation in curved spacetime \citep{Walker:1970,Frolov:2020}. The propagation of the polarization vector can therefore introduce an additional change in the observed polarization direction \citep{Corner1980,Dovciak:2008}. For a spinning black hole, frame dragging also introduces an additional geometric shift in the observed polarization direction. Thus, the polarization changes considered in this work include the geometric contribution associated with gravitational light bending, the additional geometric contribution due to frame dragging, and the polarization rotation associated with parallel transport. Separating these contributions is important for understanding the relativistic modification of the pulsar polarization profile. Although the contributions associated with black hole spin are smaller than the leading geometric shift for the systems considered here, they may become relevant in sufficiently precise polarization measurements. Gravitational light bending can therefore modify the characteristic S-shaped PPA profile predicted by the RVM.

In this work, we study the modification of pulsar polarization caused by relativistic photon propagation in a binary system containing a black hole companion. We combine the RVM description of pulsar emission with a general relativistic calculation of the photon trajectory and the parallel transport of the polarization vector along the null geodesic. This allows us to follow the polarization from the emission point to a distant observer and to quantify the geometric shift and the polarization rotation associated with parallel transport relative to our chosen reference construction. We consider both non-spinning and spinning black hole companions and examine how the resulting polarization changes depend on the binary parameters and the orientation of the pulsar spin. We also investigate the effect of relativistic spin-orbit evolution on the orbital-phase dependence of the polarization profile.

The paper is organized as follows. In Section~\ref{bema_polarization_geometry}, we describe the pulsar emission geometry and the polarization direction within the RVM. In Section~\ref{binary}, we introduce the spin-orbit dynamics and describe the binary geometry and the coordinate frames used in our calculation. In Section~\ref{definitions}, we define the different contributions to the polarization change and describe how they are quantified. In Section~\ref{subsec:bending_spin_full_GR}, we present the full general relativistic formalism used to calculate the null geodesic and parallel transport the polarization vector along the photon trajectory. In Section~\ref{numerical}, we present the numerical results for two representative systems: an intermediate-mass black hole companion with a mass of $10^5\,M_\odot$ in a relatively wide orbit and a supermassive black hole companion in a close binary. We examine the effects of gravitational light bending, parallel transport, black hole spin, and the relevant binary parameters on the observed polarization. Finally, we summarize our results and discuss their implications in Section~\ref{conclusion}.

\section{Polarization model of an Isolated Pulsar}
\label{bema_polarization_geometry}

\subsection{Pulsar Emission and Polarization Beam Geometry}

The classic ``lighthouse model'' of pulsar emission, proposed by
\citet{rc69}, provides the standard geometrical framework for understanding
the polarization properties of radio pulsars. In this model, a pulsar is
treated as a rotating magnetic dipole, and relativistic charged particles
are assumed to stream along the open magnetic field lines originating from the magnetic poles \citep{{Sturrock1971}, {Arons1983}, {Hibschman2001}, Arendt2002, Komesaroff1970, ppr05}. Due to the curvature of these magnetic field lines, the particles emit highly beamed radiation along the local tangent to the magnetic field \citep{GJ69, Deutsch1955, hpb14}. This mechanism is known as curvature radiation. The emitted radiation is highly polarized, with its
polarization direction determined by the acceleration of the radiating particles \citep{jackson98}.

\begin{figure}[htbp!]
\centering
\includegraphics[width=0.5\textwidth]{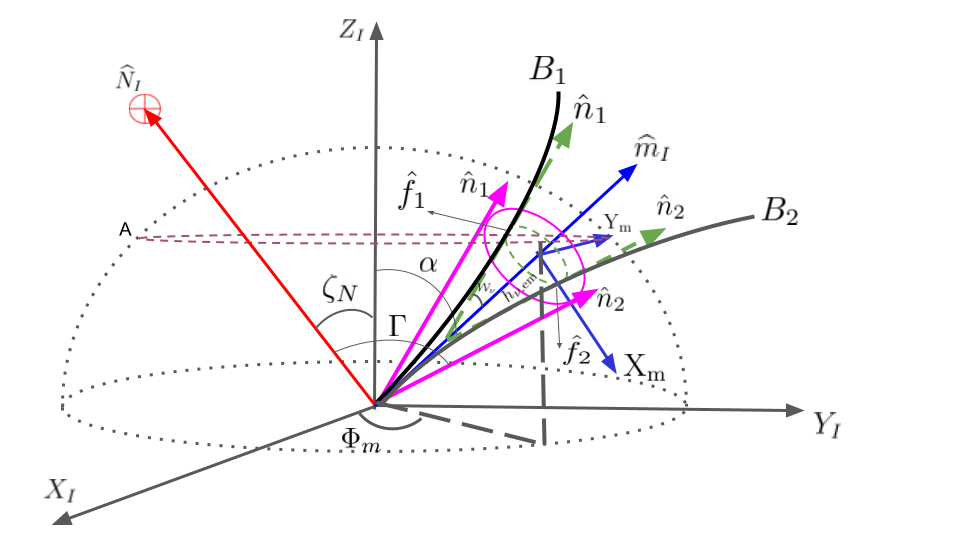}
\caption{Schematic diagram of the pulsar emission and polarization beam
geometry. The meaning of the various vectors, coordinate frames, and
angles is explained in the text.}
\label{polarization_geometry}
\end{figure}

In Fig.~\ref{polarization_geometry}, we illustrate the geometric
configuration of the pulsar emission and polarization beam. We define
a Cartesian reference frame $X_IY_IZ_I$ (hereafter the pulsar's
inertial frame or $I$-frame) with its origin at the center of the
pulsar and the spin axis aligned along the $Z_I$-axis. The LoS direction
is denoted by the unit vector $\hat{N}_I$, which lies in the $X_IZ_I$
plane and is inclined at an angle $\zeta_N$ with respect to the spin
axis. The magnetic axis, denoted by the unit vector $\hat{m}_I$, is
inclined at an angle $\alpha$ relative to the spin axis. We assume that,
at the instant $t=0$, $\hat{m}_I$ lies in the $X_IZ_I$ plane. The
rotational phase of the magnetic axis, or equivalently that of the
pulsar, is therefore defined as
$\Phi_m(t)=2\pi t/P_s$, where $P_s$ is the spin period of the pulsar. As shown in Fig.~\ref{polarization_geometry}, $\Phi_m$ is the angle between the $X_I$ axis and the projection of the magnetic axis on the $X_IY_I$ plane.
The angle between $\hat{m}_I$ and the LoS is denoted by $\Gamma$, which
varies with time. Its minimum value is known as the angle of closest
approach and is defined as
$\beta=\zeta_N-\alpha$ \citep{DB26}.

As the charged particles stream along the open magnetic field lines,
they emit radio radiation along the tangent to the field lines. The
characteristic altitude at which the radio emission originates depends
on the observing frequency $\nu$. Following \citep{KG1998,KG2003}, the
radius-to-frequency mapping for the emission height
$h_{\nu,\mathrm{em}}$ is expressed as
\begin{equation}
\label{eq:emheight}
h_{\nu,\mathrm{em}}
=
400\,\mathrm{km}
\left(\frac{\nu}{10^{9}\,\mathrm{Hz}}\right)^{-0.26}
\left(\frac{\dot{P}_{s}}{10^{-15}\,\mathrm{s\,s^{-1}}}\right)^{0.07}
\left(\frac{P_s}{\mathrm{s}}\right)^{0.30},
\end{equation}
where  $\dot{P}_s$ is the first
time derivative of the spin period. 

In Fig.~\ref{polarization_geometry}, two light-ray directions
$\hat{n}_1$ and $\hat{n}_2$ are shown, originating from the open
magnetic field lines $B_1$ and $B_2$, respectively. The directions
$\hat{n}_1$ and $\hat{n}_2$ are along the tangent to the corresponding
magnetic field lines at the emission height $h_{\nu,\mathrm{em}}$.
The collection of these tangents forms a conal emission beam \citep{lm88}. In this
simple dipolar emission model, the physical conal beam is centered on
the magnetic axis \citep{gks93, GM2001}. Since the emission height is much smaller than the
distance to the observer, the geometrical representation of the conal
beam can be translated along $\hat{m}_I$ so that its center coincides
with the origin of the $I$-frame, without affecting the geometry
relevant to the present calculation.

In Fig.~\ref{polarization_geometry}, the physical conal beam generated
at the emission height $h_{\nu,\mathrm{em}}$ is shown in green. For
visual clarity, the geometrically equivalent beam translated to the
origin of the $I$-frame is also shown, with its boundary projected onto
the plane at the emission height in magenta. The half-opening angle of
the cone, $\mathcal{W}_{\nu}$, also depends on the observing frequency.
However, there is no exact analytical expression that describes this
dependence. At an observing frequency of $1.4$~GHz, the half-opening
angle $\mathcal{W}_{\rm 1.4}$ can be parameterized as
\citep{KG1998, KG2003, gil1984, DBB23}:
\begin{equation}
\label{eq:halfopeningangleoutcone}
\mathcal{W}_{\rm 1.4}
=
5.83^\circ
\left(\frac{P_s}{\mathrm{s}}\right)^{-0.5}.
\end{equation}
The gravitational transport of the polarization vector considered in
the present work does not depend explicitly on the choice of the
half-opening angle. The value of $\mathcal{W}_{\nu}$ only determines
the range of emission directions considered in the numerical
calculations.

In this lighthouse model of pulsar emission, the emitted radiation is
highly polarized and the polarization direction is related to the
acceleration vector of the radiating particles \citep{jackson98}. In
Fig.~\ref{polarization_geometry}, the polarization vectors corresponding
to the propagation directions $\hat{n}_1$ and $\hat{n}_2$ are denoted by
$\vec{f}_1$ and $\vec{f}_2$, respectively. In the $I$-frame, the total
acceleration $\vec{a}$ of a particle moving along a rotating magnetic
field line is given by \citep{CM2010}:
\begin{equation}
\label{accelaration}
\vec{a}
=
c\frac{\partial\hat{n}}{\partial t}
+
c(\vec{v}\cdot\nabla)\hat{n}
+
c\vec{\Omega}_p\times\hat{n}
+
\vec{\Omega}_p\times
\left(\vec{\Omega}_p\times\vec{r}_0\right),
\end{equation}
where $\hat{n}$ is the unit tangent vector pointing along the
instantaneous direction of emission, $\vec{\Omega}_p$ is the angular
velocity vector of the pulsar, $\vec{v}$ is the particle velocity, and
$\vec{r}_0$ is the position vector of the emission point such that
$|\vec{r}_0|=h_{\nu,\mathrm{em}}$.

The third and fourth terms on the right-hand side of
Eq.~\eqref{accelaration} represent the Coriolis and centripetal
accelerations associated with the rotation of the pulsar,
respectively. In the low-altitude limit, where the emission height is
much smaller than the light-cylinder radius
($h_{\nu,\mathrm{em}}\ll r_{\rm LC}$), these rotational terms become
negligibly small and can be neglected. The first term represents the
explicit time dependence of the rotating magnetic-field configuration,
whereas the second term represents the acceleration due to the spatial
curvature of the magnetic field lines.

The standard Rotating Vector Model (RVM) proposed by \citet{rc69}
provides an idealized description of the variation of the polarization
direction based on the geometry of a rotating dipolar magnetic field.
To establish a simple baseline model for the present work, we restrict
our analysis to the spatial curvature of the magnetic field lines and
neglect the explicit time-dependent term. Consequently,
Eq.~\eqref{accelaration} reduces to
\begin{equation}
\label{accelaration_assume}
\vec{a}
\approx
c(\vec{v}\cdot\nabla)\hat{n}.
\end{equation}
Since the emitting particles are ultra-relativistic, we further
consider $\vec{v}\approx c\hat{n}$. More sophisticated polarimetric
models that incorporate relativistic aberration, time-delay effects, and
realistic magnetospheric current structures have also been discussed
in the literature \citep[e.g.,][]{BCW1991,WWH2014}. The present
framework is intended to provide a simple baseline model for studying
the effect of the gravitational field on the polarization direction.

\subsection{Construction of the initial direction of the polarization vector from RVM geometry}
\label{subsec:analytical_polarization_m_frame}
To evaluate the curvature acceleration given by
Eq.~\eqref{accelaration_assume}, we introduce a corotating Cartesian
reference frame $X_mY_mZ_m$ (hereafter the magnetic or $m$-frame).
The $Z_m$-axis is aligned with the magnetic axis $\hat{m}_I$, and the
angle between the $Z_m$- and $Z_I$-axes is $\alpha$. The direction of
the $X_m$-axis is chosen such that, when the $m$-frame is translated to
the origin of the $I$-frame without changing the orientation of its
axes, the projection of the $X_m$-axis onto the $X_IY_I$ plane makes an
angle $\Phi_m$ with the $X_I$-axis. The rotational phase is given by
$\Phi_m=|\vec{\Omega}_p|t$, where $t$ is the time measured from the
instant at which the magnetic axis lies in the $X_IZ_I$ plane. Following \cite{DB26}, an arbitrary vector $\vec{V}_m$ defined in the $m$-frame can be
transformed into the pulsar's inertial $I$-frame according to
\begin{equation}
\label{Transfrom}
\vec{V}_{\rm I}
=
R_{\rm Z}(\Phi_m)
R_{\rm Y}(\alpha)
\vec{V}_{\rm m},
\end{equation}
where $R_{\rm Z}(\theta)$ and $R_{\rm Y}(\theta)$ represent the
rotation matrices about the $Z$- and $Y$-axes, respectively, by an
angle $\theta$.

We consider an arbitrary light ray whose direction in the $I$-frame is
specified by its colatitude $\zeta_L$ and longitude $\Phi_L$. Here,
$\zeta_L$ is the angle between the light-ray direction and the spin
axis of the pulsar, while $\Phi_L$ is the angle between the $X_I$-axis
and the projection of the light-ray direction onto the $X_IY_I$ plane
\citep{DBB23}. The unit vector along the direction of the light ray can
therefore be written as
\begin{equation}
\hat{n}_{\rm I}
=
[
\sin\zeta_L\cos\Phi_L,
\sin\zeta_L\sin\Phi_L,
\cos\zeta_L
].
\end{equation}

The corresponding direction in the corotating $m$-frame is obtained by
applying the inverse transformation:
\begin{equation}
\label{Transfrom_tangent}
\hat{n}_{\rm m}
=
R_{\rm Y}(-\alpha)
R_{\rm Z}(-\Phi_m)
\hat{n}_{\rm I}.
\end{equation}

Assuming a pure static dipolar magnetic-field configuration, the unit
tangent vector to a magnetic field line at the emission point can be
written as
\begin{equation}
\label{tangent}
\hat{n}_m
=
\frac{1}{\mathcal{N}}
\left(
2\cos\theta_m\,\hat{r}_m
+
\sin\theta_m\,\hat{\theta}_m
\right),
\end{equation}
where $\hat{r}_m$ and $\hat{\theta}_m$ are the radial and polar unit
basis vectors, respectively. The angle $\theta_m$ denotes the polar
coordinate of the emission point in the $m$-frame, and the
normalization factor is given by
\begin{equation}
\mathcal{N}=\sqrt{1+3\cos^2\theta_m}.
\end{equation}

The unit basis vectors can be written in terms of the Cartesian basis
vectors of the $m$-frame as
\begin{align}
\label{sp_to_cart}
\hat{r}_m
&=
\sin\theta_m\cos\phi_m\,\hat{X}_m
+
\sin\theta_m\sin\phi_m\,\hat{Y}_m
+
\cos\theta_m\,\hat{Z}_m,
\\
\hat{\theta}_m
&=
\cos\theta_m\cos\phi_m\,\hat{X}_m
+
\cos\theta_m\sin\phi_m\,\hat{Y}_m
-
\sin\theta_m\,\hat{Z}_m.
\end{align}

Equations~\eqref{Transfrom_tangent} and \eqref{tangent} describe the
same light-ray direction in the $m$-frame. By comparing these two
expressions, we determine the polar and azimuthal coordinates
$(\theta_m,\phi_m)$ of the emission point corresponding to the
specified light-ray direction.

We next calculate the curvature acceleration at this emission point.
Using the ultra-relativistic approximation
$\vec{v}\approx c\hat{n}_m$ in Eq.~\eqref{accelaration_assume}, the
acceleration in the corotating $m$-frame can be written as
\begin{equation}
\vec{a}_{m}
=
c(\vec{v}\cdot\nabla)\hat{n}_m
=
c^2(\hat{n}_m\cdot\nabla)\hat{n}_m.
\end{equation}

By evaluating the spatial derivatives of the dipolar field
configuration given in Eq.~\eqref{tangent}, at the radio emission
height $|\vec{r}_0|=h_{\nu,\mathrm{em}}$, the curvature acceleration
becomes
\begin{align}
\vec{a}_{m}
=
\frac{c^2\sin\theta_m}
{\mathcal{N} h_{\nu,\mathrm{em}}}
\biggl[
&
-\left(
\frac{2\sin\theta_m}{\mathcal{N}^3}
+
\frac{\sin\theta_m}{\mathcal{N}}
\right)\hat{r}_m
\nonumber
\\
&
+
\left(
\frac{4\cos\theta_m}{\mathcal{N}^3}
+
\frac{2\cos\theta_m}{\mathcal{N}}
\right)\hat{\theta}_m
\biggr].
\end{align}

Using the expressions for the spherical basis vectors given in
Eq.~\eqref{sp_to_cart}, the acceleration vector can be expressed in
terms of the Cartesian basis vectors of the $m$-frame. The resulting
acceleration vector is then transformed into the pulsar's inertial
frame using Eq.~\eqref{Transfrom}:
\begin{equation}
\label{Transfrom_accel}
\vec{a}_{\rm I}
=
R_{\rm Z}(\Phi_m)
R_{\rm Y}(\alpha)
\vec{a}_{\rm m}.
\end{equation}

In the present simplified model, the polarization direction of the
emitted light ray is taken to be along the acceleration vector of the
radiating particles. The corresponding unit polarization vector in the
$I$-frame is therefore given by
\begin{equation}
\hat{f}_{\rm I}
=
\frac{\vec{a}_{\rm I}}
{|\vec{a}_{\rm I}|}.
\end{equation}

It is worth noting that along the exact magnetic axis
($\theta_m=0$), the spatial curvature of the dipolar magnetic field
lines vanishes and, consequently, the curvature acceleration becomes
zero. The polarization direction defined through the acceleration
vector is therefore mathematically undefined at this boundary. Since
the present model is based on curvature radiation, we restrict the
numerical calculations to regions of finite magnetic colatitude, i.e.,
$\theta_m>0$.

\section{Binary Pulsar Setup}
\label{binary}

\subsection{Orbital Geometry and Coordinate Systems}
\label{subsec:orbitalgeometrymotion}

To describe the geometry of a pulsar in a binary system, we introduce
several Cartesian coordinate systems. These frames are related to each
other by rotations and parallel translations and are used to describe
the orbital motion, the orientation of the pulsar spin axis, and the
direction of the emitted light rays. These frames are described in detail in \cite{DB26}.

We first define the $X_I^\prime Y_I^\prime Z_I^\prime$ frame, which is
obtained by translating the origin of the pulsar's inertial
$X_IY_IZ_I$ frame to the barycenter of the binary system. The two
frames have parallel axes, i.e.,
$X_I^\prime\parallel X_I$,
$Y_I^\prime\parallel Y_I$, and
$Z_I^\prime\parallel Z_I$. Consequently, the spin axis of the pulsar,
denoted by the unit vector $\widehat{S}_p$, is parallel to the
$Z_I^\prime$-axis.

We next introduce the observer-oriented $X_sY_sZ_s$ frame, hereafter
referred to as the $s$-frame. Its origin is located at the barycenter
of the binary, and the $Z_s$-axis is directed opposite to the line of
sight (LoS), i.e., from the observer towards the binary. The
$X_sY_s$ plane therefore represents the plane of the sky. The
$X_s$-axis is chosen along the line of the ascending node (AN).

The orbital motion of the binary is described in the $X_bY_bZ_b$
frame or in the $b$-frame. The origin of this frame is also at the
barycenter and the orbital plane coincides with the $X_bY_b$ plane.
The $X_b$-axis is directed towards the periastron and the $Z_b$-axis
is parallel to the orbital angular momentum. The inclination angle of
the orbit, $i$, is the angle between the $Z_s$- and $Z_b$-axes, while
$\omega$ denotes the longitude of periastron. The position of the
pulsar along the orbit is specified by its true anomaly $A_T$, and
the corresponding orbital phase is $A_T+\omega$.

\begin{figure*}[htbp]
\centering
\includegraphics[width=0.8\textwidth]{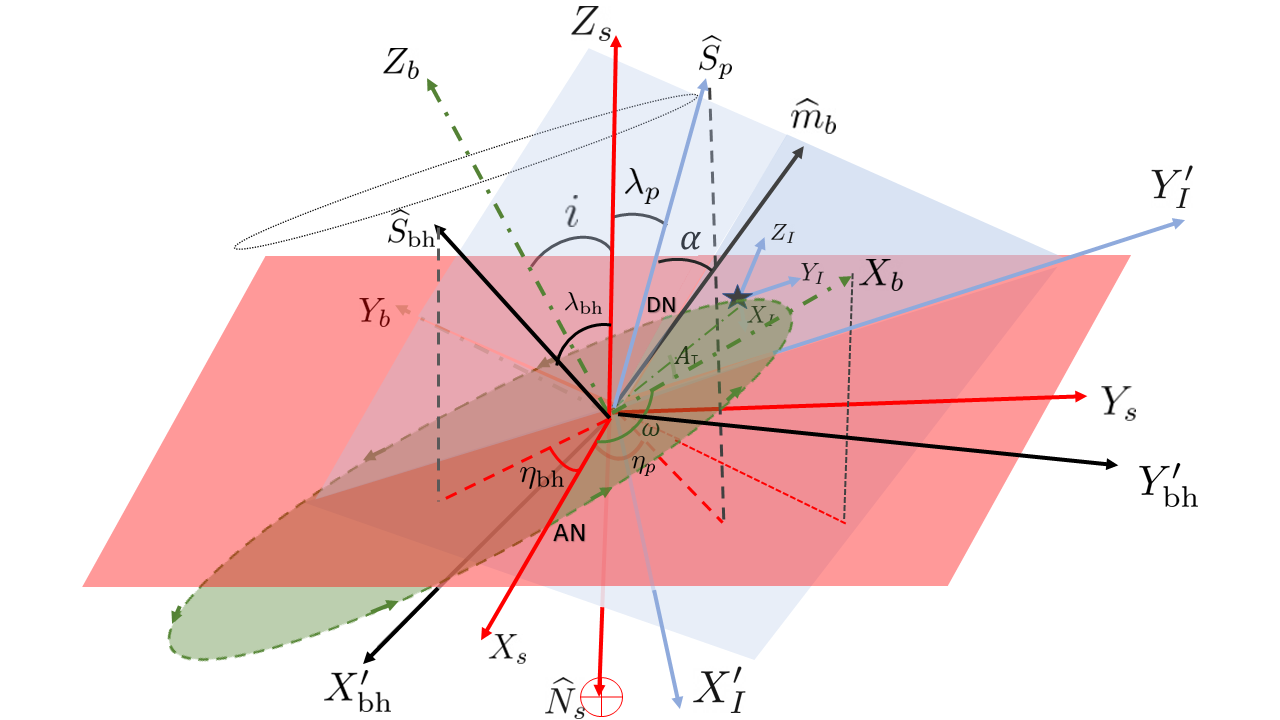}
\caption{The orbital geometry of a pulsar (marked by a $\star$) in a binary system. The sky plane, corresponding to the ${\rm X_s}{\rm Y_s}$ plane, is shown in light red, whereas the orbital plane, corresponding to the ${\rm X_b Y_b}$ plane, is shown in light green. The blue plane denotes the ${\rm X_I^{\prime} Y_I^{\prime}}$ plane. The unit vectors along the spin axes of the pulsar and the companion are represented by $\widehat{S}_p$ and $\widehat{S}_{\rm bh}$, respectively, and are shown after being shifted parallel to the barycenter. The ascending and descending nodes are labeled as AN and DN, respectively. Here, $\omega$ denotes the longitude of periastron, $A_T$ the true anomaly, and $i$ the inclination angle. The angles $\lambda_p$, $\eta_p$, $\lambda_{\rm bh}$, and $\eta_{\rm bh}$ are defined in the text. This figure is taken from \citep{DB25}.}
\label{frames}
\end{figure*}

Figure~\ref{frames} illustrates the relative orientation of these
coordinate systems. The plane of the sky, represented by the
$X_sY_s$ plane, and the orbital plane, represented by the $X_bY_b$
plane, are shown together with the translated inertial plane
$X_I^\prime Y_I^\prime$. The ascending and descending nodes are
marked by AN and DN, respectively, and the pulsar is indicated by a
star.

For convenience, we also introduce two additional coordinate systems
centered on the two members of the binary. The $X_pY_pZ_p$ frame is
centered on the pulsar and is referred to as the $p$-frame, while the
$X_cY_cZ_c$ frame is centered on the companion and is referred to as the
$c$-frame. Both frames are obtained from the $b$-frame by parallel
translations along the line joining the two bodies. Hence, their axes
remain parallel to those of the $b$-frame. As a result, any direction
vector has the same components in the $b$-, $p$-, and $c$-frames.

For a rotating companion, we additionally define a companion-fixed
$X_{\rm bh}Y_{\rm bh}Z_{\rm bh}$ frame, hereafter referred to as the
bh-frame. The $Z_{\rm bh}$-axis is aligned with the spin axis of the
companion $\widehat{S}_{\rm bh}$, while the $Y_{\rm bh}$-axis is
chosen to lie in the plane of the sky. A translated
$X_{\rm bh}^\prime Y_{\rm bh}^\prime Z_{\rm bh}^\prime$ frame is
obtained by shifting the bh-frame origin to the barycenter without
changing the orientation of its axes. The $c$- and bh-frames therefore
have the same origin but, in general, different orientations. The
$Z_c$-axis is parallel to the orbital angular momentum, whereas the
$Z_{\rm bh}$-axis is aligned with the companion spin.

For a pulsar-BH binary, the mass of the black hole is expected
to be much larger than that of the pulsar. Consequently, the
barycenter lies close to the center of the black hole and the $b$- and
$c$-frames are nearly coincident. This approximation is useful for
describing the propagation of the emitted radiation in the
gravitational field of the companion.

The orientation of the pulsar spin axis with respect to the observer
is specified by two angles, $\lambda_p$ and $\eta_p$. The angle
$\lambda_p$ describes the inclination of the pulsar spin axis with
respect to the $Z_s$-axis, while $\eta_p$ specifies the orientation of
its projection onto the sky plane measured from the $X_s$-axis. The
corresponding orientation of the companion spin is described by
$\lambda_{\rm bh}$ and $\eta_{\rm bh}$. In particular,
$\eta_{\rm bh}$ determines the orientation of the $Y_{\rm bh}$-axis
with respect to the sky frame.

The position of the pulsar in the $b$-frame is obtained from the
orbital elements. Let $\vec{r}_b$ denote the position vector of the
pulsar with respect to the barycenter and $\widehat{r}_b$ the
corresponding unit vector. The distance between the pulsar and the
companion can then be written as
\begin{equation}
\label{eq:separation}
|\vec{r}_c|
=
|\vec{r}_{\rm bh}|
=
\left(1+\frac{M_p}{M_c}\right)|\vec{r}_b|,
\end{equation}
where $M_p$ and $M_c$ denote the masses of the pulsar and the
companion, respectively. Since the $b$- and $c$-frames are parallel,
the corresponding unit vectors satisfy
$\widehat{r}_c=\widehat{r}_b$.

The position vector in the companion spin-aligned frame is obtained
through the following sequence of rotations:
\begin{equation}
\label{eq:r_in_bh_frame}
\vec{r}_{\rm bh}
=
R_y(-\lambda_{\rm bh})
R_z(-\eta_{\rm bh})
R_x(i)
R_z(\omega)
\vec{r}_c ,
\end{equation}
while its representation in the sky frame is
\begin{equation}
\label{eq:r_in_s_frame}
\vec{r}_{s}
=
R_x(i)
R_z(\omega)
\vec{r}_b .
\end{equation}

The unit vectors required in the subsequent calculation can be
written directly in terms of the corresponding angular coordinates as
\begin{equation}
\label{eq:Sbh_with_angles}
\widehat{S}_{\rm bh}
=
[
\cos\eta_{\rm bh}\sin\lambda_{\rm bh},
\sin\eta_{\rm bh}\sin\lambda_{\rm bh},
\cos\lambda_{\rm bh}
],
\end{equation}
\begin{equation}
\label{eq:rb_with_angles}
\widehat{r}_{b}
=
[
\cos A_T,
\sin A_T,
0
],
\end{equation}
and
\begin{equation}
\label{eq:Ns_vector}
\widehat{N}_{s}
=
[0,0,-1].
\end{equation}

We now introduce a ray-dependent reference frame, referred to as the
$T$-frame and denoted by $X_TY_TZ_T$. This frame is constructed using
the position and initial propagation direction of an individual light
ray. The $Y_T$-axis is chosen along $\widehat{r}_b$, while the
$Z_T$-axis is directed along
$\widehat{r}_b\times\widehat{n}_b$, where $\widehat{n}_b$ is the
initial direction of the light ray in the $b$-frame.

The initial propagation direction of a light ray is first obtained in
the pulsar's inertial frame from the emission
geometry discussed in Section~\ref{bema_polarization_geometry}. It is
then transformed into the $b$-frame as
\begin{equation}
\label{eq:n_in_b_frame}
\widehat{n}_{b}
=
R_z(-\omega)
R_x(-i)
R_z(\eta_p)
R_y(\lambda_p)
\widehat{n}_{\rm I}.
\end{equation}

With the above choice of the $T$-frame, the pulsar lies along the
$Y_T$-axis. This construction is particularly convenient for tracing
the light ray because the initial position and propagation direction
then lie in a common plane.

The angle between the pulsar spin axis and the orbital angular
momentum is denoted by $\xi_p$. From the geometry of the system, this
angle can be expressed in terms of the orientation angles introduced
above as
\begin{equation}
\label{eq:xip}
\xi_p
=
\cos^{-1}
\left(
\cos\lambda_p\cos i
-
\sin\lambda_p\sin\eta_p\sin i
\right).
\end{equation}

\subsection{Orbital and Spin Evolution}
\label{subsec:orbitandspinevolution}

The orientation of the pulsar and the companion does not necessarily
remain fixed during the evolution of a relativistic binary. In
addition to the secular evolution of the orbital parameters due to
gravitational-wave emission, relativistic spin-orbit coupling, and possibly spin-spin interaction, can
produce precession of the spin axes and changes in the orbital
orientation. These effects can therefore modify the geometry of the
pulsar beam and, consequently, the direction of the emitted radiation.

For the orbital evolution, we consider the secular changes in the
orbital period and eccentricity caused by gravitational-wave emission.
The corresponding expressions for $\dot{P}_b$ and $\dot{e}$ are
adopted from the standard treatment of gravitational radiation
reaction \citep{peters1964,Lorimer2004}. Since the present calculation
tracks the binary over a finite interval, these secular changes are
implemented by updating the orbital parameters with time.

The spin evolution is treated separately. In the present work, we
consider systems in which the spin axis of the black hole is aligned
with the orbital angular momentum. Under this assumption, the
black-hole spin does not undergo an appreciable change in orientation.
However, the pulsar spin can precess around the orbital angular
momentum due to relativistic spin-orbit coupling.

The evolution of the pulsar spin vector can be written as
\begin{equation}
\frac{d\overrightarrow{\mathcal{S}}_{p}}{dt}
=
\overrightarrow{\Omega}_{\mathcal{S}_p}
\times
\overrightarrow{\mathcal{S}}_{p},
\end{equation}
where $\overrightarrow{\mathcal{S}}_p$ is the pulsar spin vector and
$\overrightarrow{\Omega}_{\mathcal{S}_p}$ is the corresponding
precession frequency. Up to the order considered here, we write
\begin{equation}
\label{eq:spin_prec}
\overrightarrow{\Omega}_{\mathcal{S}_p}
=
\overrightarrow{\Omega}_{\rm 1PN_p}
+
\overrightarrow{\Omega}_{\rm LT_p},
\end{equation}
where the first term represents the standard relativistic spin
precession and the second term describes the Lense-Thirring
contribution.

Following \citep{BarkerOconnell1975}, the leading spin-precession
term is written as
\begin{equation}
\label{eq:spinprecfreq1PN}
\overrightarrow{\Omega}_{\rm 1PN_p}
=
A_{\rm 1PN_p}\,\widehat{L},
\end{equation}
where $\widehat{L}$ is the unit vector along the orbital angular
momentum and
\begin{equation}
\label{eq:spinprec1PN}
A_{\rm 1PN_p}
=
\left(\frac{G}{c^3}\right)^{2/3}
\frac{\sigma_b^{5/3}}{1-e^2}
\frac{M_c(4M_p+3M_c)}
{2(M_p+M_c)^{4/3}} .
\end{equation}
Here, $e$ is the orbital eccentricity and
\begin{equation}
\sigma_b=\frac{2\pi}{P_b}.
\end{equation}

The Lense-Thirring contribution to the pulsar spin precession is
given by
\begin{equation}
\label{eq:spinprecfreqLT}
\overrightarrow{\Omega}_{\rm LT_p}
=
A_{\rm LT_p}
\left[
\widehat{S}_{\rm bh}
-
3
(\widehat{L}\cdot\widehat{S}_{\rm bh})
\widehat{L}
\right].
\end{equation}
Under the assumption adopted here that
$\widehat{S}_{\rm bh}\parallel\widehat{L}$, this expression reduces to
\begin{equation}
\overrightarrow{\Omega}_{\rm LT_p}
=
-2A_{\rm LT_p}\widehat{L},
\end{equation}
where
\begin{equation}
\label{eq:spinprecLT}
A_{\rm LT_p}
=
\frac{G}{c^3}
\tilde{a}
\frac{\sigma_b^2}{(1-e^2)^{3/2}}
\frac{M_c^2}{2(M_p+M_c)}.
\end{equation}
The dimensionless parameter $\tilde{a}$ describes the spin of the
black hole and is related to its angular momentum by
\begin{equation}
\mathcal{S}_{\rm bh}
=
\tilde{a}\frac{GM_c^2}{c}.
\end{equation}
The positive and negative values of $\tilde{a}$ correspond to the prograde and
retrograde spin orientations, respectively, with
$|\tilde{a}|\leq1$.

The spin-orbit coupling also contributes to the secular advance of
the periastron. We write the total rate of periastron advance as
\begin{equation}
\label{eq:per_adv}
\dot{\omega}
=
\dot{\omega}_{\rm PN}
+
\dot{\omega}_{\rm LT,p}
+
\dot{\omega}_{\rm LT,c},
\end{equation}
where the first term denotes non-spinning post-Newtonian
and the remaining terms arise from the spin-orbit coupling of the
pulsar and the companion.

The post-Newtonian contribution is written as
\begin{equation}
\label{eq:per_advPN}
\dot{\omega}_{\rm PN}
=
\frac{3\beta_0^2\sigma_b}{1-e^2}
\left[
1+\beta_0^2 f_0
\right],
\end{equation}
with
\begin{equation}
\label{eq:beta0}
\beta_0
=
\frac{(GM_T\sigma_b)^{1/3}}{c},
\end{equation}
where
\begin{equation}
M_T=M_p+M_c.
\end{equation}
The mass-ratio factor $f_0$ is given by
\begin{multline}
\label{eq:f0p}
f_0
=
\frac{1}{1-e^2}
\left(
\frac{39}{4}\mathscr{X}_p^2
+
\frac{27}{4}\mathscr{X}_c^2
+
15\mathscr{X}_p\mathscr{X}_c
\right)
\\
-
\left(
\frac{13}{4}\mathscr{X}_p^2
+
\frac{1}{4}\mathscr{X}_c^2
+
\frac{13}{3}\mathscr{X}_p\mathscr{X}_c
\right),
\end{multline}
where
\begin{equation}
\mathscr{X}_p=\frac{M_p}{M_T},
\qquad
\mathscr{X}_c=\frac{M_c}{M_T}.
\end{equation}

The spin-orbit contributions to the periastron advance are as follows:
\begin{subequations}
\begin{equation}
\label{eq:per_advLTp}
\dot{\omega}_{\rm LT,p}
=
-
\frac{3\beta_0^3\sigma_b}{1-e^2}
\left[
\beta_{sp}g_{sp}
\right],
\end{equation}
and
\begin{equation}
\label{eq:per_advLTc}
\dot{\omega}_{\rm LT,c}
=
-
\frac{3\beta_0^3\sigma_b}{1-e^2}
\left[
\tilde{a}g_{sc}
\right].
\end{equation}
\label{eq:per_advLT}
\end{subequations}

The dimensionless spin parameter of the pulsar is
\begin{equation}
\label{eq:per_advbetasp}
\beta_{sp}
\frac{c\mathcal{S}_p}{GM_p^2}
=
\frac{2\pi c I_p}{GM_p^2P_s},
\end{equation}
where
\begin{equation}
\mathcal{S}_p
=
\frac{2\pi I_p}{P_s}
\end{equation}
is the magnitude of the pulsar's spin angular momentum. Here, $I_p$
denotes the moment of inertia of the pulsar. Its value depends on the
mass, spin period, and equation of state of the neutron star
\citep[e.g.,][]{bagchi2010}. For a canonical $1.8\,M_\odot$ pulsar,
we adopt $I_p=1.8\times10^{45}\ {\rm g\,cm^2}$.

Following \citep{bagchi2017}, the geometrical factor associated with
the pulsar spin is
\begin{equation}
\label{eq:per_advgsp}
g_{sp}
=
\frac{
\mathscr{X}_p
(4\mathscr{X}_p+3\mathscr{X}_c)
}{
6(1-e^2)^{1/2}\sin^2 i
}
\left[
(3\sin^2 i-1)\cos\xi_p
+
\cos i\cos\lambda_p
\right].
\end{equation}
The angle $\xi_p$ is the angle between the pulsar spin axis and the
orbital angular momentum, as defined in
Section~\ref{subsec:orbitalgeometrymotion}.

For the companion, the alignment
$\widehat{S}_{\rm bh}\parallel\widehat{L}$ simplifies the corresponding
geometrical factor to
\begin{equation}
\label{eq:per_adgsc}
g_{sc}
=
\frac{
\mathscr{X}_c
(4\mathscr{X}_c+3\mathscr{X}_p)
}{
3(1-e^2)^{1/2}
}.
\end{equation}

The Lense-Thirring interaction can also modify the orbital
inclination. Since the black-hole spin is assumed to be parallel to
the orbital angular momentum, its contribution to the evolution of
$i$ vanishes. The remaining contribution from the pulsar spin is
given by \citep{damourschafer1988} as
\begin{equation}
\label{eq:idot}
\frac{di}{dt}
=
\frac{\sigma_b^2}
{2(M_p+M_c)c^2}
\sum_{j={\rm p,bh}}
\vec{\mathcal{S}}_j\cdot\widehat{\Upsilon}
=
\frac{\pi I_p\sigma_b^2}
{(M_p+M_c)c^2}
\widehat{\mathcal{S}}_{\rm p}
\cdot
\widehat{\Upsilon},
\end{equation}
where $\widehat{\Upsilon}$ is a unit vector along the line of nodes.
The evolution of $i$ changes the orientation of the orbital angular
momentum and therefore also affects the direction about which the
pulsar spin precesses.

The evolution of the pulsar spin modifies its orientation with
respect to the observer. In particular, the angles $\lambda_p$ and
$\eta_p$ become time dependent. Through Eq.~\eqref{eq:xip}, these
changes also modify $\xi_p$ and consequently the spin-dependent
contribution to the periastron advance.

For sufficiently compact binaries, higher-order spin and orbital
coupling terms may become important \citep{barker1982, majar2009, majar2012}. Such effects are particularly
relevant for systems containing a millisecond pulsar orbiting a
supermassive black hole.

\section{Polarization Rotation Induced by Gravitational Bending}
\label{definitions}

The light rays emitted by the pulsar propagate through the curved
spacetime generated by the companion. The curvature of spacetime
affects both the propagation time and the direction of the light rays.
These effects become most significant near superior conjunction,
where the pulsar lies closest to the line of sight to the companion.
For an edge-on orbit, this configuration corresponds to
$i=90^\circ$ and $A_T+\omega=90^\circ$. The change in the propagation
direction of the light rays gives rise to latitudinal and longitudinal
bending delays \citep{sch90, dk1995, RL06, Rafikov2005}. These delays arise because the
latitude and longitude of a light ray reaching the observer after
gravitational bending differ from those of the corresponding ray in
the absence of gravitational bending \citep{DBB23}.

According to the Rotating Vector Model (RVM), the polarization direction of the radiation emitted by the pulsar depends on the direction of the emitted light ray relative to the magnetic axis \citep{lpp94}. Gravitational bending changes the emission direction required for the ray to reach the observer, and consequently, the polarization direction. In addition to this geometric effect, the polarization vector is parallel transported along the null geodesic in curved spacetime. Since the initial light ray is not directed toward the observer, its polarization vector at the emission point does not, in general, lie in the observer's sky plane, whereas the final polarization vector does. Therefore, the initial and final polarization vectors cannot be directly compared to separate the effect of parallel transport. To distinguish the geometric contribution from the additional polarization rotation associated with propagation in curved spacetime, we construct a reference final polarization vector by connecting the initial and final ray directions through a sequence of rotations. The difference between the reference final polarization direction and the direction obtained by parallel transport defines the polarization transport rotation relative to our chosen reference construction.

To quantify these effects, we consider the polarization vectors in the $X_sY_s$ sky plane, defined in Section~\ref{subsec:orbitalgeometrymotion}. Let $\widehat{f}_{\rm sky,s}$ denote the polarization vector of the unbent fiducial ray at the reference rotational phase $\Phi_m=0$. This vector lies in the $X_sY_s$ plane and defines the fiducial polarization direction. We denote the initial polarization vector of a ray emitted from a different location within the pulsar emission beam and subsequently deflected towards the observer by $\widehat{f}^{\prime}_{\rm sky,s}$. In general, this vector does not lie in the observer's sky plane. We construct the reference final polarization vector $\widehat{f}^{\prime}_{\rm sky,f}$ by connecting the initial and final ray directions through a sequence of rotations. This construction preserves the polarization components parallel and perpendicular to the corresponding initial and final propagation planes of the light ray.\footnote{We define the propagation plane of the light ray as the plane containing the pulsar, the companion, and the light ray. This choice is well defined in Schwarzschild spacetime but not globally in Kerr spacetime. Defining a global reference plane in Kerr spacetime would require constructing a local triad along the null geodesic. Our choice provides a practical reference convention, although it does not guarantee a globally defined propagation plane \citep{Brodutch:2011}.}

\begin{figure*}[htbp!]
\centering
\includegraphics[width=\textwidth]{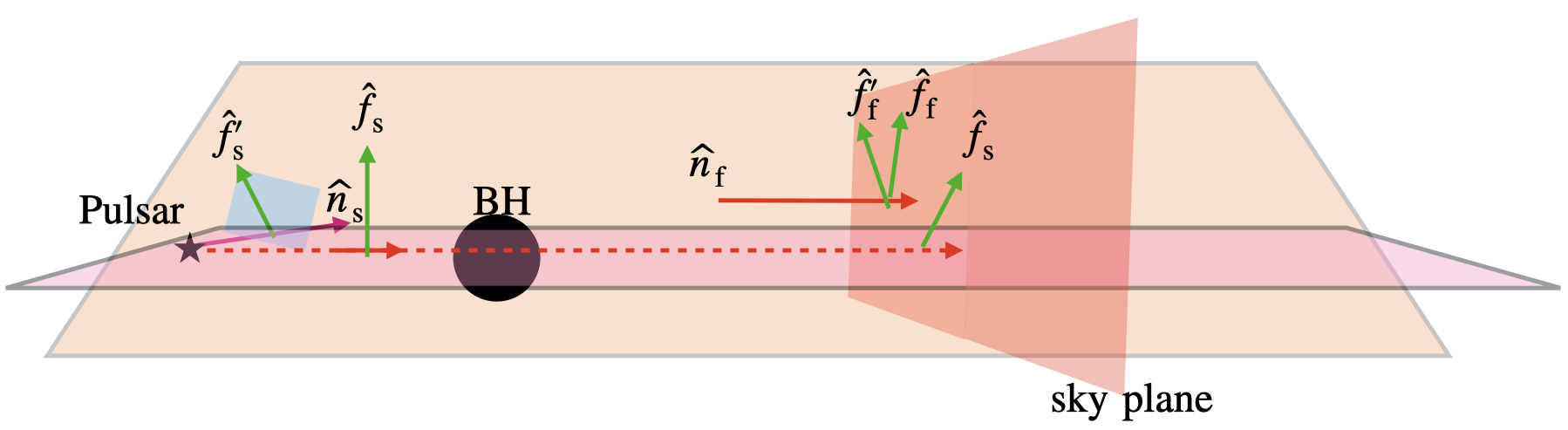}
\caption{Schematic diagram illustrating the geometry and the polarization
vectors used in this work. The pulsar is marked by a star, and the
companion black hole is denoted by BH. The red plane represents the
observer's sky plane. The fiducial unbent light ray is shown by the
dashed line, with its polarization vector $\widehat{f}_{\rm sky,s}$
lying in the sky plane. The initial propagation direction of the light
ray that reaches the observer after gravitational bending is denoted
by $\widehat{n}_s$, and its corresponding polarization vector is
$\widehat{f}'_{\rm sky,s}$. The blue plane is perpendicular to
$\widehat{n}_s$ and contains $\widehat{f}'_{\rm sky,s}$. The initial
and final propagation planes are shown in magenta and orange,
respectively. The initial plane of propagation contains the pulsar the companion and the initial light ray
while the final plane of propagation contains the pulsar the companion and the final light ray direction
$\widehat{n}_{\rm f}$. The reference polarization vector constructed
using the sequence of rotations described in the text is denoted by
$\widehat{f}'_{\rm sky,f}$, whereas $\widehat{f}_{\rm sky,f}$ denotes
the polarization vector obtained by parallel transport along the
curved-spacetime null geodesic. Both final polarization vectors lie in
the observer's sky plane, allowing their polarization angles to be
compared.}
\label{fig:schematic_pol_vecs}
\end{figure*}

The reference vector $\widehat{f}^{\prime}_{\rm sky,f}$ provides a geometric reference for comparing the final polarization directions. It is obtained by connecting the initial and final ray directions through a sequence of rotations, rather than by defining a reference triad at each point along the null geodesic. We denote the final polarization vector obtained by parallel transport along the curved-spacetime null geodesic by $\widehat{f}_{\rm sky,f}$. Both final polarization vectors lie in the observer's sky plane, allowing their PPAs to be defined and compared. The details of the reference construction are described in Section~\ref{subsec:polarization_transport}. The geometry and the polarization vectors used in these definitions are illustrated schematically in Fig.~\ref{fig:schematic_pol_vecs}.

We define the polarization position angle (PPA) of the fiducial
unbent ray as
\begin{equation}
\label{eq:psi_s}
\psi_s
=
\operatorname{atan2}
\left(
\widehat{f}_{{\rm sky,s}_y},
\widehat{f}_{{\rm sky,s}_x}
\right),
\end{equation}
where $\operatorname{atan2}(y,x)$ denotes the four-quadrant inverse
tangent function. The PPA of the final polarization vector obtained
from the reference construction is
\begin{equation}
\label{eq:psi_s_prime}
\psi'_f
=
\operatorname{atan2}
\left(
\widehat{f}^{\prime}_{{\rm sky,f}_y},
\widehat{f}^{\prime}_{{\rm sky,f}_x}
\right).
\end{equation}

The difference between these two angles characterizes the
geometric polarization shift associated with the change in the
emission geometry caused by gravitational bending. We define this
quantity as
\begin{equation}
\label{eq:psi_geo}
\Delta\psi_{\rm geo}
=
\psi_s-\psi'_f.
\end{equation}
Here, the reference construction provides the basis
for determining the change associated with the different emission
geometry, without including the effect of polarization transport
in curved spacetime.

For the bent ray, the final polarization vector obtained by parallel
transport in curved spacetime is denoted by
$\widehat{f}_{\rm sky,f}$. Its PPA is
\begin{equation}
\label{eq:psi_o}
\psi_f
=
\operatorname{atan2}
\left(
\widehat{f}_{{\rm sky,f}_y},
\widehat{f}_{{\rm sky,f}_x}
\right).
\end{equation}
We define the total change in the polarization direction between
the fiducial unbent ray and the final bent ray as the
\textit{total relativistic polarization rotation},
$\Delta\psi_{\rm tot}$:
\begin{equation}
\label{eq:psi_tot}
\Delta\psi_{\rm tot}
=
\psi_s-\psi_f.
\end{equation}

The total change includes the geometric contribution and the
additional change associated with polarization transport in curved
spacetime. We therefore define the contribution associated with
curved-spacetime transport as
\begin{equation}
\label{eq:psi_pt}
\Delta\psi_{\rm pt}
=
\Delta\psi_{\rm tot}
-
\Delta\psi_{\rm geo}.
\end{equation}
Thus, $\Delta\psi_{\rm pt}$ measures the difference between the
final polarization directions obtained from the curved-spacetime
parallel transport and the reference construction.
This quantity is related to the polarization rotation associated
with propagation in a gravitational field, often referred to as
gravitational Faraday rotation \citep{Ishihara1987, Sereno2005, NouriZonoz1999}. However, polarization rotation in
a gravitational field depends on the choice of the local
polarization basis, and a geometric contribution associated with
the reference frame can also be present \citep{Brodutch:2011}.
Although our reference construction explicitly preserves the
polarization components parallel and perpendicular to the
corresponding propagation planes, we therefore do not interpret
$\Delta\psi_{\rm pt}$ as a pure gravitational Faraday rotation.
Instead, within the present construction, it represents a
contribution containing both geometric and gravitomagnetic
effects. A complete separation of these contributions would
require tracking the rotation of the polarization basis along the
entire null geodesic, for example by defining an appropriate
inertial or reference polarization triad at each point along the
trajectory \citep{Brodutch:2011}. For a non-rotating black hole
companion, $\Delta\psi_{\rm pt}=0$ in our setup, consistent with
the symmetry of the Schwarzschild spacetime.

The spin of the companion also affects the propagation of light
rays through the frame-dragging effect, thereby modifying their
directions \citep{BE22}. Light rays propagating in directions
co-rotating with the companion spin experience a different amount
of bending from those propagating in counter-rotating directions.
This difference modifies both the latitudinal and longitudinal
bending delays, giving rise to the frame-dragging bending delays
\citep{DB25}. Consequently, the companion spin can also modify the
observed polarization direction.

To quantify the spin-induced change in the geometric polarization
shift, we compare the results for a spinning companion with those
for a non-spinning companion. The latter corresponds to
$\tilde{a}=0$, where $\tilde{a}$ is the dimensionless spin
parameter of the companion. We define the spin-induced change in
the geometric contribution as the
\textit{frame-dragging geometric polarization shift},
$\Delta\psi_{\rm FD,geo}$:
\begin{equation}
\label{eq:psi_FD_geo}
\Delta\psi_{\rm FD,geo}
=
\left(\Delta\psi_{\rm geo}\right)_{\tilde{a}}
-
\left(\Delta\psi_{\rm geo}\right)_{0},
\end{equation}
where $\left(\Delta\psi_{\rm geo}\right)_{\tilde{a}}$ and
$\left(\Delta\psi_{\rm geo}\right)_{0}$ denote the geometric
polarization shifts for spinning and non-spinning companions,
respectively.

\section{Relativistic Light Propagation and Polarization Transport}
\label{subsec:bending_spin_full_GR}

\subsection{Relativistic Null Geodesics and Lensing}
\label{subsec:null_geodesics}

To describe the propagation of the radiation emitted by the pulsar in
the gravitational field of the companion, we solve the null geodesic
equations in the Kerr spacetime. The gravitational contribution of the pulsar to the spacetime is neglected as we are assuming a significantly more massive companion. We use Boyer--Lindquist coordinates
$x^\mu$, where $x^0=ct$, $x^1=r$, $x^2=\theta$, and $x^3=\phi$, and
trace the light ray from the emission point to an observer at
asymptotic infinity. The coordinate systems and their relative
orientations are defined in Section~\ref{subsec:orbitalgeometrymotion}.
In the following, the subscripts ``T'', ``bh'', ``sky'', and ``I'' denote
the ray-centered T-frame, the companion spin-aligned frame, sky frame and pulsar inertial frame. The subscript ``s'' and ``f'' denote the source,
and asymptotic infinity, respectively.

Following \citep{chandrasekhar84, kerr63, pro16, GSL20}, the equations governing the null geodesics in
the Kerr spacetime can be written in terms of separated radial and
latitudinal integrals as
\begin{equation}
\label{geodesic11}
I_r=G_\theta,
\end{equation}
and
\begin{equation}
\label{geodesic22}
\phi_f-\phi_s=I_\phi+D_\lambda G_\phi.
\end{equation}
The radial and latitudinal effective potentials are given by
\begin{subequations}
\label{parameters_geodesic}
\begin{align}
\label{R_rparticle}
\mathcal{R}(r)
={}&
r^4+
\left[
\left(\tilde{a}\frac{GM_c}{c^2}\right)^2
-D_\lambda^2-D_q
\right]r^2
\nonumber\\
&+
2\left(\frac{GM_c}{c^2}\right)
\left[
D_q+
\left\{
D_\lambda-
\left(\tilde{a}\frac{GM_c}{c^2}\right)
\right\}^2
\right]r
\nonumber\\
&-
\left(\tilde{a}\frac{GM_c}{c^2}\right)^2D_q
-
\left(\frac{M_c^2c^4}{E^2}\right)r^2\Delta ,
\end{align}
\begin{align}
\label{theta_thetaparticle}
\Theta(\theta)
={}&
D_q+
\left(\tilde{a}\frac{GM_c}{c^2}\right)^2\cos^2\theta
-D_\lambda^2\cot^2\theta
\nonumber\\
&-
\left(\frac{M_c^2c^4}{E^2}\right)
\left(\tilde{a}\frac{GM_c}{c^2}\right)^2
\cos^2\theta ,
\end{align}
\begin{equation}
\label{I_r}
I_r=
\fint_r
\frac{dr}
{{\rm sign}(p_r)\sqrt{\mathcal{R}(r)}},
\end{equation}
\begin{equation}
\label{I_phi}
I_\phi=
\fint_r
\frac{
\left(\tilde{a}\frac{GM_c}{c^2}\right)
\left[
(2GM_c/c^2)r-
D_\lambda\left(\tilde{a}\frac{GM_c}{c^2}\right)
\right]
}
{
{\rm sign}(p_r)\sqrt{\mathcal{R}(r)}\Delta
}
\,dr,
\end{equation}
\begin{equation}
\label{G_theta}
G_\theta=
\fint_\theta
\frac{d\theta}
{{\rm sign}(p_\theta)\sqrt{\Theta(\theta)}},
\end{equation}
\begin{equation}
\label{G_phi}
G_\phi=
\fint_\theta
\frac{\csc^2\theta}
{{\rm sign}(p_\theta)\sqrt{\Theta(\theta)}}
\,d\theta,
\end{equation}
\begin{equation}
\label{D_lambda}
D_\lambda=\frac{cL_z}{E},
\end{equation}
\begin{equation}
\label{D_q}
D_q=\frac{\mathcal{L}}{E^2}.
\end{equation}
\end{subequations}

Here, $E$, $L_z$, and $\mathcal{L}$ are the conserved energy, axial
angular momentum, and Carter constant of the photon, respectively.
The quantities $D_\lambda$ and $D_q$ are the corresponding normalized
constants of motion. The mass of the companion is denoted by $M_c$,
and its spin is described by the dimensionless parameter $\tilde{a}$,
which is related to its angular momentum through
\begin{equation}
\mathcal{S}_{\rm bh}
=
\tilde{a}\frac{GM_c^2}{c}.
\end{equation}
The photon four-momentum is denoted by
$p_\mu=(p_t,p_r,p_\theta,p_\phi)$.

The metric functions appearing in the above equations are
\begin{equation}
\label{def_deltakerr}
\Delta
=
r^2-
2\frac{GM_c}{c^2}r+
\left(\tilde{a}\frac{GM_c}{c^2}\right)^2,
\end{equation}
\begin{equation}
\label{def_rhokerr}
\rho^2
=
r^2+
\left(\tilde{a}\frac{GM_c}{c^2}\right)^2\cos^2\theta,
\end{equation}
\begin{equation}
\label{def_sigmakerr}
\Sigma^2
=
\left[
r^2+
\left(\tilde{a}\frac{GM_c}{c^2}\right)^2
\right]^2
-
\left(\tilde{a}\frac{GM_c}{c^2}\right)^2
\Delta\sin^2\theta,
\end{equation}
and
\begin{equation}
\label{def_Omegakerr}
\Omega
=
\frac{
2\left(\frac{GM_c}{c^2}\right)r
\left(\tilde{a}\frac{GM_c}{c^2}\right)
}
{\Sigma^2}.
\end{equation}

The constants of motion are determined from the initial conditions of
the light ray at the emission point. As described in
Section~\ref{subsec:orbitalgeometrymotion}, the initial direction of
the ray is specified in the T-frame. In this frame,
$p_{\theta T,s}=0$. If $\chi$ denotes the angle between the orbital
position vector $\vec{r}_b$ and the initial propagation direction
$\widehat{n}_b$, then
\begin{equation}
\frac{p_{\phi T,s}}{p_{tT,s}}
=
-\sin\chi .
\end{equation}

We define the normalized angular momentum vector at the source as
\begin{equation}
\vec{\tilde{L}}_{T,s}
=
\frac{\vec{L}_{T,s}}
{|\vec{r}_b|p_{tT,s}}.
\end{equation}
Its Cartesian components in the T-frame are
\begin{subequations}
\label{eq:cartesianangularmomentumtildeTframe}
\begin{equation}
\tilde{L}_{X_{T,s}}
=
-
\left[
\frac{p_{\phi T,s}}{p_{tT,s}}
\cos\theta_{T,s}\cos\phi_{T,s}
+
\frac{p_{\theta T,s}}{p_{tT,s}}
\sin\phi_{T,s}
\right],
\end{equation}
\begin{equation}
\tilde{L}_{Y_{T,s}}
=
-
\left[
\frac{p_{\phi T,s}}{p_{tT,s}}
\cos\theta_{T,s}\sin\phi_{T,s}
-
\frac{p_{\theta T,s}}{p_{tT,s}}
\cos\phi_{T,s}
\right],
\end{equation}
\begin{equation}
\tilde{L}_{Z_{T,s}}
=
\frac{p_{\phi T,s}}{p_{tT,s}}
\sin\theta_{T,s}.
\end{equation}
\end{subequations}

At the emission point, the construction of the T-frame gives
$\theta_{T,s}=90^\circ$ and $\phi_{T,s}=90^\circ$. The normalized
angular momentum vector can then be transformed into the companion
spin-aligned frame through
\begin{align}
\vec{\tilde{L}}_{{\rm bh}',s}
={}&
R_y(-\lambda_{\rm bh})
R_z(-\eta_{\rm bh})
R_x(i)
R_z(\omega)
\nonumber\\
&\times
R_z(\varphi_{\rm Tb})
R_y(\vartheta_{\rm Tb})
\vec{\tilde{L}}_{T,s},
\end{align}
where $\vartheta_{\rm Tb}$ is the angle between the $Z_T$ and $Z_b$
axes and $\varphi_{\rm Tb}$ specifies the relative orientation of the
$Y_T$ and $Y_b$ axes \citep{DBB23}. Since the bh- and
${\rm bh}'$-frames are parallel,
\begin{equation}
\vec{\tilde{L}}_{{\rm bh},s}
=
\vec{\tilde{L}}_{{\rm bh}',s}.
\end{equation}

The components of the angular momentum vector determine the initial
momentum ratios in the companion frame:
\begin{subequations}
\label{eq:cartesianangularmomentumtilde_bhframeLinearMomentum}
\begin{equation}
\frac{p_{\theta_{\rm bh},s}}{p_{t_{\rm bh},s}}
=
\tilde{L}_{Y_{\rm bh},s}\cos\phi_{\rm bh,s}
-
\tilde{L}_{X_{\rm bh},s}\sin\phi_{\rm bh,s},
\end{equation}
\begin{equation}
\frac{p_{\phi_{\rm bh},s}}{p_{t_{\rm bh},s}}
=
\frac{\tilde{L}_{Z_{\rm bh},s}}
{\sin\theta_{\rm bh,s}}.
\end{equation}
\end{subequations}
Because the transformation between the T- and bh-frames consists only
of spatial rotations, the temporal component of the photon momentum
remains unchanged:
\begin{equation}
p_{t_{\rm bh},s}=p_{tT,s}.
\end{equation}

Following \citep{Yang_2013}, the constants of motion can then be
calculated from the initial momentum components:
\begin{subequations}
\begin{equation}
\label{Dlambda_calculate}
D_\lambda
=
\frac{
\sin\theta_{\rm bh}
(p_{\phi_{\rm bh}}/p_{t_{\rm bh}})
}
{
-\sqrt{\Delta}\rho^2/\Sigma^2
+
\Omega\sin\theta_{\rm bh}
(p_{\phi_{\rm bh}}/p_{t_{\rm bh}})
},
\end{equation}
\begin{multline}
\label{Dq_calculate}
D_q
=
\left[
\frac{D_\lambda^2}{\sin^2\theta_{\rm bh}}
-
\left(\tilde{a}\frac{GM_c}{c^2}\right)^2
\right]\cos^2\theta_{\rm bh} \\
+
\left[
\frac{p_{\theta_{\rm bh}}}{p_{t_{\rm bh}}}
(1-D_\lambda\Omega)
\right]^2
\frac{\Sigma^2}{\Delta}.
\end{multline}
\end{subequations}

Evaluating these quantities at the emission point fixes the constants
$D_\lambda$ and $D_q$ for each light ray. The null geodesic can then be
integrated from source to infinity by analytical expressions of elliptic integrals as suggested by \cite{KDL20} and \cite{GSL20} to determine the trajectory of the ray and its
direction at asymptotic infinity. The final angular coordinates
$\theta_{{\rm bh},f}$ and $\phi_{{\rm bh},f}$ give the asymptotic
propagation direction
\begin{equation}
\label{n_bh_f}
\widehat{n}_{{\rm bh},f}
=
[
\cos\phi_{{\rm bh},f}\sin\theta_{{\rm bh},f},
\sin\phi_{{\rm bh},f}\sin\theta_{{\rm bh},f},
\cos\theta_{{\rm bh},f}
].
\end{equation}

The final propagation direction is transformed from the companion
frame to the pulsar inertial frame according to
\begin{equation}
\label{nIinft}
\widehat{n}_{I,f}
=
R_y(-\lambda_p)
R_z(-\eta_p)
R_z(\eta_{\rm bh})
R_y(\lambda_{\rm bh})
\widehat{n}_{{\rm bh},f}.
\end{equation}

\subsection{Parallel Transport of the Polarization Vector}
\label{subsec:polarization_transport}

The gravitationally bent ray changes not only its propagation direction
but also the polarization direction measured by the observer. For each
light ray, the initial polarization vector is determined from the
emission geometry described in Section~\ref{bema_polarization_geometry}.
We denote this vector in the pulsar inertial frame by
$\widehat{f}_{I,s}$. To construct the corresponding flat-spacetime
reference polarization at the observer, we first transform
$\widehat{f}_{I,s}$ to the $T$-frame according to
\begin{multline}
\widehat{f}_{T,s}
=
R_y(-\vartheta_{\rm Tb})
R_z(-\varphi_{\rm Tb})
R_z(-\omega)
R_x(-i)
\\
\times
R_z(\eta_p)
R_y(\lambda_p)
\widehat{f}_{I,s}.
\end{multline}
Let $\widehat{n}_{T,s}$ and $\widehat{n}_{T,f}$ denote the initial and
final directions of the light ray in the $T$-frame. Their polar and
azimuthal angles are denoted by
$(\theta_{T,s},\phi_{T,s})$ and
$(\theta_{T,f},\phi_{T,f})$, respectively. In Schwarzschild spacetime,
the light ray remains in the $X_TY_T$ plane, and hence
$\theta_{T,s}=\theta_{T,f}=90^\circ$. In Kerr spacetime, the
propagation plane of the light ray does not remain fixed. We therefore
construct the corresponding flat-spacetime reference polarization at
the final ray direction using
\begin{equation}
\label{eq:f_T_flat}
\widehat{f}^{\prime}_{T,f}
=
R_z(\phi_{T,f})
R_y(\theta_{T,f})
R_y(-\theta_{T,s})
R_z(-\phi_{T,s})
\widehat{f}_{T,s}.
\end{equation}
This transformation connects the initial and final ray directions
while preserving the polarization components relative to the
corresponding propagation planes. In particular, we define the
initial plane of propagation using $\widehat{\boldsymbol r}_T
\times\widehat{\boldsymbol n}_{T,s}$ and the final plane using
$\widehat{\boldsymbol n}_{T,f}\times\widehat{\boldsymbol r}_T$,
where $\widehat{\boldsymbol r}_T$ is the unit vector from the
companion to the pulsar in the $T$-frame. We explicitly verify that
the components of $\widehat{f}_{T,s}$ in the initial propagation plane
and perpendicular to it are equal to the corresponding components of
$\widehat{f}^{\prime}_{T,f}$ in the final propagation plane and
perpendicular to it. Thus, Eq.~\eqref{eq:f_T_flat} provides the
flat-spacetime reference construction used to determine the geometric
polarization shift. It does not involve an explicit integration of
the parallel-transport equation along the photon trajectory.

The reference polarization vector $\widehat{f}^{\prime}_{T,f}$ is
then transformed to the observer sky frame used to define the
geometric polarization shift in Section~\ref{definitions}.

We next calculate the parallel transport of the polarization vector
along the null geodesic in the curved spacetime of the companion.
For this purpose, the initial polarization vector
$\widehat{f}_{I,s}$ is first transformed into the companion
spin-aligned frame according to
\begin{equation}
\label{f_bh}
\widehat{f}_{\rm bh,s}
=
R_y(-\lambda_{\rm bh})
R_z(-\eta_{\rm bh})
R_z(\eta_p)
R_y(\lambda_p)
\widehat{f}_{I,s}.
\end{equation}
The polarization vector is then expressed in the spherical coordinate
basis associated with the companion frame. The Cartesian components of
the polarization vector in the companion frame, denoted by
$f_{{\rm bh},{s_x}}$, $f_{{\rm bh},{s_y}}$, and
$f_{{\rm bh},{s_z}}$, are transformed into the corresponding
spherical components as
\begin{align}
f_{({\rm bh,s})r}
={}&
f_{{\rm bh},{s_x}}\cos\phi_{{\rm bh},s}\sin\theta_{{\rm bh},s}
+
f_{{\rm bh},{s_y}}\sin\phi_{{\rm bh},s}\sin\theta_{{\rm bh},s}
\nonumber\\
&+
f_{{\rm bh},{s_z}}\cos\theta_{{\rm bh},s},
\end{align}
\begin{align}
f_{({\rm bh,s})\theta}
={}&
f_{{\rm bh},{s_x}}\cos\phi_{{\rm bh},s}\cos\theta_{{\rm bh},s}
+
f_{{\rm bh},{s_y}}\sin\phi_{{\rm bh},s}\cos\theta_{{\rm bh},s}
\nonumber\\
&-
f_{{\rm bh},{s_z}}\sin\theta_{{\rm bh},s},
\end{align}
and
\begin{equation}
f_{({\rm bh,s})\phi}
=
-f_{{\rm bh},{s_x}}\sin\phi_{{\rm bh},s}
+
f_{{\rm bh},{s_y}}\cos\phi_{{\rm bh},s},
\end{equation}
where $\theta_{{\rm bh},s}$ and $\phi_{{\rm bh},s}$ are the polar and
azimuthal positions of the pulsar in the companion frame.

At the emission point, the polarization direction is specified in the
local spatial frame of the companion. We therefore choose the
corresponding polarization four-vector to have a vanishing temporal
component \citep{Corner1980},
\begin{equation}
\label{eq:f_bh_initial}
f_{({\rm bh,s})\,\mu}
=
\left(
0,
f_{({\rm bh,s})r},
f_{({\rm bh,s})\theta},
f_{({\rm bh,s})\phi}
\right).
\end{equation}
This choice represents the polarization as a purely spatial vector in
the local companion frame. The initial polarization vector is required
to be transverse to the propagation direction, and we explicitly
verify that
\begin{equation}
\label{eq:initial_transverse}
\widehat{f}_{\rm bh,s}
\cdot
\widehat{n}_{\rm bh,s}
=
0.
\end{equation}

The initial polarization four-vector is transformed from the companion
frame to the Boyer--Lindquist coordinate basis using the transformation
matrix $\tilde{E}^{\mu}{}_{\nu}$ defined in
Appendix~\ref{Appendix}:
\begin{equation}
\label{f_in_BL}
f_{({\rm BL,s})\nu}
=
\tilde{E}^{\mu}{}_{\nu}
f_{({\rm bh,s})\,\mu}.
\end{equation}
The initial polarization four-vector also satisfies
\begin{equation}
\label{eq:initial_lf}
l^\mu f_{({\rm BL})\mu}
=
0,
\end{equation}
where $l^\mu$ is the null vector tangent to the photon trajectory.
Equations~\eqref{eq:initial_transverse} and
\eqref{eq:initial_lf} provide consistency checks on the construction
of the initial polarization state.

To describe the parallel transport of the polarization vector along
the null geodesic, we introduce the null tetrad
$(l^\mu,e_2^\mu,e_3^\mu,k^\mu)$ associated with the photon trajectory.
The explicit expressions for the tetrad vectors are given in
Appendix~\ref{Appendix}. Here, $l^\mu$ is tangent to the null
geodesic and represents the photon propagation direction, while
$e_2^\mu$ and $e_3^\mu$ span the two-dimensional screen transverse to
the propagation direction. The vector $k^\mu$ completes the null
tetrad.

Following \citep{VA2025}, the parallel transport of the polarization
vector can be described through its projections onto the tetrad:
\begin{subequations}
\label{eq:condition1}
\begin{align}
\left(
l^\mu\tilde{E}^{\nu}{}_{\mu}
\right)_f
f_{({\rm bh},f)\nu}
&=
\left(
l^\mu\tilde{E}^{\nu}{}_{\mu}
\right)_s
f_{({\rm bh},s)\nu}
=0,
\label{eq:conda}
\\
\left(
e_2^\mu\tilde{E}^{\nu}{}_{\mu}
\right)_f
f_{({\rm bh},f)\nu}
&=
\left(
e_2^\mu\tilde{E}^{\nu}{}_{\mu}
\right)_s
f_{({\rm bh},s)\nu}
=C_1,
\label{eq:condb}
\\
\left(
e_3^\mu\tilde{E}^{\nu}{}_{\mu}
\right)_f
f_{({\rm bh},f)\nu}
&=
\left(
e_3^\mu\tilde{E}^{\nu}{}_{\mu}
\right)_s
f_{({\rm bh},s)\nu}
=C_2.
\label{eq:condc}
\end{align}
\end{subequations}
Here, $C_1$ and $C_2$ are constants associated with the parallel
transport of the polarization vector along the null geodesic. The
subscripts ``s'' and ``f'' denote quantities evaluated at the emission
point and at asymptotic infinity, respectively. The first of
Eqs.~\eqref{eq:condition1} expresses the transversality of the
polarization vector with respect to the photon propagation direction,
while the second and third equations specify its two independent
components in the transverse screen. Solving these equations (see
Appendix~\ref{Appendix}) gives the polarization vector at asymptotic
infinity.

At asymptotic infinity, the spacetime is locally flat and the
polarization vector is transverse to the final propagation direction.
We therefore express the final polarization vector in terms of its
spherical components, $f_{\theta,f}$ and $f_{\phi,f}$. The corresponding
normalized polarization direction in the companion Cartesian frame is
\begin{equation}
\label{eq:f_bh_final}
\widehat{f}_{{\rm bh},f}
=
\frac{
f_{\theta,f}\,e_{\theta,f}
+
f_{\phi,f}\,e_{\phi,f}
}{
\left|
f_{\theta,f}\,e_{\theta,f}
+
f_{\phi,f}\,e_{\phi,f}
\right|
},
\end{equation}
where
\begin{equation}
e_{\theta,f}
=
[
\cos\theta_{{\rm bh},f}\cos\phi_{{\rm bh},f},
\cos\theta_{{\rm bh},f}\sin\phi_{{\rm bh},f},
-\sin\theta_{{\rm bh},f}
],
\end{equation}
and
\begin{equation}
e_{\phi,f}
=
[
-\sin\phi_{{\rm bh},f},
\cos\phi_{{\rm bh},f},
0
].
\end{equation}

The final polarization direction is transformed into the observer
sky-plane coordinate system according to
\begin{equation}
\label{eq:f_sky_final}
\widehat{f}_{{\rm sky},f}
=
R_z(\eta_{\rm bh})
R_y(\lambda_{\rm bh})
\widehat{f}_{{\rm bh},f},
\end{equation}
and into the pulsar inertial frame according to
\begin{equation}
\label{eq:f_I_final}
\widehat{f}_{I,f}
=
R_y(-\lambda_p)
R_z(-\eta_p)
R_z(\eta_{\rm bh})
R_y(\lambda_{\rm bh})
\widehat{f}_{{\rm bh},f}.
\end{equation}

As a further consistency check, the final polarization vector obtained
from the parallel-transport calculation remains transverse to the
final propagation direction:
\begin{equation}
\label{eq:final_transverse}
\widehat{f}_{I,f}
\cdot
\widehat{n}_{I,f}
=
0.
\end{equation}
Thus, the initial and final polarization states satisfy the required
transversality conditions, while the parallel-transport calculation
determines the change in the polarization direction accumulated along
the curved null geodesic.

The initial polarization vector of the bent ray and the corresponding
final polarization vectors from the flat-spacetime reference
construction and the curved-spacetime parallel-transport calculation
are then used to determine the geometric polarization shift
$\Delta\psi_{\rm geo}$ and the polarization transport rotation
$\Delta\psi_{\rm pt}$, respectively, as defined in
Section~\ref{definitions}.

\section{Numerical Study}
\label{numerical}

We implement the general relativistic framework developed in the
previous sections to quantitatively investigate the effects of
gravitational light bending and polarization transport on the
polarization direction of a pulsar in a binary system. In particular,
we study the variations of the polarization rotation angles defined in
Section~\ref{definitions} as a function of the orbital phase and
investigate the corresponding changes in the classical polarization
``S-curve'', which is governed by the intrinsic emission geometry
\citep{rc69}.

\subsection{Polarization Rotation in a Pulsar--Intermediate-Mass Black Hole Binary}
\label{intmass}

Since no pulsar--BH binary has yet been confirmed
observationally, we consider a hypothetical binary configuration for
our numerical calculations. Following our previous works
\citep{DBB23,DB25}, we adopt the same values for most of the system
parameters. These include the pulsar mass $M_p$, spin period $P_s$,
first time derivative of the spin period $\dot{P}_s$, orbital period
$P_b$, and orbital eccentricity $e$. The geometrical parameters
$\lambda_p$, $\eta_p$, $\lambda_{\rm bh}$, and $\eta_{\rm bh}$ are
defined in Section~\ref{subsec:orbitalgeometrymotion}. The complete
set of parameters adopted in the present numerical calculations is
listed in Table~\ref{tab:PSRintBH}.

We consider a companion intermediate-mass black hole (IMBH) with mass
$M_c=10^5\,M_\odot$ and an orbital inclination of $i=85^\circ$. The
parameters $\eta_p$, $\lambda_p$, $\lambda_{\rm bh}$, $\eta_{\rm bh}$,
$\alpha$, and $\beta$ are not directly determined from observations
and are therefore chosen for the present numerical study. As
discussed in Section~\ref{subsec:orbitandspinevolution}, the angles
$\lambda_p$ and $\eta_p$ evolve slowly with time due to the geodetic
precession of the pulsar spin.

For the binary configuration considered here, the initial values of
$\lambda_p$ and $\eta_p$ are chosen to be $130^\circ$ and
$45^\circ$, respectively. Using Eq.~\ref{eq:spin_prec}, we find that
over one orbital period of $0.5$~yr, the values of $\lambda_p$ and
$\eta_p$ decrease by $0.015^\circ$ and $0.01^\circ$, respectively.
The orbital precession discussed in
Section~\ref{subsec:orbitandspinevolution} also causes changes in the
orbital parameters $\omega$, $P_b$, and $i$. Over one orbital period,
the change in $\omega$ is $0.04^\circ$, while the changes in $i$ and
$P_b$ are $2.28\times10^{-13}\,^\circ$ and
$4.95\times10^{-7}$~s, respectively. For the parameters adopted in
Table~\ref{tab:PSRintBH}, the effects of the companion spin on these
quantities are very small. The magnetic inclination angle and the
impact angle are chosen to be $\alpha=49^\circ$ and $\beta=1^\circ$,
respectively, and are kept fixed throughout the numerical study. For
the adopted pulsar emission geometry, $\lambda_p$ and the angle
$\zeta_N$ between the pulsar spin axis and the line of sight are
related through $\lambda_p=180^\circ-\zeta_N$ \citep{DBB23}.

To determine the initial direction of the polarization vector, we
calculate the emission height by fixing the observing frequency at
$\nu=1.4$~GHz. Using Eq.~\eqref{eq:emheight}, we obtain
$h_{1.4\,{\rm em}}=437.19$~km. The corresponding half-opening angle
of the emission cone, calculated using
Eq.~\eqref{eq:halfopeningangleoutcone}, is
$\mathcal{W}_{1.4}=4.51^\circ$.

\begin{table}[htbp]
\centering
\caption{Parameters adopted for the hypothetical pulsar-IMBH binary considered in the numerical calculations.}
\label{tab:PSRintBH}
\begin{tabular}{ll}
\hline
Parameters & Value \\
\hline
$P_{\rm s}$ [s] & 1.67 \\
$\dot{P}_{\rm s}$ [$\mathrm{s\,s}^{-1}$] & $1.38 \times 10^{-15}$ \\
$P_{\rm b}$ [years] & 0.5 \\
$e$ & 0.35 \\
$\omega$ [deg] & 73.804 \\
$i$ [deg] & 85 \\
$M_p$ [$M_{\odot}$] & 1.8 \\
$M_c$ [$M_{\odot}$] & $10^5$ \\
$\alpha$ [deg] & 49.0 \\
$\beta$ [deg] & 1.0 \\
$\eta_p$ [deg] & 45.0 \\
$\lambda_p$ [deg] & 130.0 \\
$\eta_{\rm bh}$ [deg] & $-90.0$ \\
$\lambda_{\rm bh}$ [deg] & 85 \\
$\tilde{a}$ & 0.0 (or 0.5) \\
\hline
\end{tabular}
\end{table}

\begin{figure}[htbp]
    \centering
    \includegraphics[width=\linewidth]{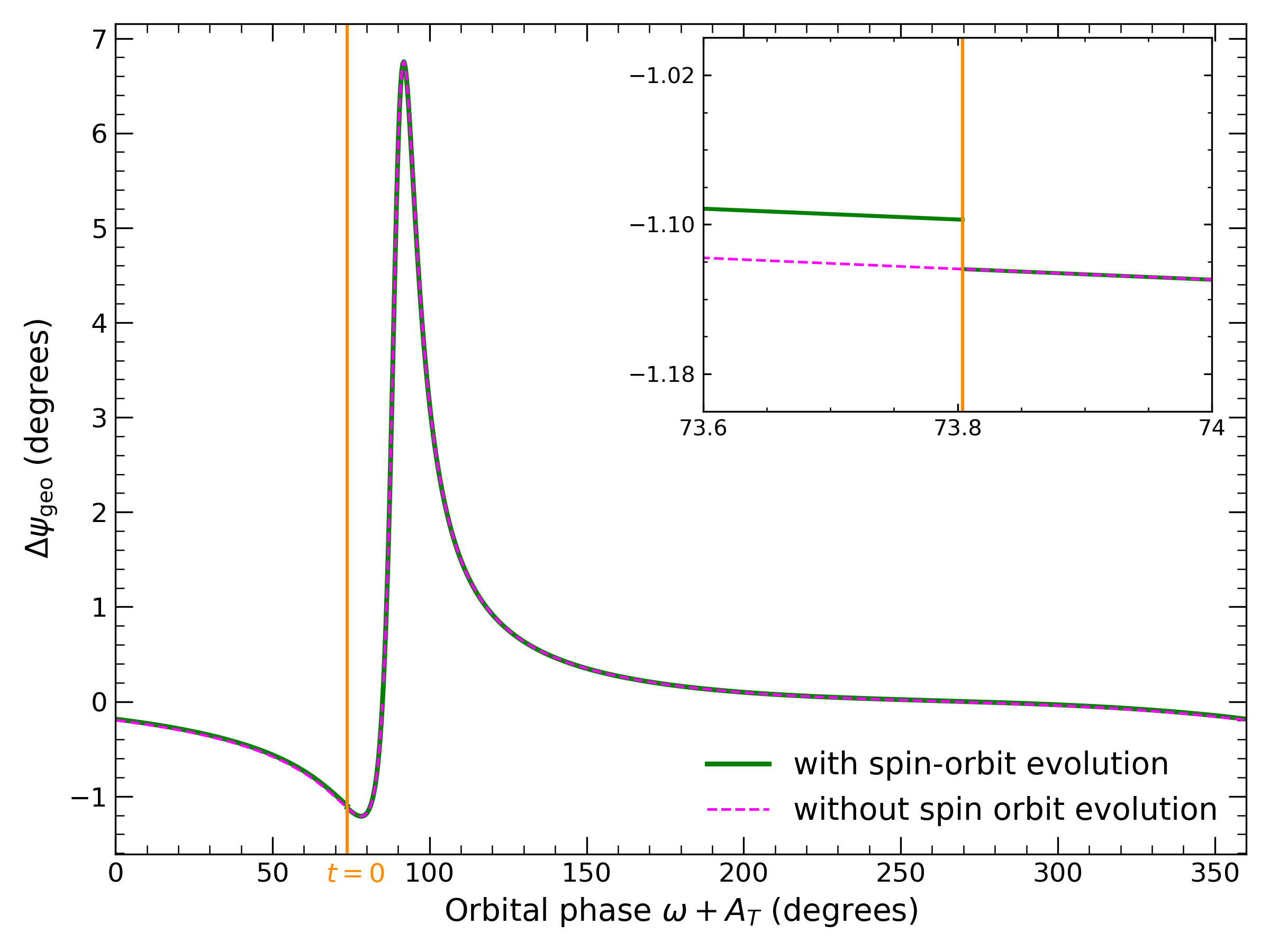}
    \caption{Variation of the geometric polarization shift
    $\Delta\psi_{\rm geo}$ with orbital phase $A_T+\omega$ for a
    hypothetical pulsar--IMBH binary with a
    non-spinning companion ($\tilde{a}=0$). The dashed magenta curve
    represents the result obtained without including spin--orbit
    precession, whereas the solid green curve includes the evolution
    of the pulsar and orbital geometry due to spin--orbit coupling.
    The vertical reference line indicates the initial orbital phase at
    $t=0$, corresponding to $A_T+\omega=73.804^\circ$.}
    \label{fig:psi_geo_intsmbh}
\end{figure}

We first consider the case in which the companion is non-spinning,
i.e., $\tilde{a}=0$. In this case, the polarization transport
rotation vanishes in our construction, $\Delta\psi_{\rm pt}=0$,
and the polarization shift is determined by the change in the
propagation direction caused by gravitational light bending.
Figure~\ref{fig:psi_geo_intsmbh} shows the variation of the
geometric polarization shift $\Delta\psi_{\rm geo}$ with orbital
phase $A_T+\omega$ for the hypothetical pulsar--IMBH binary considered here. The orbital phase, in degrees,
is plotted along the abscissa, and $\Delta\psi_{\rm geo}$, also in
degrees, is plotted along the ordinate. We find that the magnitude
of $\Delta\psi_{\rm geo}$ reaches approximately $7^\circ$ near
$A_T+\omega=90^\circ$. This occurs because the gravitational
bending of the light rays becomes strongest as the pulsar approaches
superior conjunction.

In Fig.~\ref{fig:psi_geo_intsmbh}, the initial longitude of
periastron is chosen to be $\omega=73.804^\circ$. Therefore, at
$t=0$, the pulsar is located at the orbital phase
$A_T+\omega=73.804^\circ$, as indicated by the vertical reference
line. Because of spin--orbit coupling, the pulsar spin and orbital
angular momentum vectors precess, causing the orientation of the
pulsar spin and the orbital geometry to evolve with time. As a
result, the longitude of periastron changes slowly, and the value
of $A_T+\omega$ at the same orbital position differs slightly
from its initial value after one orbital period. This evolution
appears as a gradual shift of the polarization rotation curves
relative to the vertical reference line.

The dashed magenta curve represents the calculation without
spin--orbit precession, whereas the solid green curve includes
the evolution of the pulsar spin and orbital geometry due to
spin--orbit coupling. Without spin--orbit precession, the
geometrical configuration repeats after each orbital period, and
the polarization shift returns to its initial values in successive
orbits. When spin--orbit precession is included, the changing
orientation of the pulsar spin modifies the emission geometry.
Consequently, the polarization shift does not necessarily return
to the same value after one orbital period. Since the companion
is non-spinning in this case, $\Delta\psi_{\rm pt}$ remains zero
in our construction.

\begin{figure*}[htbp]
\centering
\begin{subfigure}{0.48\textwidth}
\centering
    \includegraphics[width=\linewidth]{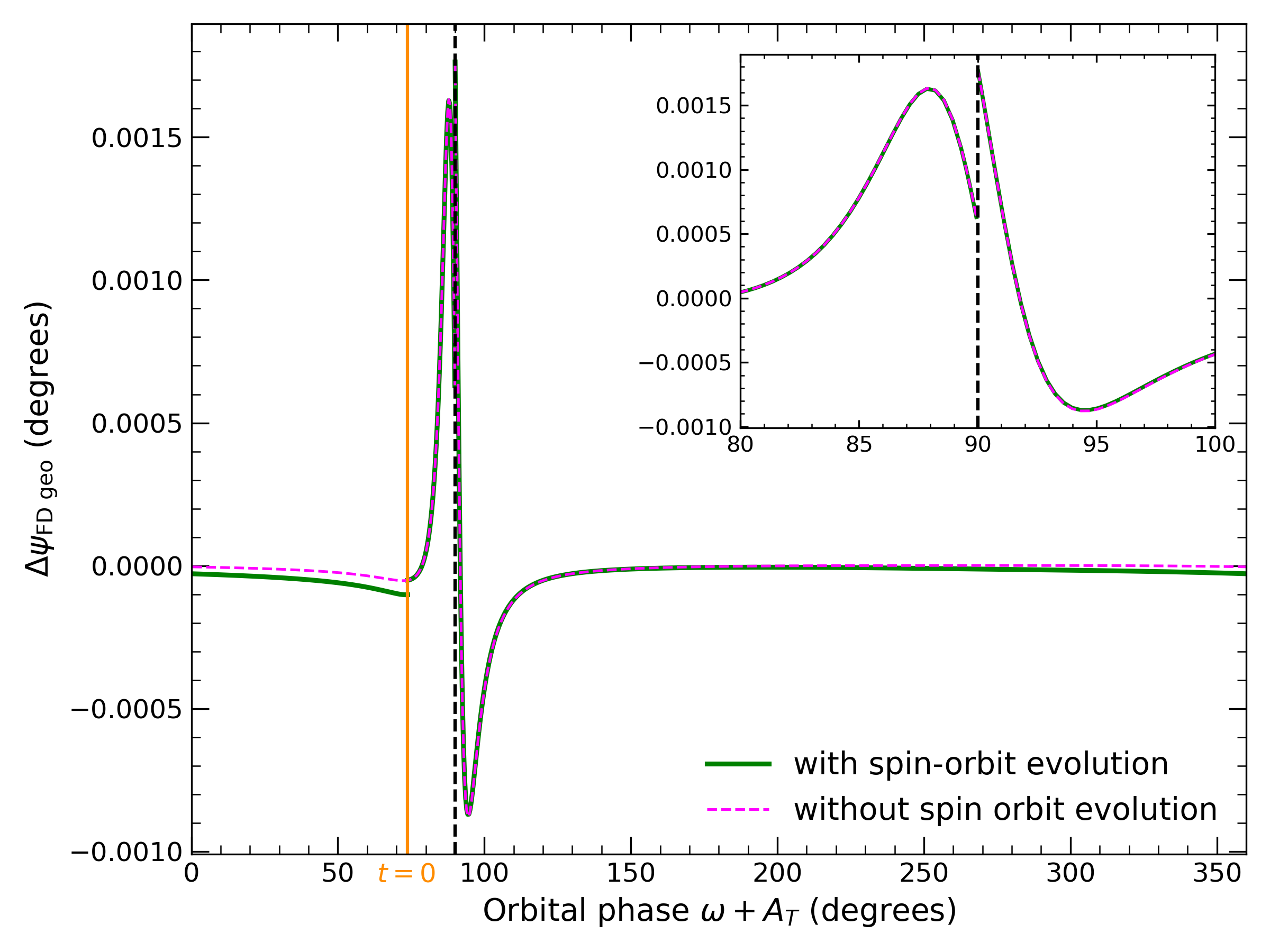}
    \caption{Frame-dragging contribution to the geometric polarization shift $\Delta \psi_{\rm FD, geo}$.}
    \label{fig:psi_fdgeo_intsmbh}
\end{subfigure}
\hfill
\begin{subfigure}{0.48\textwidth}
\centering
    \includegraphics[width=\linewidth]{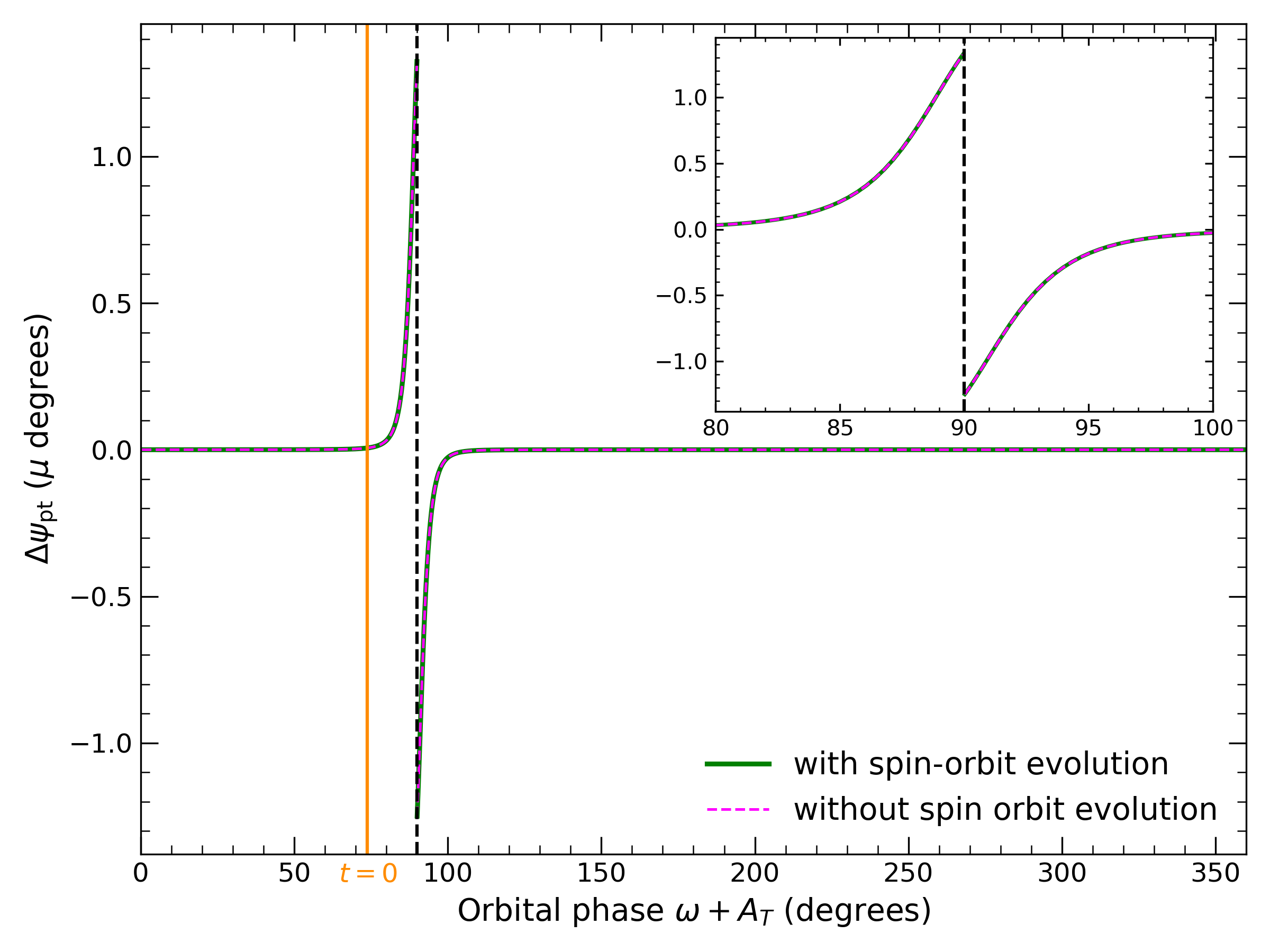}
    \caption{Frame-dragging contribution to the polarization transport rotation $\Delta \psi_{\rm pt}$.}
    \label{fig:psi_pt_intsmbh}
\end{subfigure}
\caption{Spin-induced changes in the polarization rotation for a
hypothetical pulsar--IMBH binary with
$\tilde{a}=0.5$. The left panel shows the spin-induced geometric
polarization shift $\Delta\psi_{\rm FD,geo}$, while the right panel
shows the corresponding change in the polarization transport
rotation $\Delta\psi_{\rm pt}$. The dashed magenta curves represent
calculations performed without including spin--orbit precession,
whereas the solid green curves include the evolution of the pulsar
and orbital geometry due to spin--orbit coupling.}
\label{fig:psi_spin_effects}
\end{figure*}

We next consider a spinning companion with $\tilde{a}=0.5$.
The companion's spin affects the propagation of light through
frame dragging and also influences polarization transport along
the curved-spacetime null geodesic. We therefore examine
separately the spin-induced change in the geometric polarization
shift, $\Delta\psi_{\rm FD,geo}$, and the polarization transport
rotation, $\Delta\psi_{\rm pt}$. The left and right panels of
Fig.~\ref{fig:psi_spin_effects} show these quantities, respectively,
as functions of orbital phase $A_T+\omega$. The orbital phase,
in degrees, is plotted along the abscissa, while
$\Delta\psi_{\rm FD,geo}$, in degrees, and
$\Delta\psi_{\rm pt}$, in microdegrees, are plotted along the
ordinates of the left and right panels, respectively. For the
hypothetical binary considered here, the magnitude of
$\Delta\psi_{\rm FD,geo}$ is of the order of a few millidegrees,
whereas $\Delta\psi_{\rm pt}$ is of the order of a few
microdegrees. Thus, the spin-induced change in the geometric
polarization shift is considerably larger than the polarization
transport rotation.

A sharp discontinuity is visible at $A_T+\omega=90^\circ$ in
both panels of Fig.~\ref{fig:psi_spin_effects}. This feature
arises from the change in the orientation of the photon trajectory
relative to the spin axis of the companion. During one half of
the orbit, the deflected light rays follow prograde trajectories,
whereas during the other half, they follow retrograde trajectories.
The transition between these configurations produces the
discontinuity in the frame-dragging contribution. As derived in
\citet{DB25}, the boundary between these configurations satisfies
\begin{equation}
\label{eq:discontinuity_condition}
\widehat{N}_{s} \cdot
\left(\widehat{r}_{s} \times
\widehat{\mathcal{S}}_{\rm bh}\right) = 0.
\end{equation}

The vertical reference lines in
Figs.~\ref{fig:psi_geo_intsmbh} and
\ref{fig:psi_spin_effects} indicate the initial orbital phase,
$A_T+\omega=73.804^\circ$. Spin--orbit
precession produces a gradual shift of the curves relative to this
reference line. The resulting change in $\Delta\psi_{\rm pt}$,
however, is too small to be visible in Fig.~\ref{fig:psi_spin_effects}.

\begin{figure*}[htbp!]
\centering
\begin{subfigure}{0.3\textwidth}
    \centering
    \includegraphics[width=\linewidth]{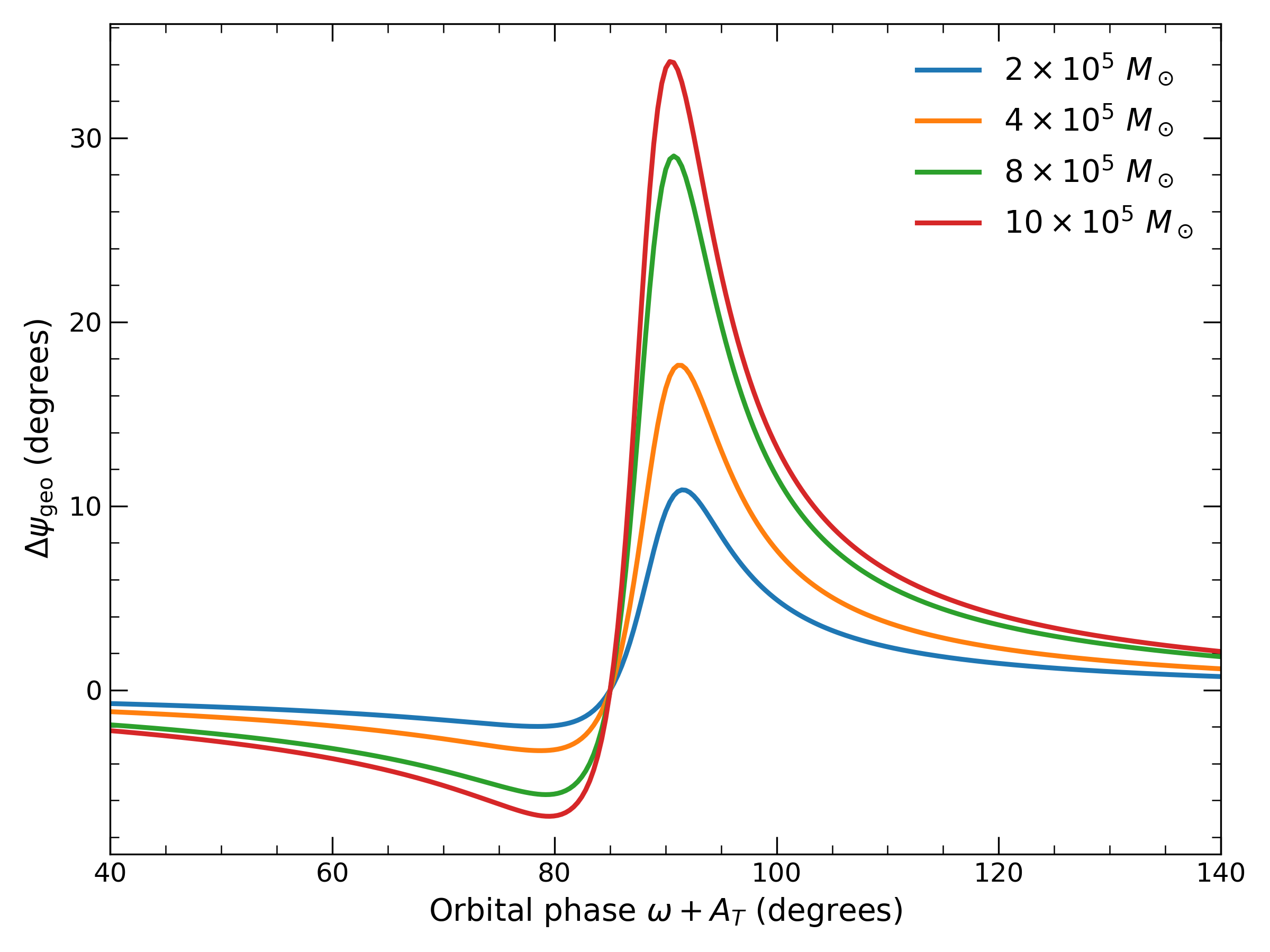}
    \caption{$\Delta\psi_{\rm geo}$ for different values of $M_c$.}
    \label{fig:diff_mass_psi_geo}
\end{subfigure}
\hfill
\begin{subfigure}{0.3\textwidth}
    \centering
    \includegraphics[width=\linewidth]{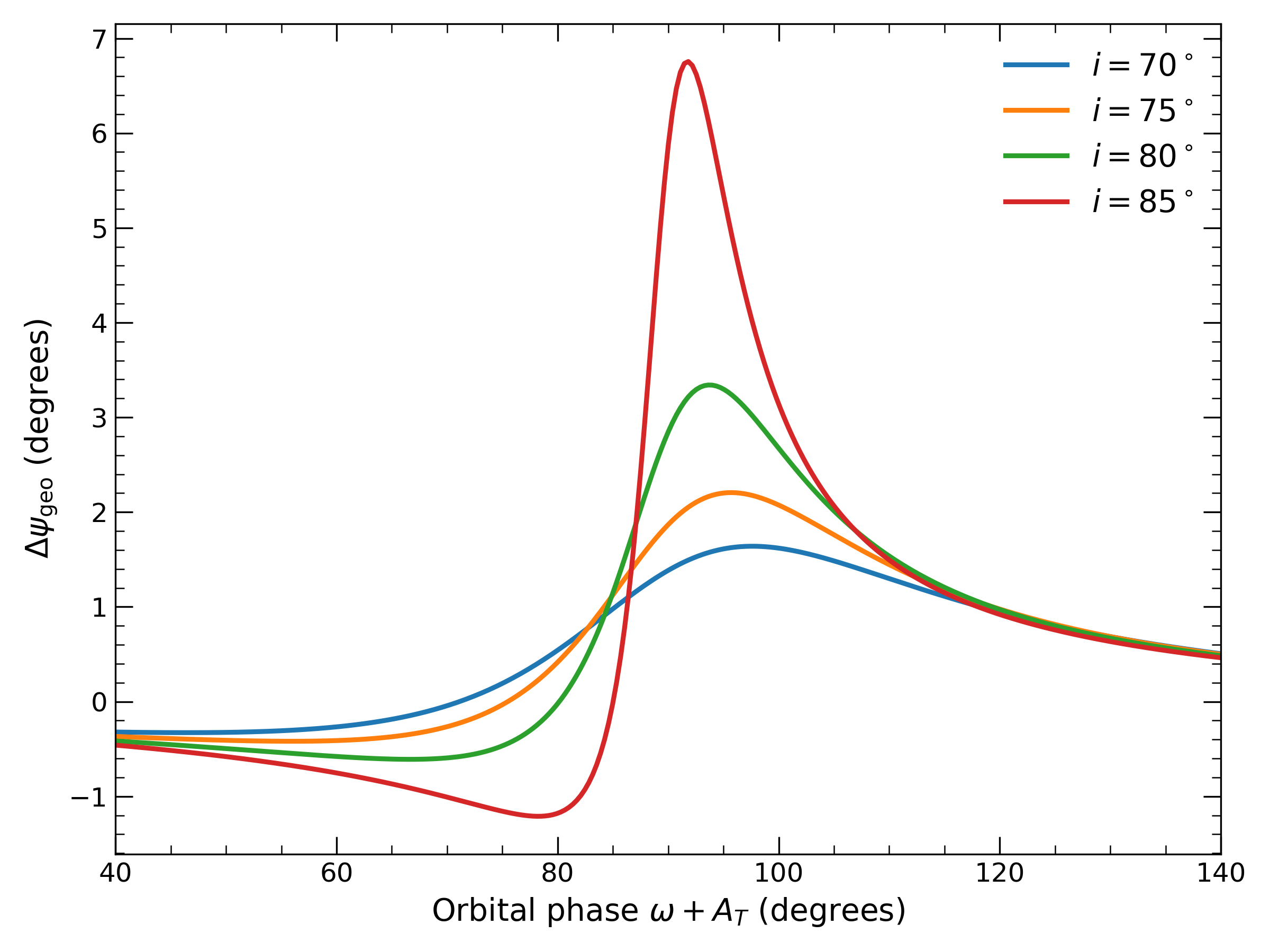}
    \caption{$\Delta\psi_{\rm geo}$ for different values of $i$.}
    \label{fig:diff_inclination_psi_geo}
\end{subfigure}
\hfill
\begin{subfigure}{0.3\textwidth}
    \centering
    \includegraphics[width=\linewidth]{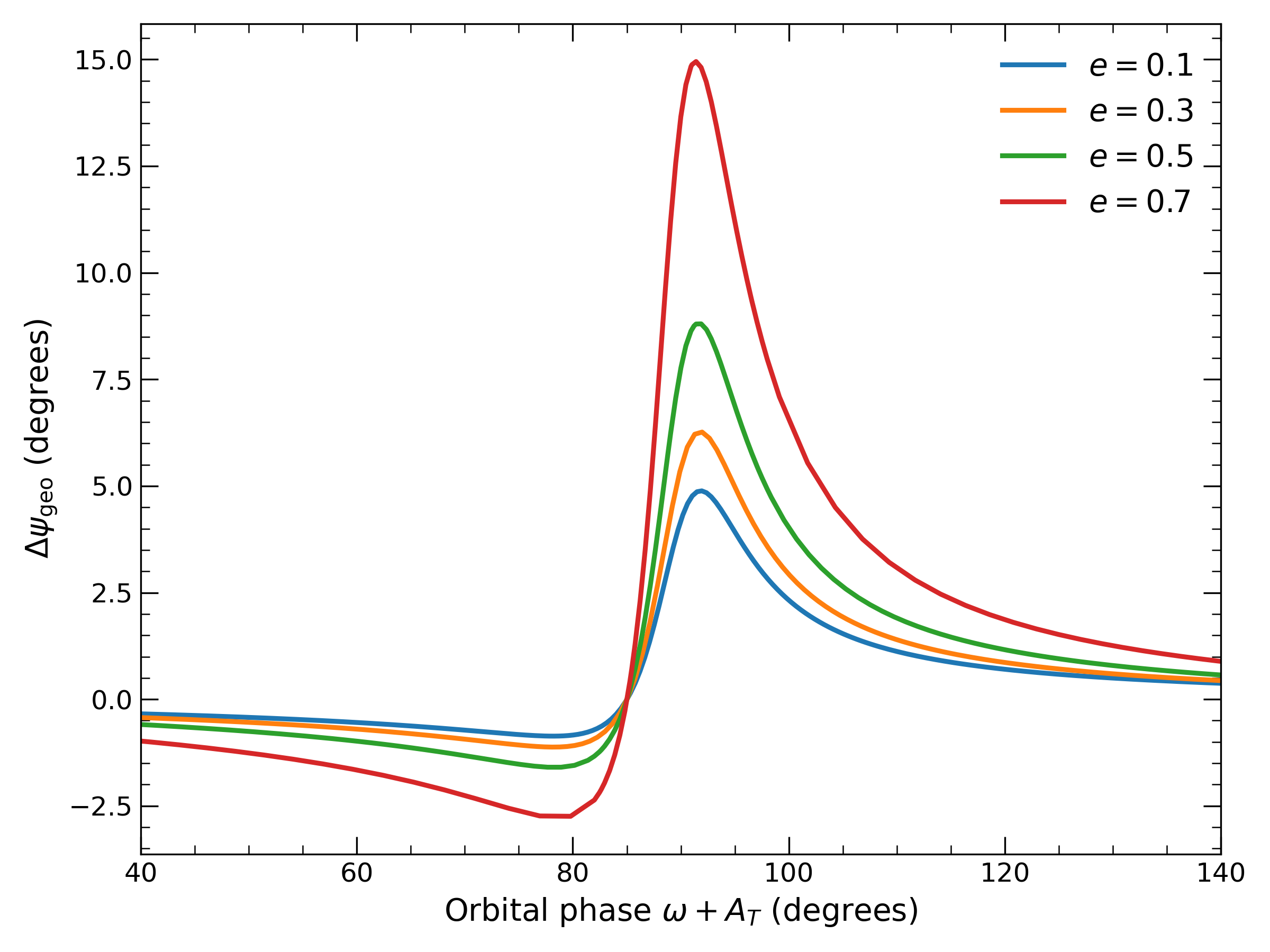}
    \caption{$\Delta\psi_{\rm geo}$ for different values of $e$.}
    \label{fig:diff_e_psi_geo}
\end{subfigure}

\vspace{0.2cm}

\begin{subfigure}{0.3\textwidth}
    \centering
    \includegraphics[width=\linewidth]{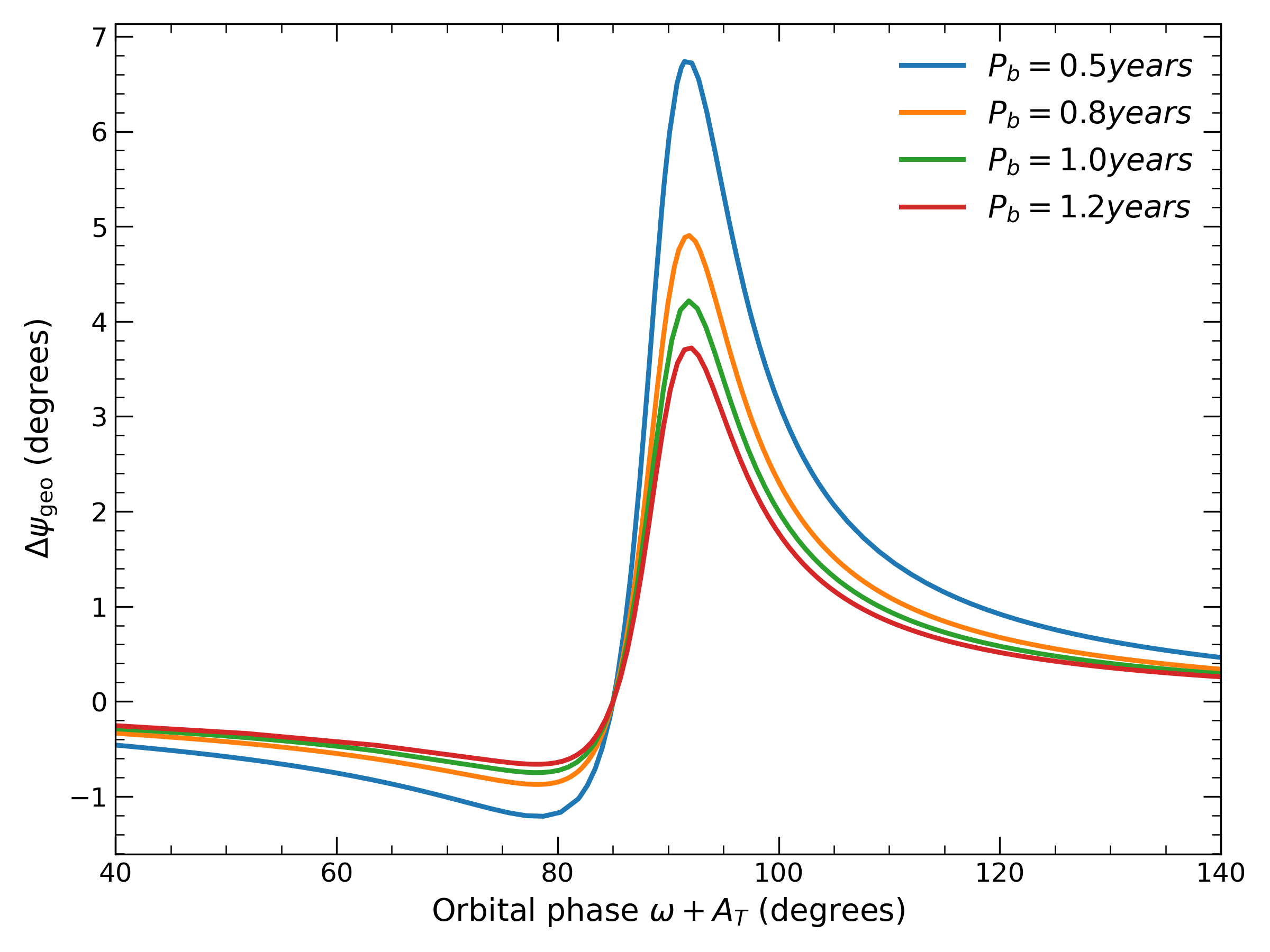}
    \caption{$\Delta\psi_{\rm geo}$ for different values of $P_b$.}
    \label{fig:diff_period_psi_geo}
\end{subfigure}
\hfill
\begin{subfigure}{0.3\textwidth}
    \centering
    \includegraphics[width=\linewidth]{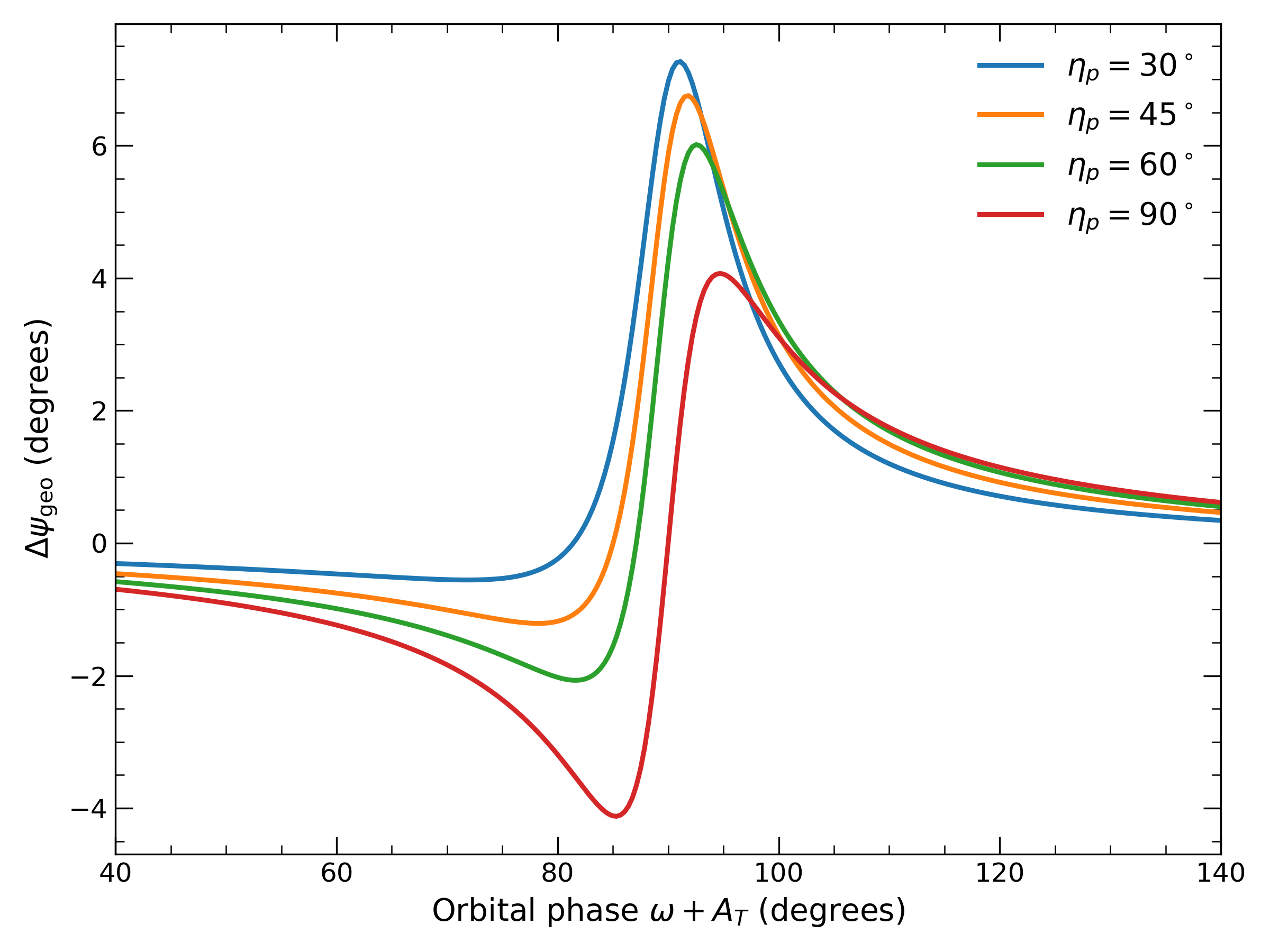}
    \caption{$\Delta\psi_{\rm geo}$ for positive values of $\eta_p$.}
    \label{fig:diff_eta_psi_geo}
\end{subfigure}
\hfill
\begin{subfigure}{0.3\textwidth}
    \centering
    \includegraphics[width=\linewidth]{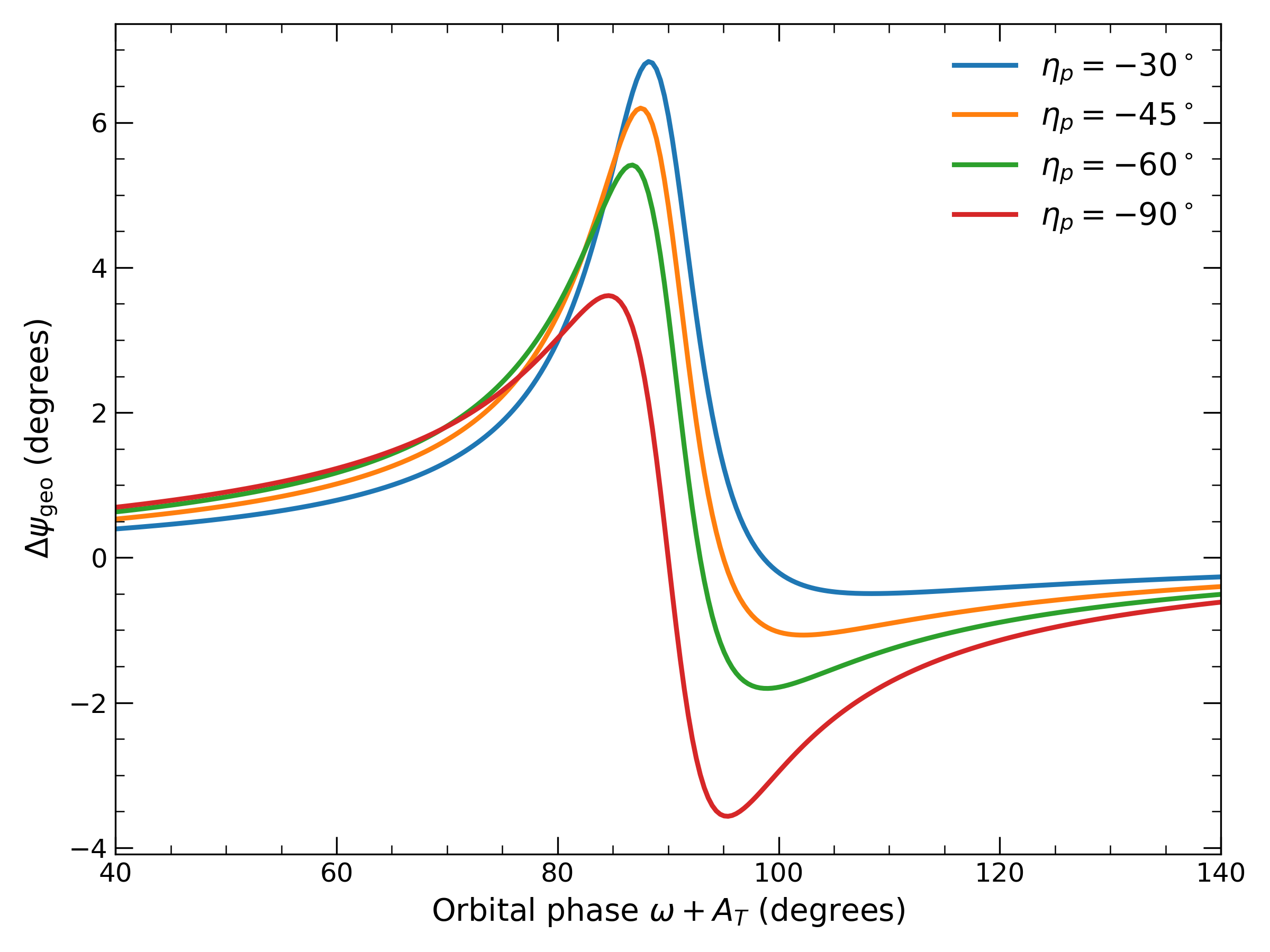}
    \caption{$\Delta\psi_{\rm geo}$ for negative values of $\eta_p$.}
    \label{fig:diff_eta_neg_psi_geo}
\end{subfigure}

\caption{Dependence of the geometric polarization shift
$\Delta\psi_{\rm geo}$ on the binary parameters for the hypothetical
pulsar--IMBH binary. In each panel, one
parameter is varied while the remaining parameters are fixed at the
values given in Table~\ref{tab:PSRintBH}. The panels show the
dependence on the companion mass $M_c$, orbital inclination $i$,
orbital eccentricity $e$, orbital period $P_b$, and positive and
negative values of the pulsar spin orientation parameter $\eta_p$.
Spin--orbit coupling is neglected.}
\label{diff_param_psi_geo}
\end{figure*}

\begin{figure*}[htbp]
\centering
\begin{subfigure}{0.3\textwidth}
    \centering
    \includegraphics[width=\linewidth]{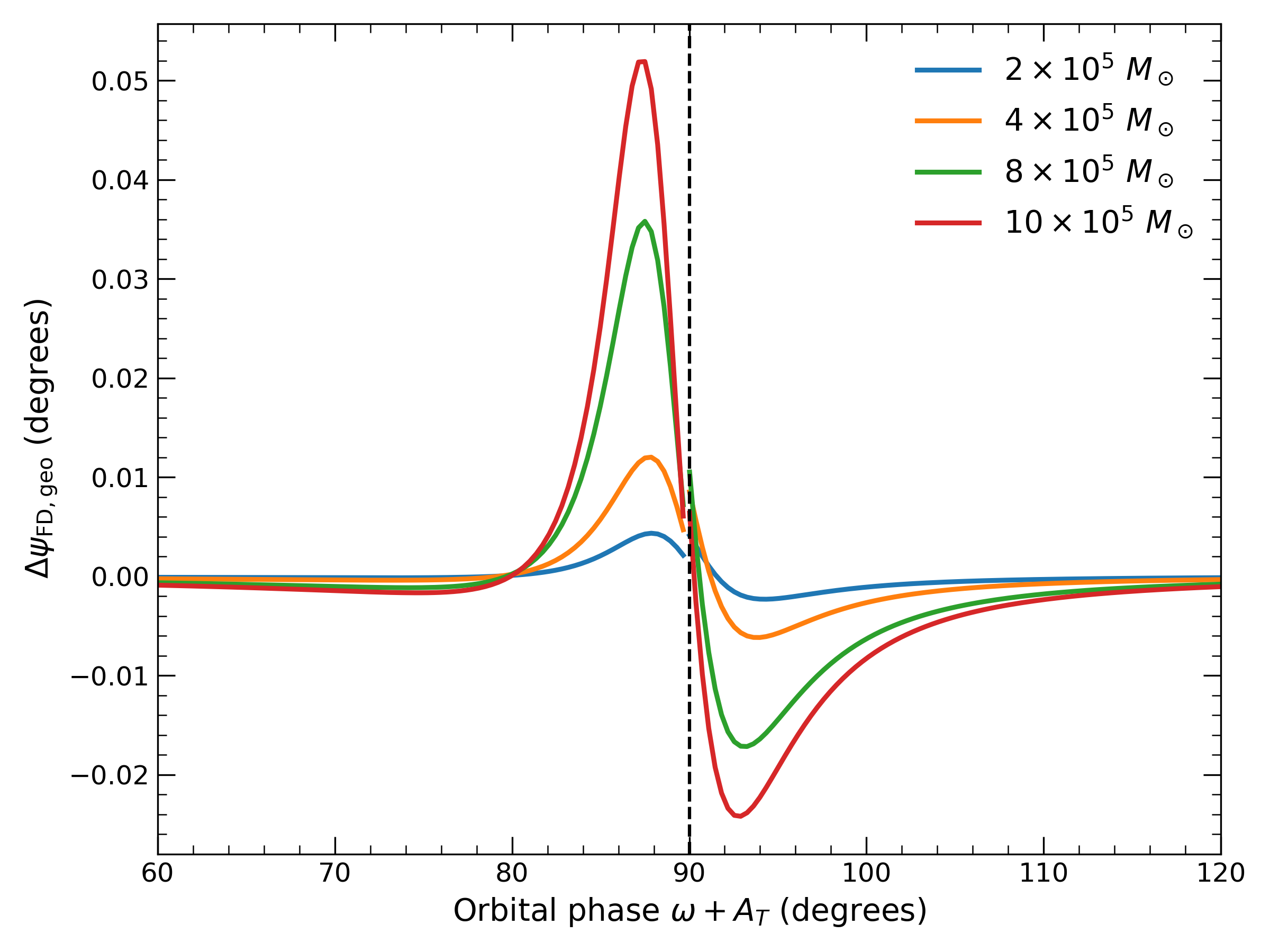}
    \caption{$\Delta\psi_{\rm FD,geo}$ for different values of $M_c$.}
    \label{fig:diff_mass_psi_fd}
\end{subfigure}
\hfill
\begin{subfigure}{0.3\textwidth}
    \centering
    \includegraphics[width=\linewidth]{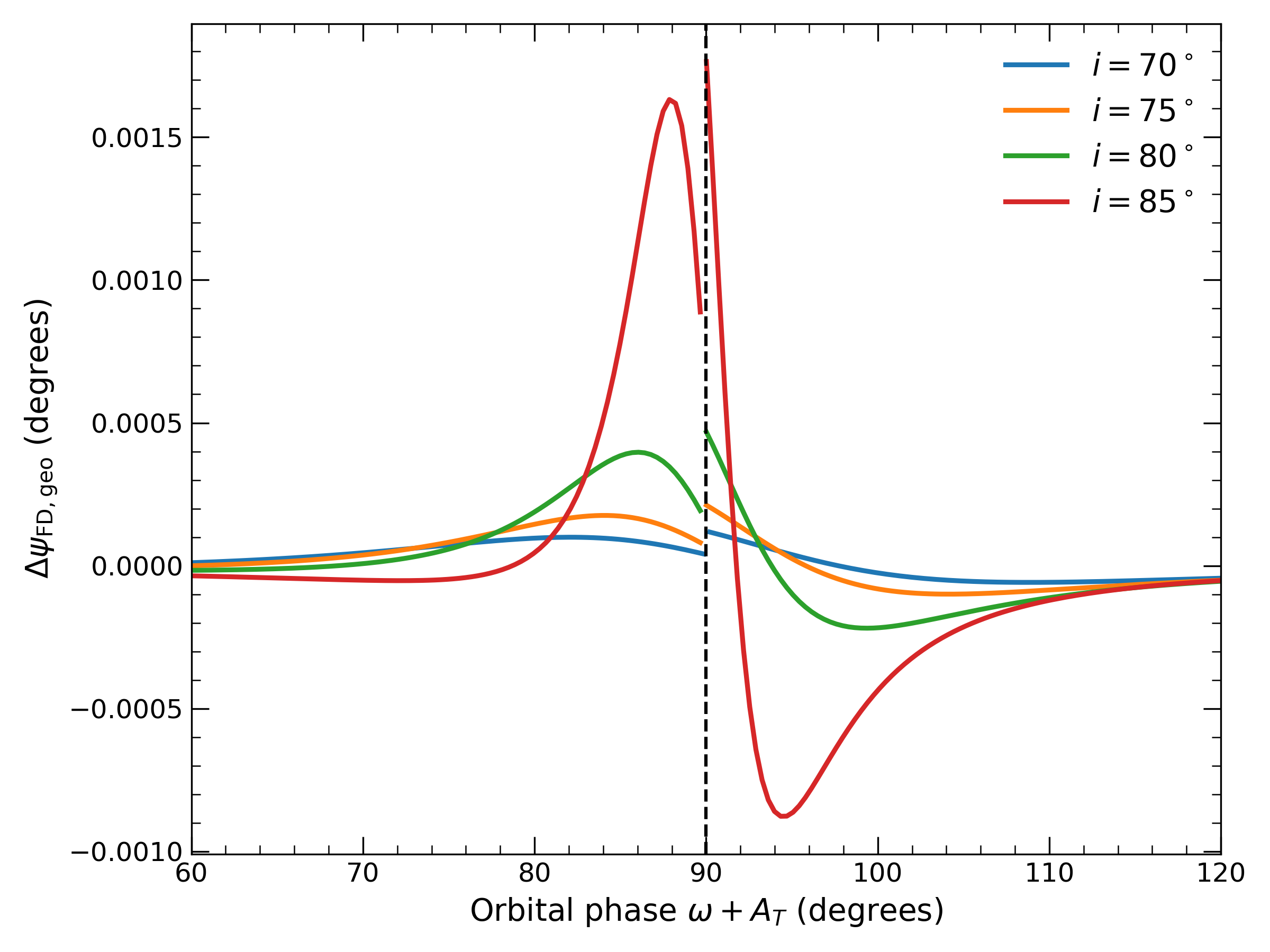}
    \caption{$\Delta\psi_{\rm FD,geo}$ for different values of $i$.}
    \label{fig:diff_inclination_psi_fd}
\end{subfigure}
\hfill
\begin{subfigure}{0.3\textwidth}
    \centering
    \includegraphics[width=\linewidth]{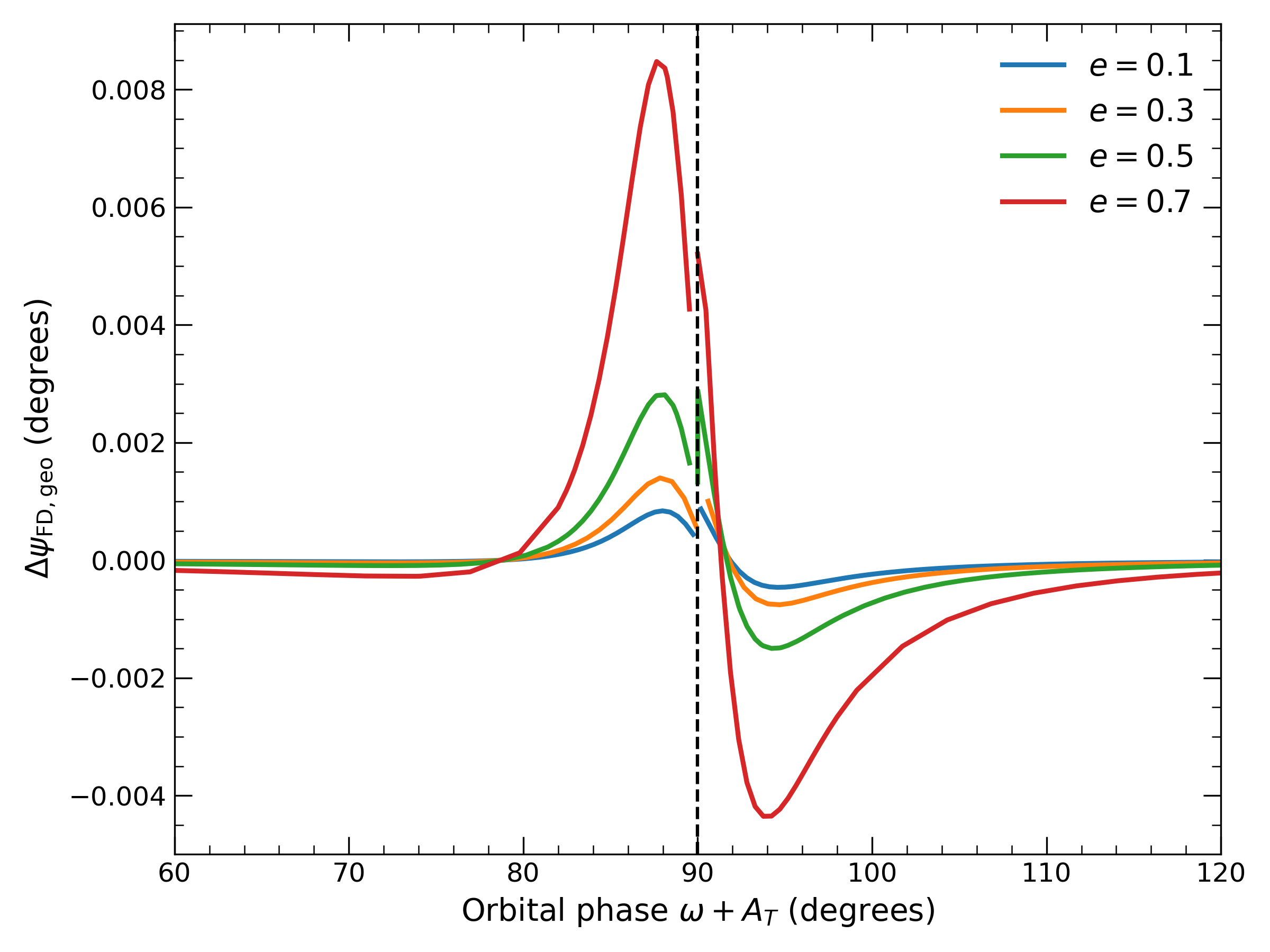}
    \caption{$\Delta\psi_{\rm FD,geo}$ for different values of $e$.}
    \label{fig:diff_e_psi_fd}
\end{subfigure}

\vspace{0.2cm}

\begin{subfigure}{0.3\textwidth}
    \centering
    \includegraphics[width=\linewidth]{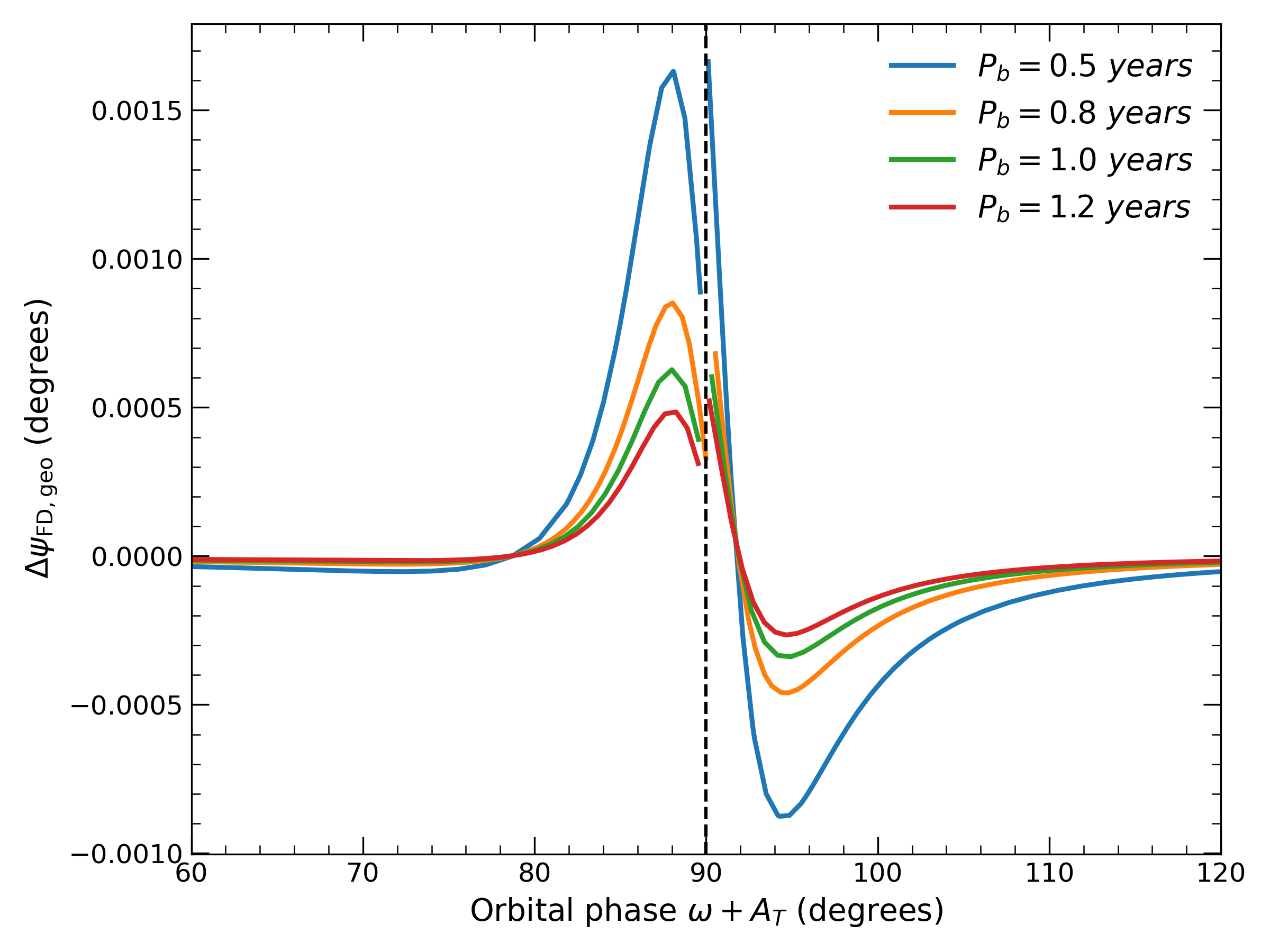}
    \caption{$\Delta\psi_{\rm FD,geo}$ for different values of $P_b$.}
    \label{fig:diff_period_psi_fd}
\end{subfigure}
\hfill
\begin{subfigure}{0.3\textwidth}
    \centering
    \includegraphics[width=\linewidth]{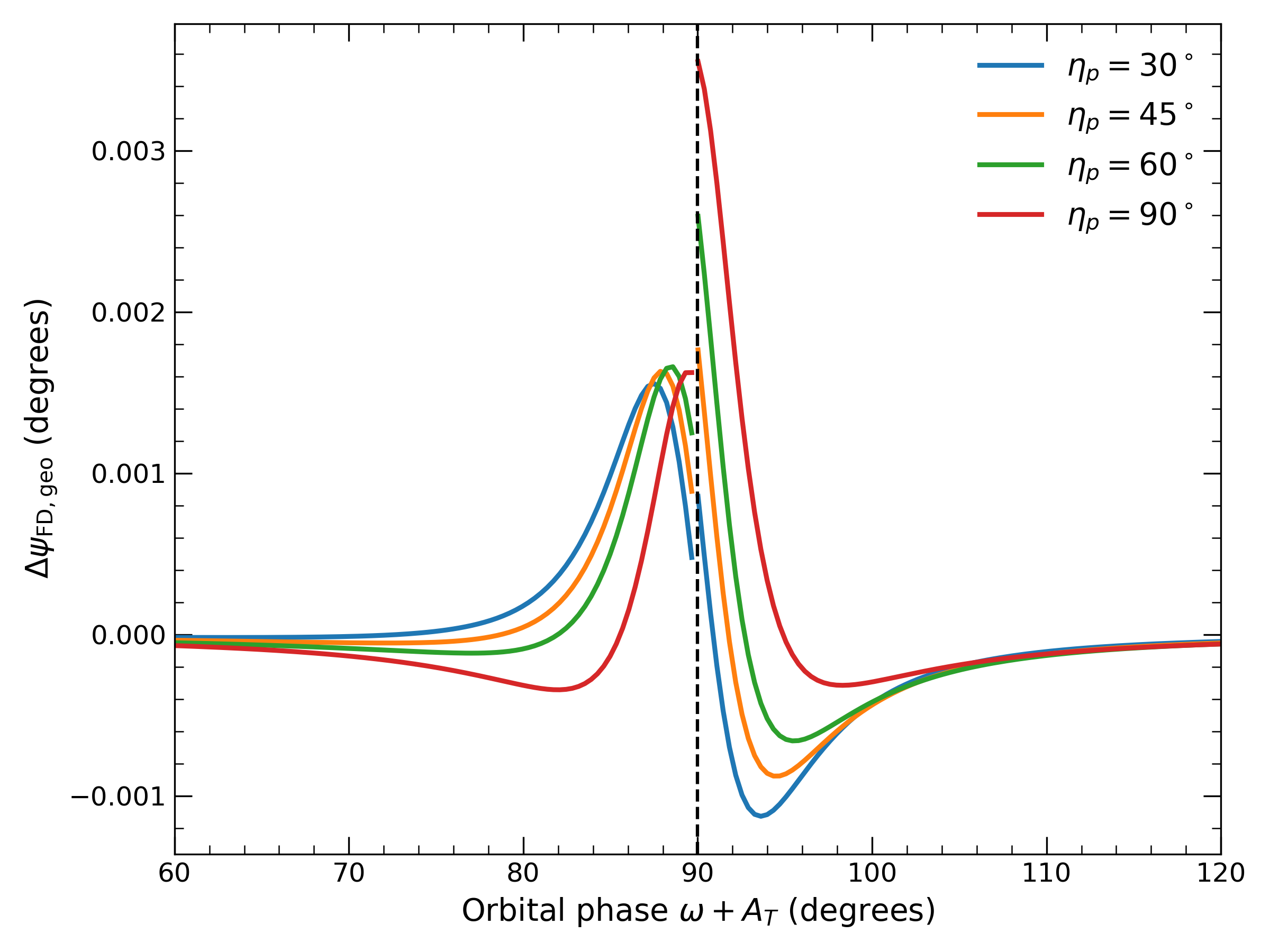}
    \caption{$\Delta\psi_{\rm FD,geo}$ for positive values of $\eta_p$.}
    \label{fig:diff_eta_psi_fd}
\end{subfigure}
\hfill
\begin{subfigure}{0.3\textwidth}
    \centering
    \includegraphics[width=\linewidth]{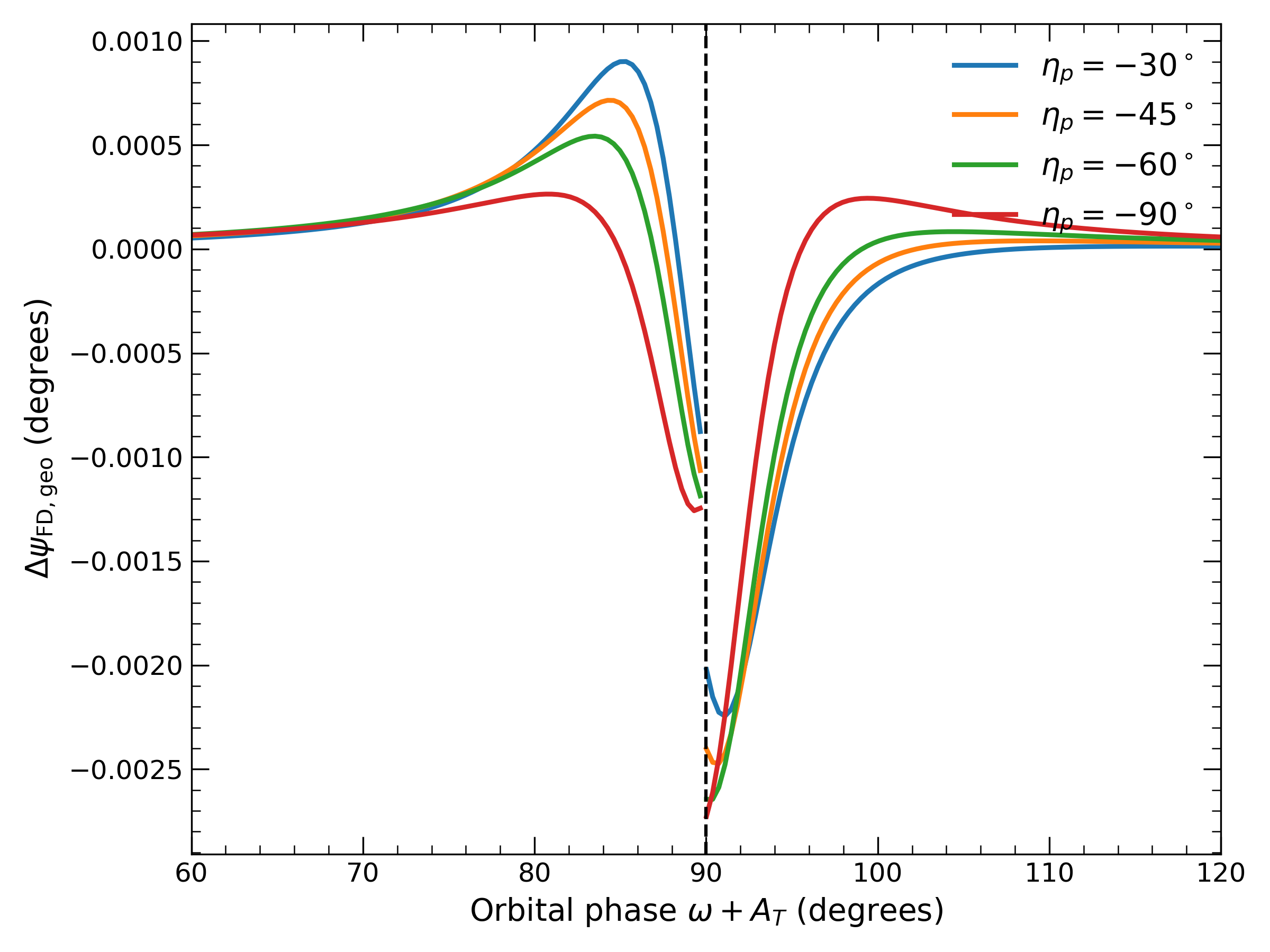}
    \caption{$\Delta\psi_{\rm FD,geo}$ for negative values of $\eta_p$.}
    \label{fig:diff_eta_neg_psi_fd}
\end{subfigure}

\caption{Dependence of the frame-dragging contribution to the
geometric polarization shift, $\Delta\psi_{\rm FD,geo}$, on the
binary parameters for the hypothetical pulsar--IMBH binary. In each panel, one parameter is varied while
the remaining parameters are fixed at the values given in
Table~\ref{tab:PSRintBH}. The panels show the dependence on
$M_c$, $i$, $e$, $P_b$, and positive and negative values of
$\eta_p$. Spin--orbit coupling is neglected.}
\label{diff_param_psi_fd}
\end{figure*}

\begin{figure*}[htbp]
\centering
\begin{subfigure}{0.3\textwidth}
    \centering
    \includegraphics[width=\linewidth]{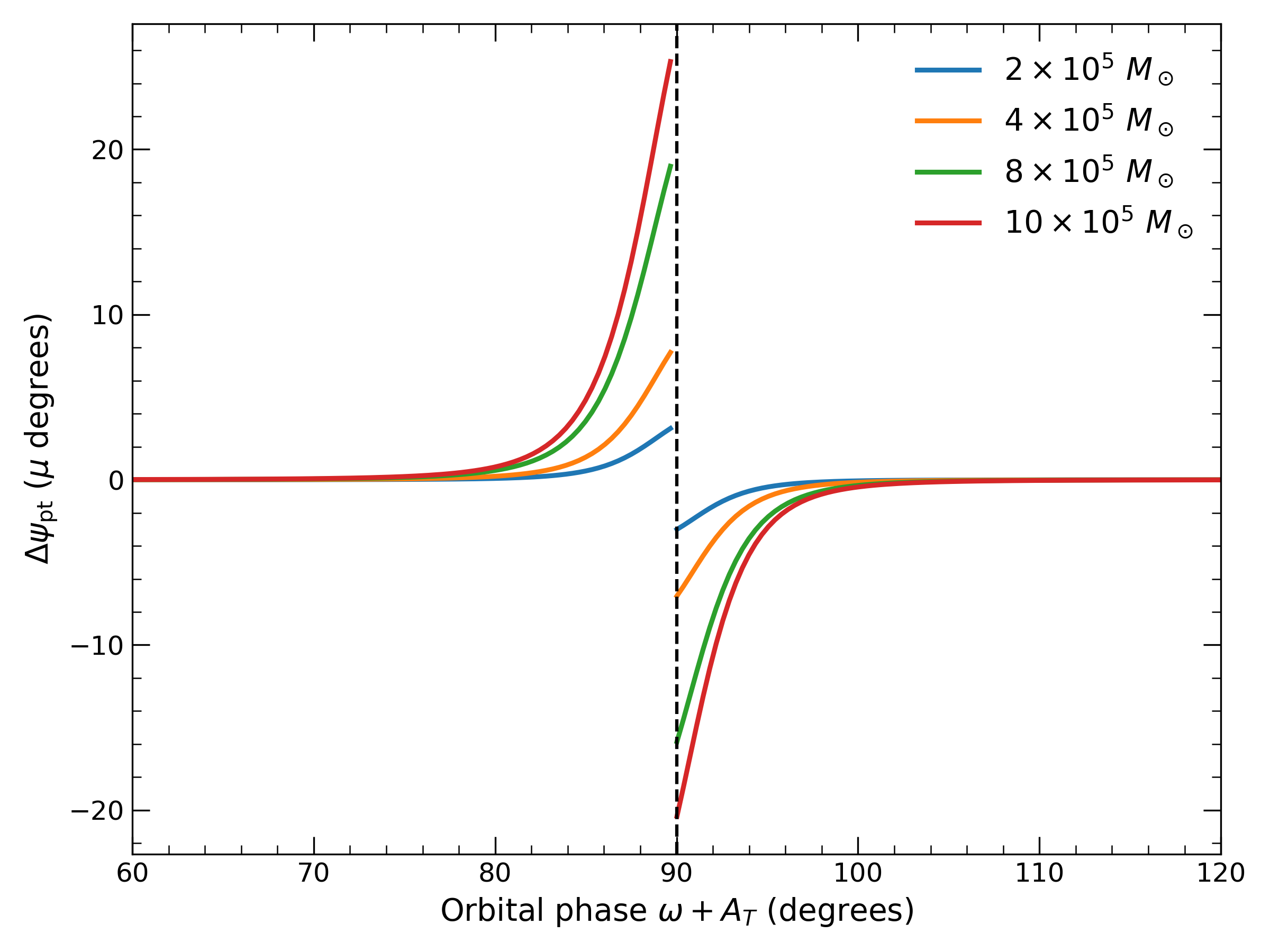}
    \caption{$\Delta\psi_{\rm pt}$ for different values of $M_c$.}
    \label{fig:diff_mass_psi_pt}
\end{subfigure}
\hfill
\begin{subfigure}{0.3\textwidth}
    \centering
    \includegraphics[width=\linewidth]{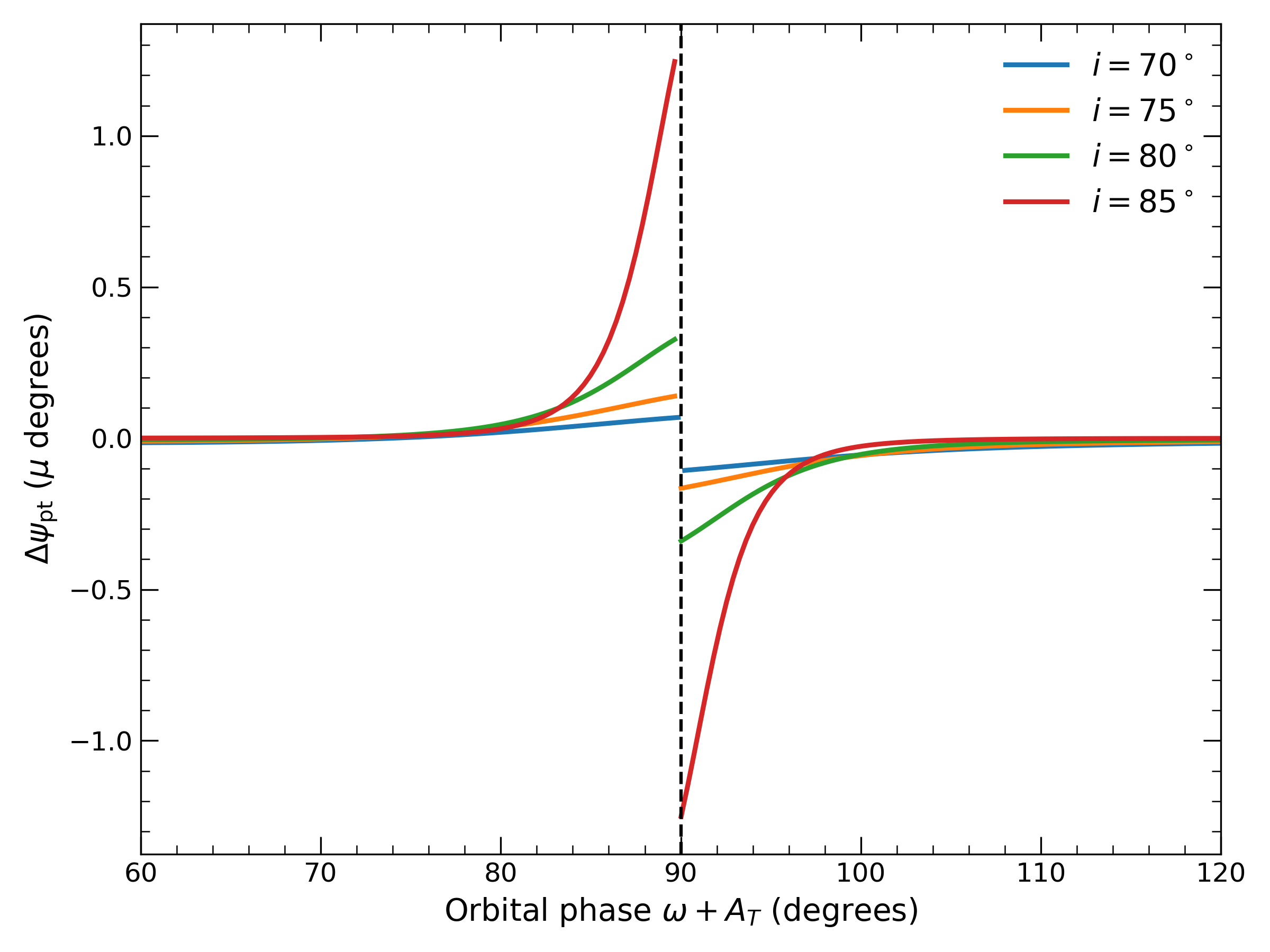}
    \caption{$\Delta\psi_{\rm pt}$ for different values of $i$.}
    \label{fig:diff_inclination_psi_pt}
\end{subfigure}
\hfill
\begin{subfigure}{0.3\textwidth}
    \centering
    \includegraphics[width=\linewidth]{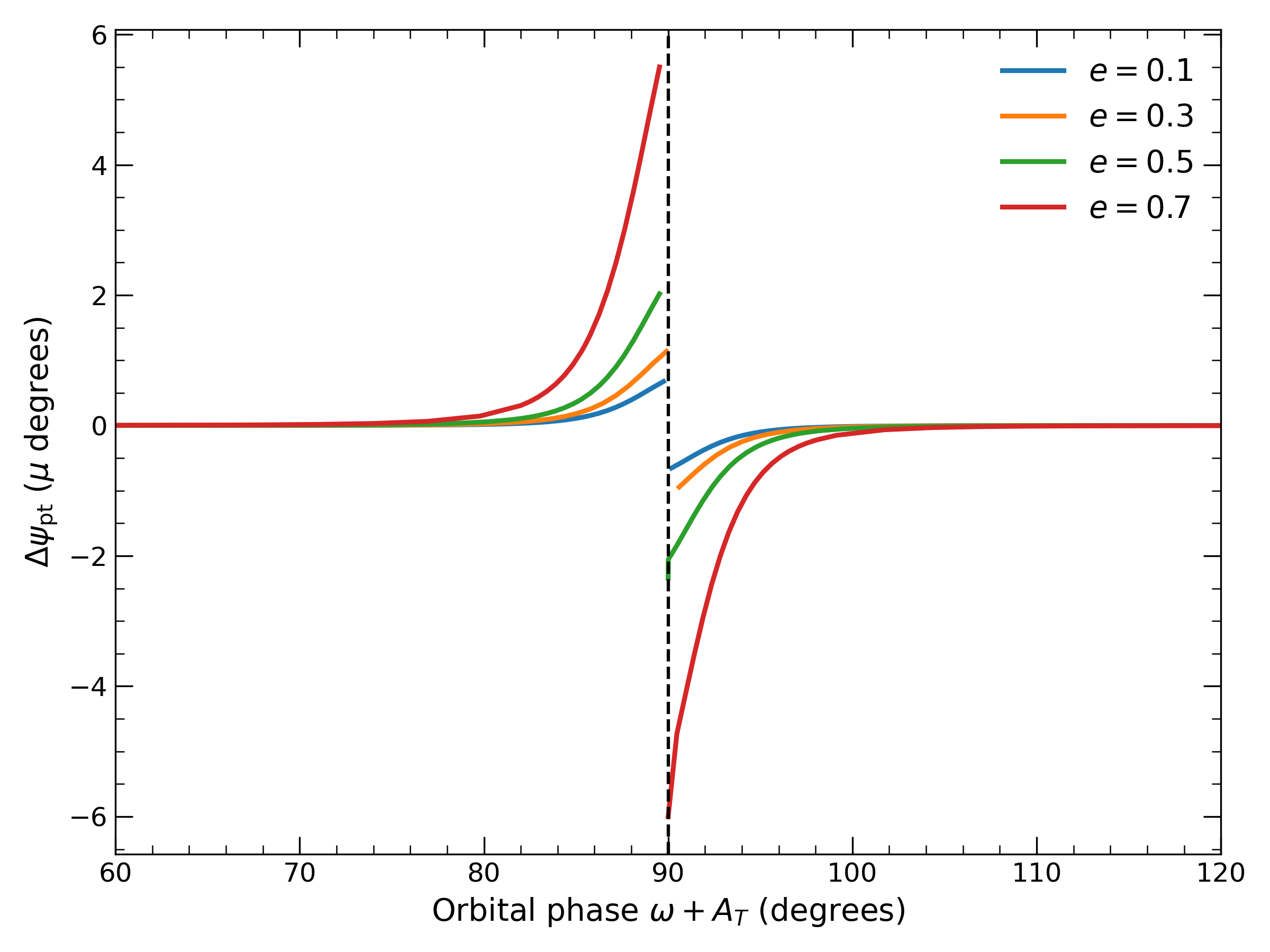}
    \caption{$\Delta\psi_{\rm pt}$ for different values of $e$.}
    \label{fig:diff_e_psi_pt}
\end{subfigure}

\vspace{0.2cm}

\begin{subfigure}{0.3\textwidth}
    \centering
    \includegraphics[width=\linewidth]{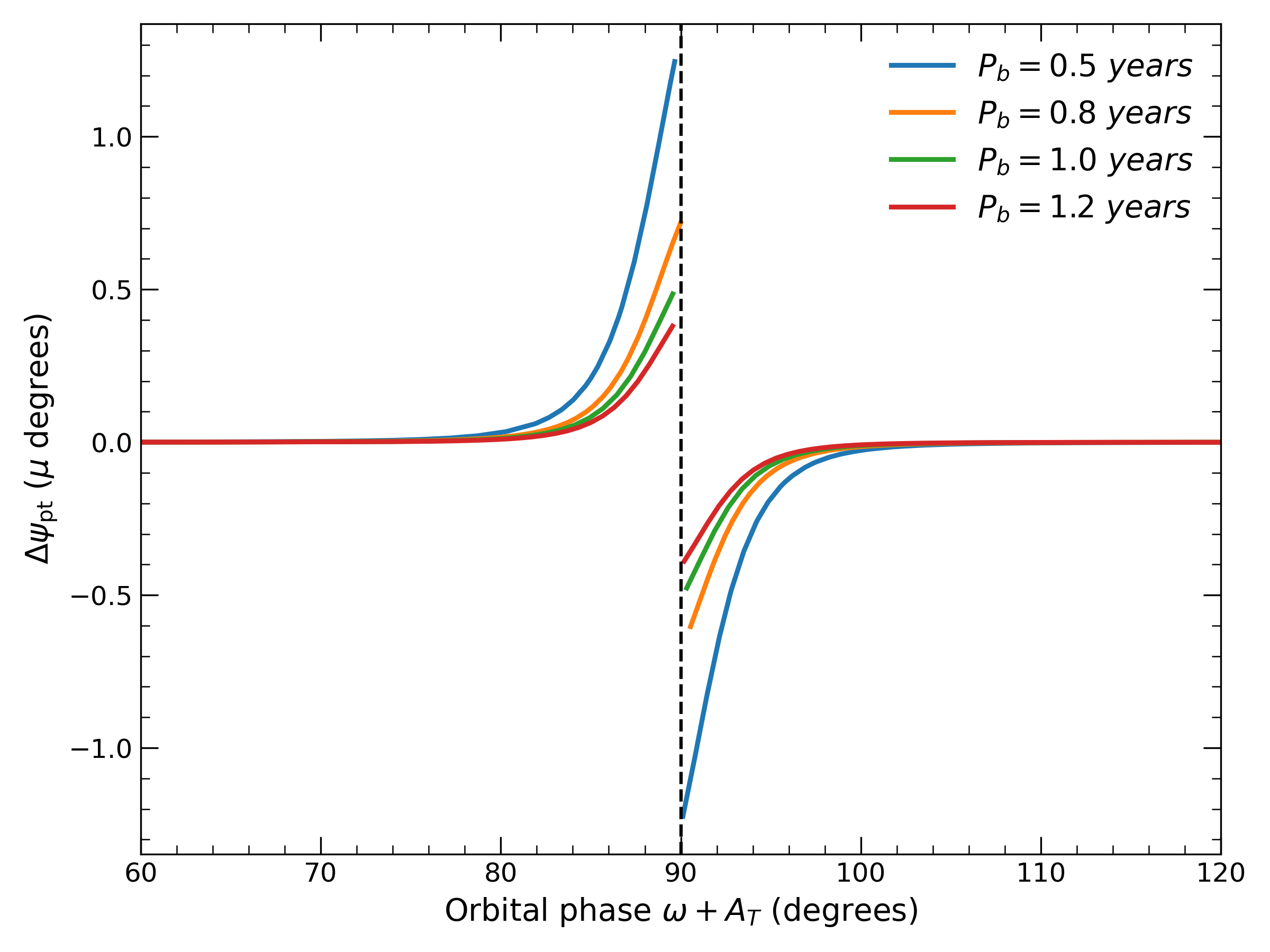}
    \caption{$\Delta\psi_{\rm pt}$ for different values of $P_b$.}
    \label{fig:diff_period_psi_pt}
\end{subfigure}
\hfill
\begin{subfigure}{0.3\textwidth}
    \centering
    \includegraphics[width=\linewidth]{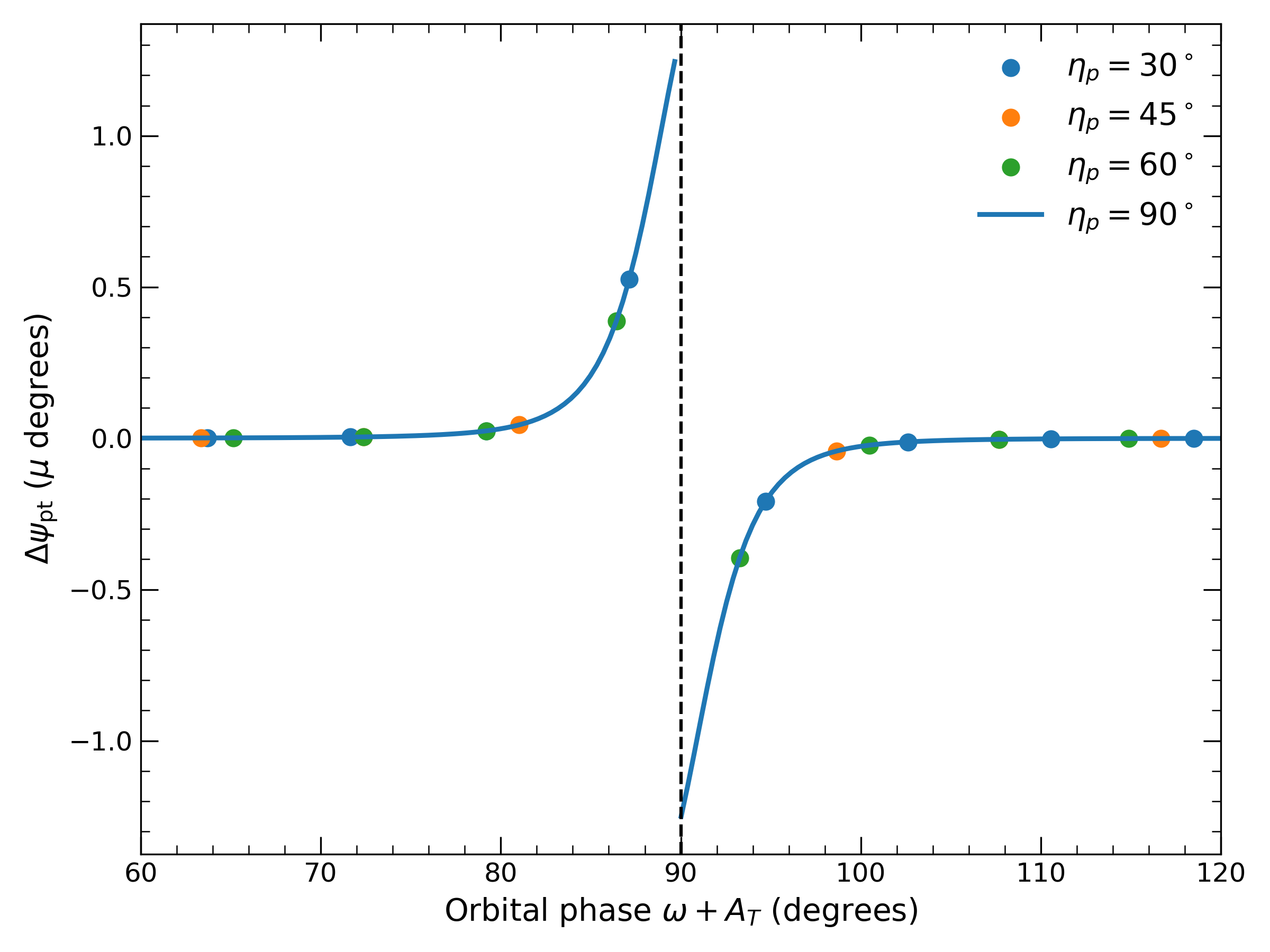}
    \caption{$\Delta\psi_{\rm pt}$ for positive values of $\eta_p$.}
    \label{fig:diff_eta_psi_pt}
\end{subfigure}
\hfill
\begin{subfigure}{0.3\textwidth}
    \centering
    \includegraphics[width=\linewidth]{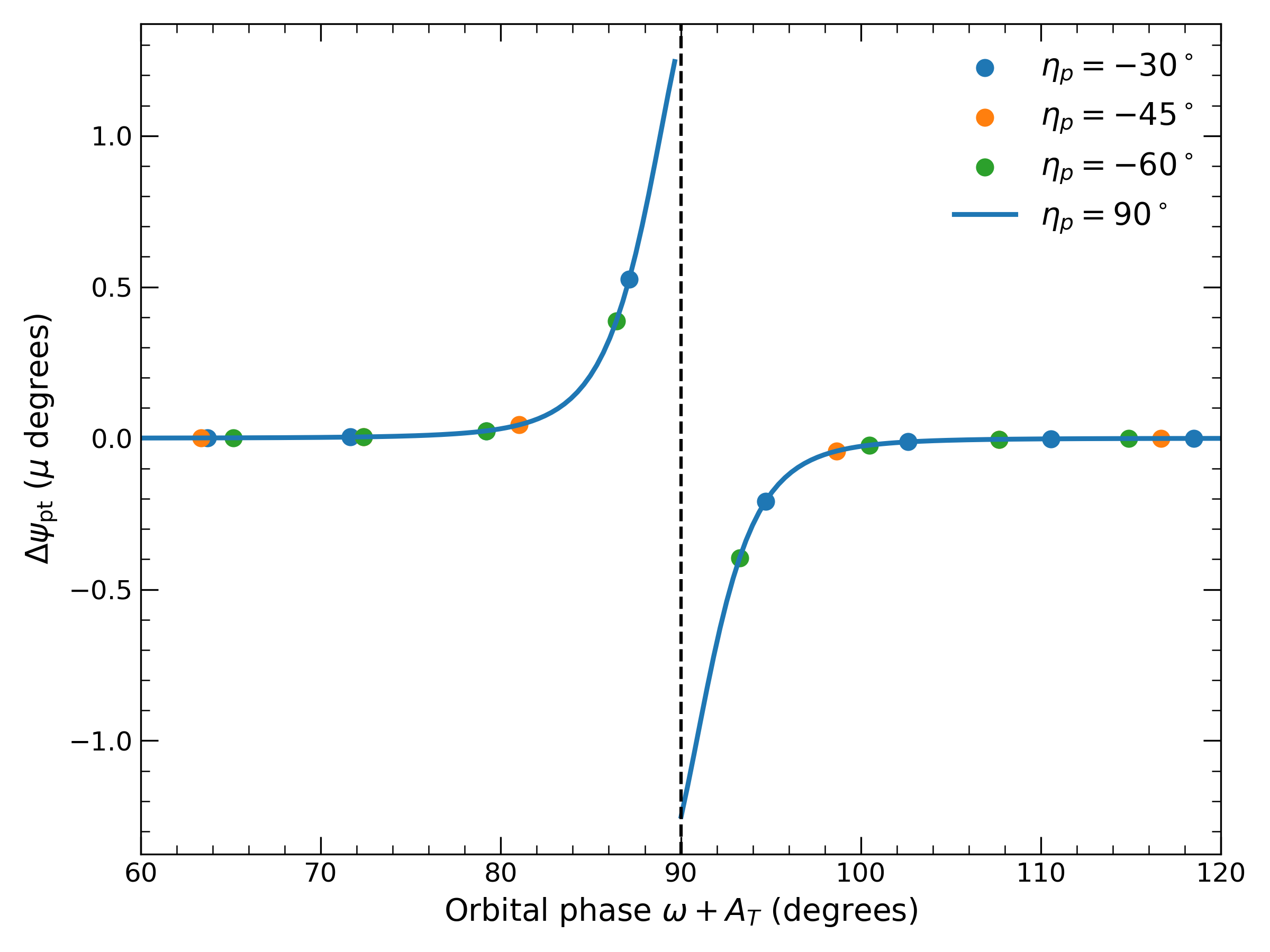}
    \caption{$\Delta\psi_{\rm pt}$ for negative values of $\eta_p$.}
    \label{fig:diff_eta_neg_psi_pt}
\end{subfigure}

\caption{Dependence of the polarization transport rotation
$\Delta\psi_{\rm pt}$ on the binary parameters for the hypothetical
pulsar--IMBH binary. In each panel, one
parameter is varied while the remaining parameters are fixed at
the values given in Table~\ref{tab:PSRintBH}. The panels show the
dependence on $M_c$, $i$, $e$, $P_b$, and positive and negative
values of $\eta_p$. Spin--orbit coupling is neglected.}
\label{diff_param_psi_pt}
\end{figure*}

We now investigate the dependence of the polarization rotations on
the binary parameters $M_c$, $P_b$, $e$, $i$, and $\eta_p$.
Figures~\ref{diff_param_psi_geo},
\ref{diff_param_psi_fd}, and \ref{diff_param_psi_pt} show the
variations of $\Delta\psi_{\rm geo}$,
$\Delta\psi_{\rm FD,geo}$, and $\Delta\psi_{\rm pt}$, respectively,
when one binary parameter is varied while the remaining parameters
are fixed at the values given in Table~\ref{tab:PSRintBH}. Since
the spin--orbit interaction is weak for the system considered here,
we neglect its effect in this parameter study.

Figures~\ref{fig:diff_mass_psi_geo}, \ref{fig:diff_mass_psi_fd},
and \ref{fig:diff_mass_psi_pt} show the variations of
$\Delta\psi_{\rm geo}$, $\Delta\psi_{\rm FD,geo}$, and
$\Delta\psi_{\rm pt}$, respectively, for companion masses
$M_c=2\times10^5$, $4\times10^5$, $8\times10^5$, and
$10^6\,M_\odot$. The magnitudes of all three quantities increase
with increasing companion mass, reflecting the stronger gravitational
bending and the associated changes in the polarization direction.

Figures~\ref{fig:diff_inclination_psi_geo},
\ref{fig:diff_inclination_psi_fd}, and
\ref{fig:diff_inclination_psi_pt} show the corresponding variations
for orbital inclinations $i=70^\circ$, $75^\circ$, $80^\circ$, and
$85^\circ$. The magnitudes of the polarization rotations increase
with increasing inclination. As the orbit becomes more edge-on, the
light rays pass closer to the companion near superior conjunction,
resulting in stronger gravitational bending.

The effect of orbital eccentricity is shown in
Figs.~\ref{fig:diff_e_psi_geo}, \ref{fig:diff_e_psi_fd}, and
\ref{fig:diff_e_psi_pt} for $e=0.1$, $0.3$, $0.5$, and $0.7$.
The magnitudes of $\Delta\psi_{\rm geo}$,
$\Delta\psi_{\rm FD,geo}$, and $\Delta\psi_{\rm pt}$ increase with
increasing eccentricity. For fixed orbital period and component
masses, a higher eccentricity brings the pulsar closer to the
companion near periastron, leading to stronger gravitational bending
\citep{DBB23}.

Figures~\ref{fig:diff_period_psi_geo},
\ref{fig:diff_period_psi_fd}, and
\ref{fig:diff_period_psi_pt} show the variations for orbital periods
$P_b=0.5$, $0.8$, $1.0$, and $1.2$ yr. The magnitudes of the
polarization rotations decrease with increasing orbital period.
For fixed component masses, a longer orbital period corresponds to
a wider orbit, increasing the characteristic separation between the
pulsar and the companion and reducing the gravitational bending.

We finally examine the dependence on the pulsar spin orientation
parameter $\eta_p$. Figures~\ref{fig:diff_eta_psi_geo},
\ref{fig:diff_eta_neg_psi_geo}, \ref{fig:diff_eta_psi_fd},
\ref{fig:diff_eta_neg_psi_fd}, \ref{fig:diff_eta_psi_pt}, and
\ref{fig:diff_eta_neg_psi_pt} show the results for positive and
negative values of $\eta_p$. The variations of
$\Delta\psi_{\rm geo}$ and $\Delta\psi_{\rm FD,geo}$ for positive
and negative $\eta_p$ are approximately symmetric about an orbital
phase of $90^\circ$. This dependence arises because the latitude
and longitude of the light ray reaching the observer after
gravitational bending depend on the orientation of the pulsar spin
\citep{DBB23}. Consequently, the geometric polarization shift and
its spin-induced change also depend on $\eta_p$. However,
$\Delta\psi_{\rm pt}$ shows no dependence on pulsar spin orientation angle $\eta_p$.

We also find that the orbital phase at which the geometric
polarization shift reaches its maximum or minimum depends on
$\eta_p$ and $i$, although it remains close to
$A_T+\omega=90^\circ$. We do not investigate the dependence on
$\lambda_p$ here because, for substantially different values of
$\lambda_p$, the pulsar would not be visible to the observer
\citep{DBB23}.

\begin{figure*}[htbp]
\centering
\begin{subfigure}{0.32\textwidth}
    \centering
    \includegraphics[width=\linewidth]{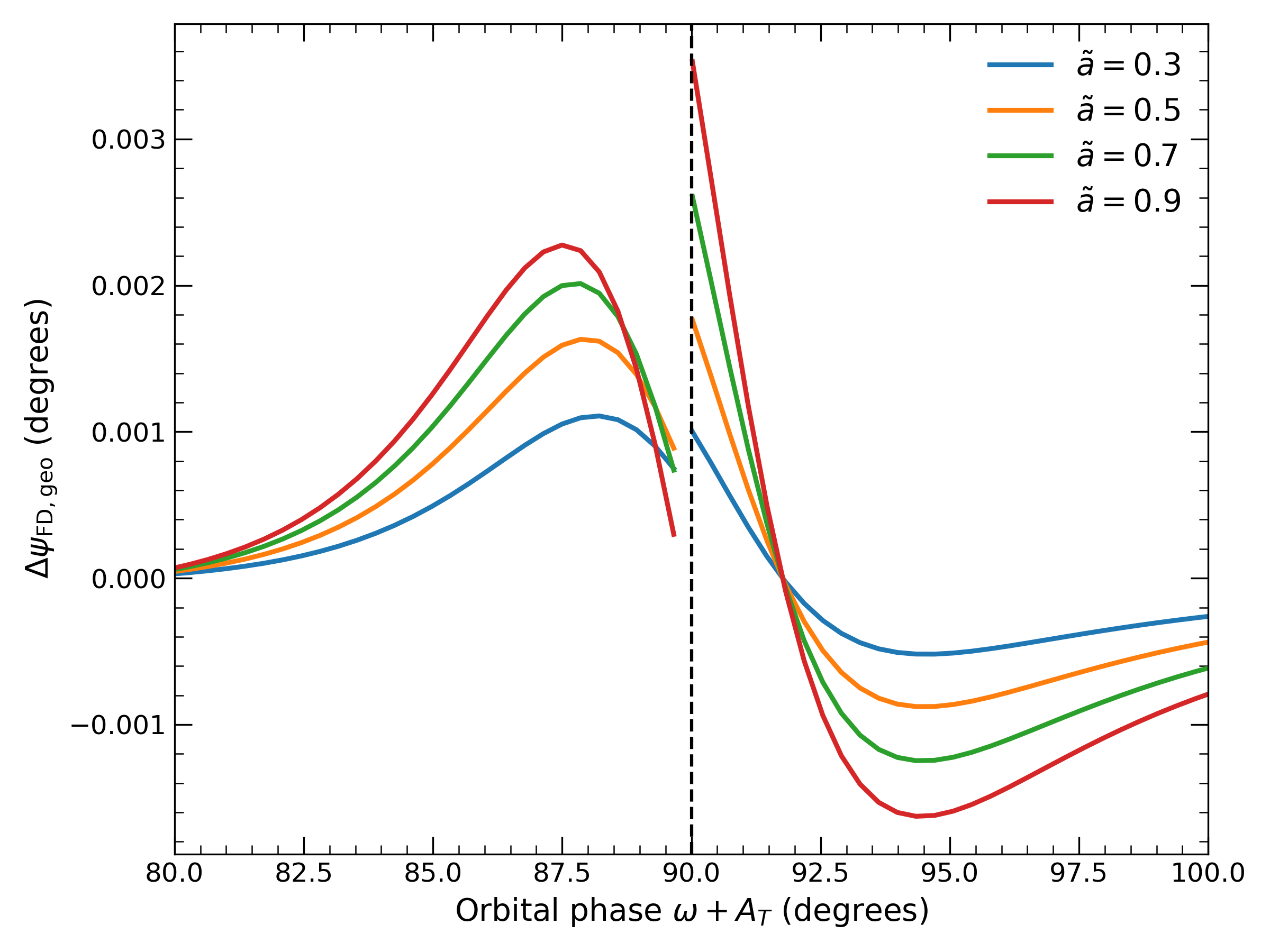}
    \caption{$\Delta\psi_{\rm FD,geo}$ for different values of $\tilde{a}$.}
    \label{fig:diff_spin_fdgeo}
\end{subfigure}
\hfill
\begin{subfigure}{0.32\textwidth}
    \centering
    \includegraphics[width=\linewidth]{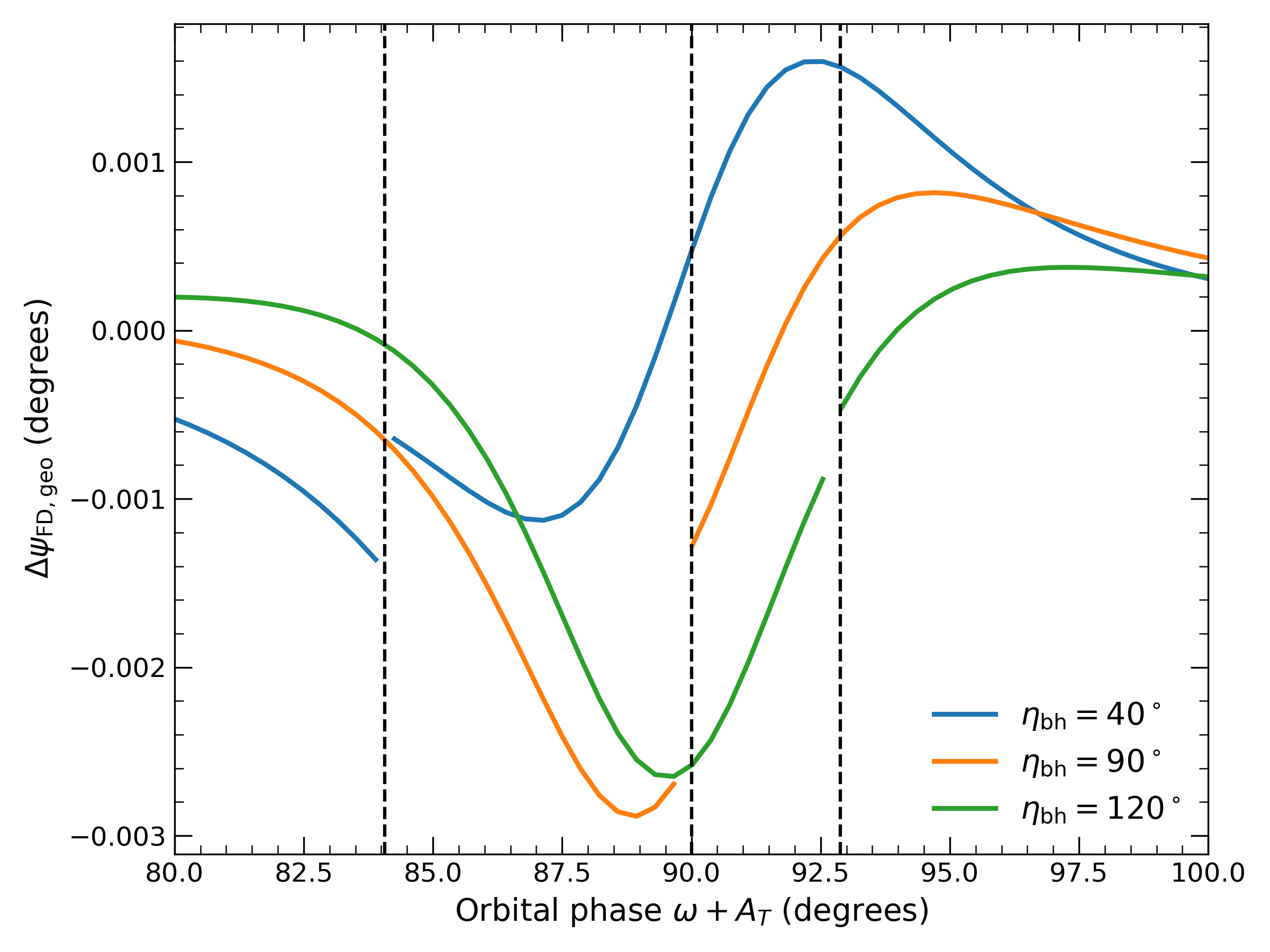}
    \caption{$\Delta\psi_{\rm FD,geo}$ for different values of $\eta_{\rm bh}$.}
    \label{fig:diff_etabh_fd}
\end{subfigure}
\hfill
\begin{subfigure}{0.32\textwidth}
    \centering
    \includegraphics[width=\linewidth]{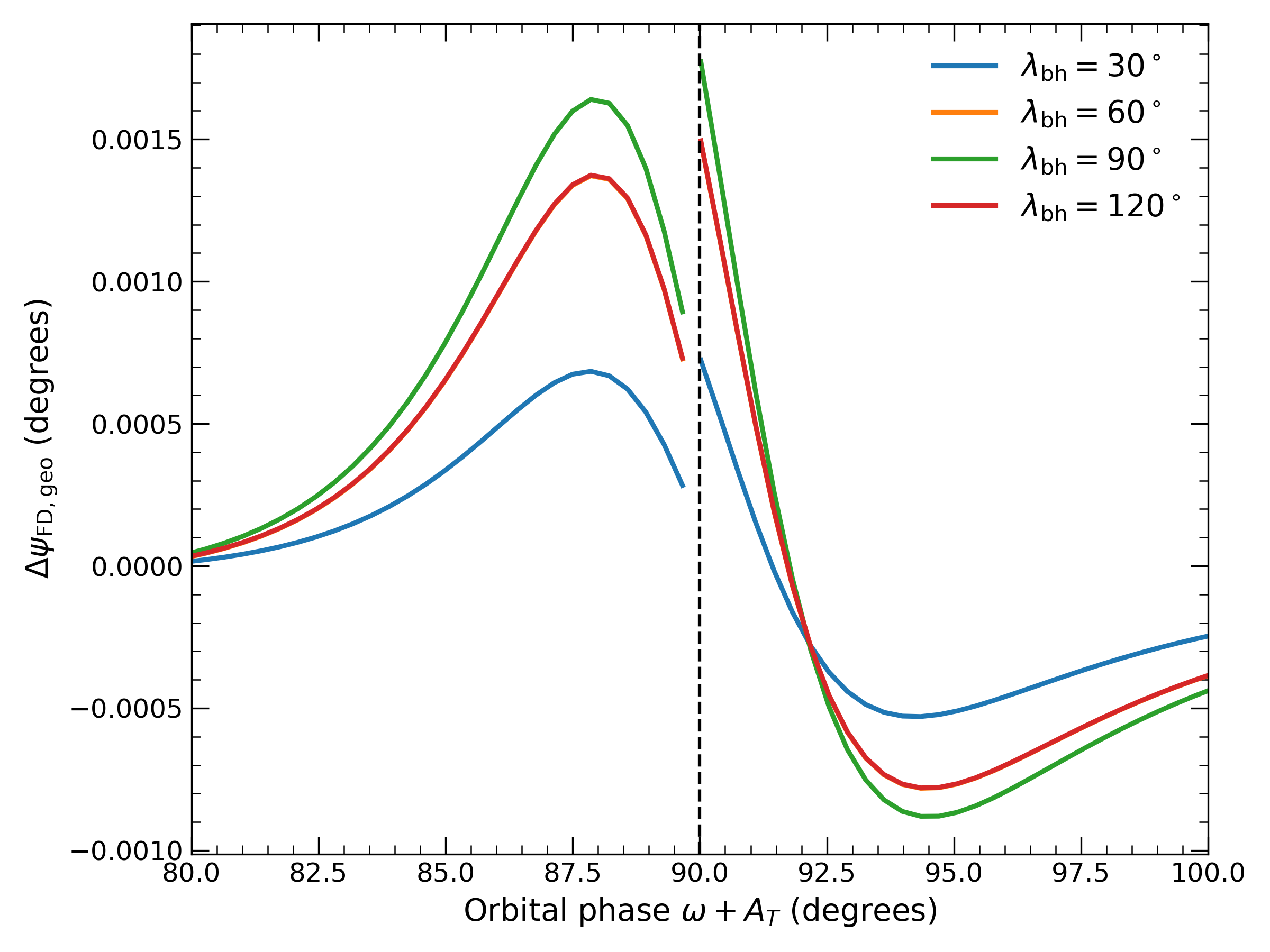}
    \caption{$\Delta\psi_{\rm FD,geo}$ for different values of $\lambda_{\rm bh}$.}
    \label{fig:diff_lamdabh_fd}
\end{subfigure}

\caption{Dependence of the frame-dragging contribution to the
geometric polarization shift, $\Delta\psi_{\rm FD,geo}$, on the
companion spin parameter $\tilde{a}$ and the spin orientation angles
$\eta_{\rm bh}$ and $\lambda_{\rm bh}$ for the hypothetical
pulsar--IMBH binary.}
\label{fig:diff_fdgeo}
\end{figure*}

\begin{figure*}[htbp]
\centering
\begin{subfigure}{0.32\textwidth}
    \centering
    \includegraphics[width=\linewidth]{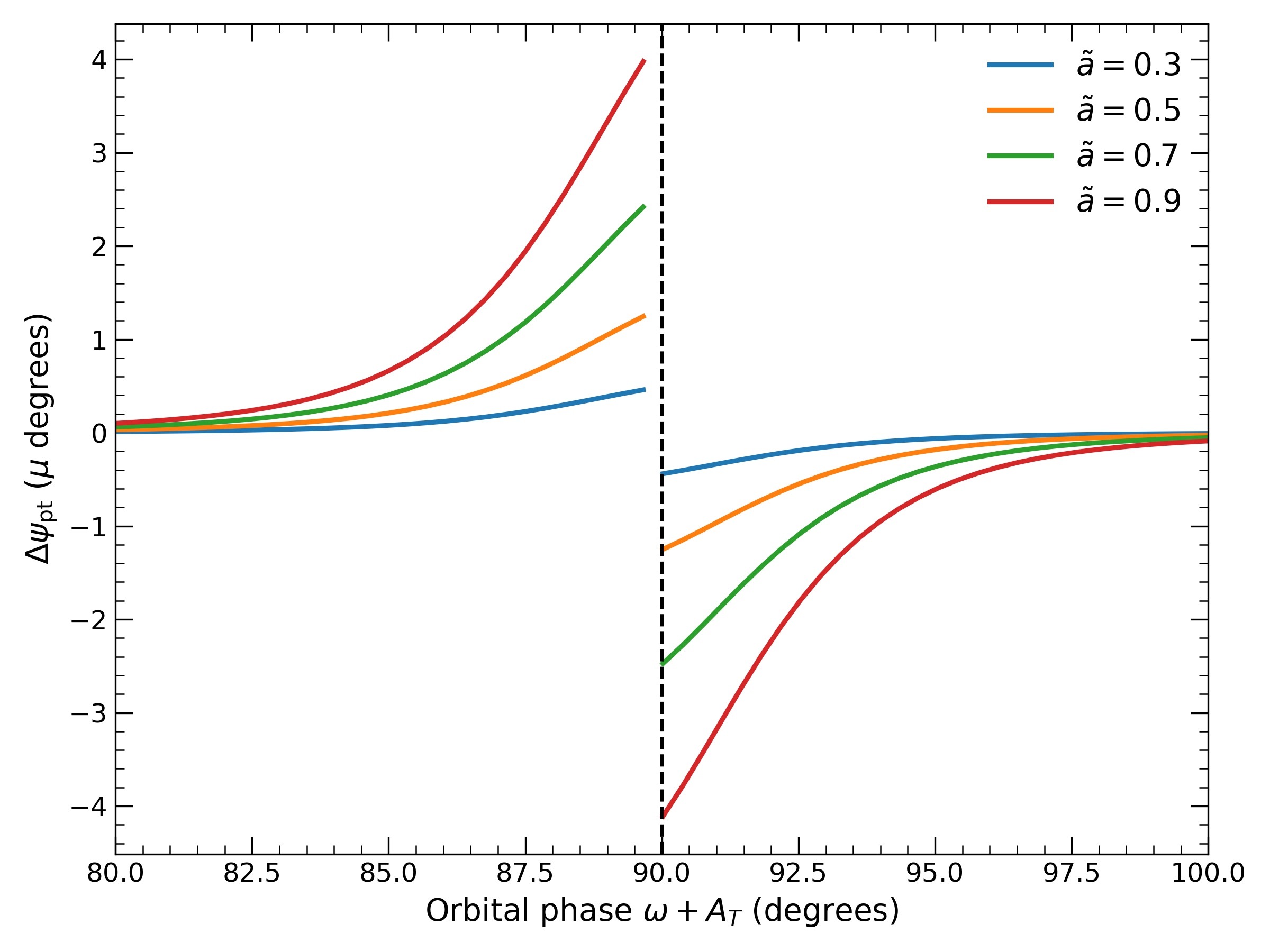}
    \caption{$\Delta\psi_{\rm pt}$ for different values of $\tilde{a}$.}
    \label{fig:diff_spin_pt}
\end{subfigure}
\hfill
\begin{subfigure}{0.32\textwidth}
    \centering
    \includegraphics[width=\linewidth]{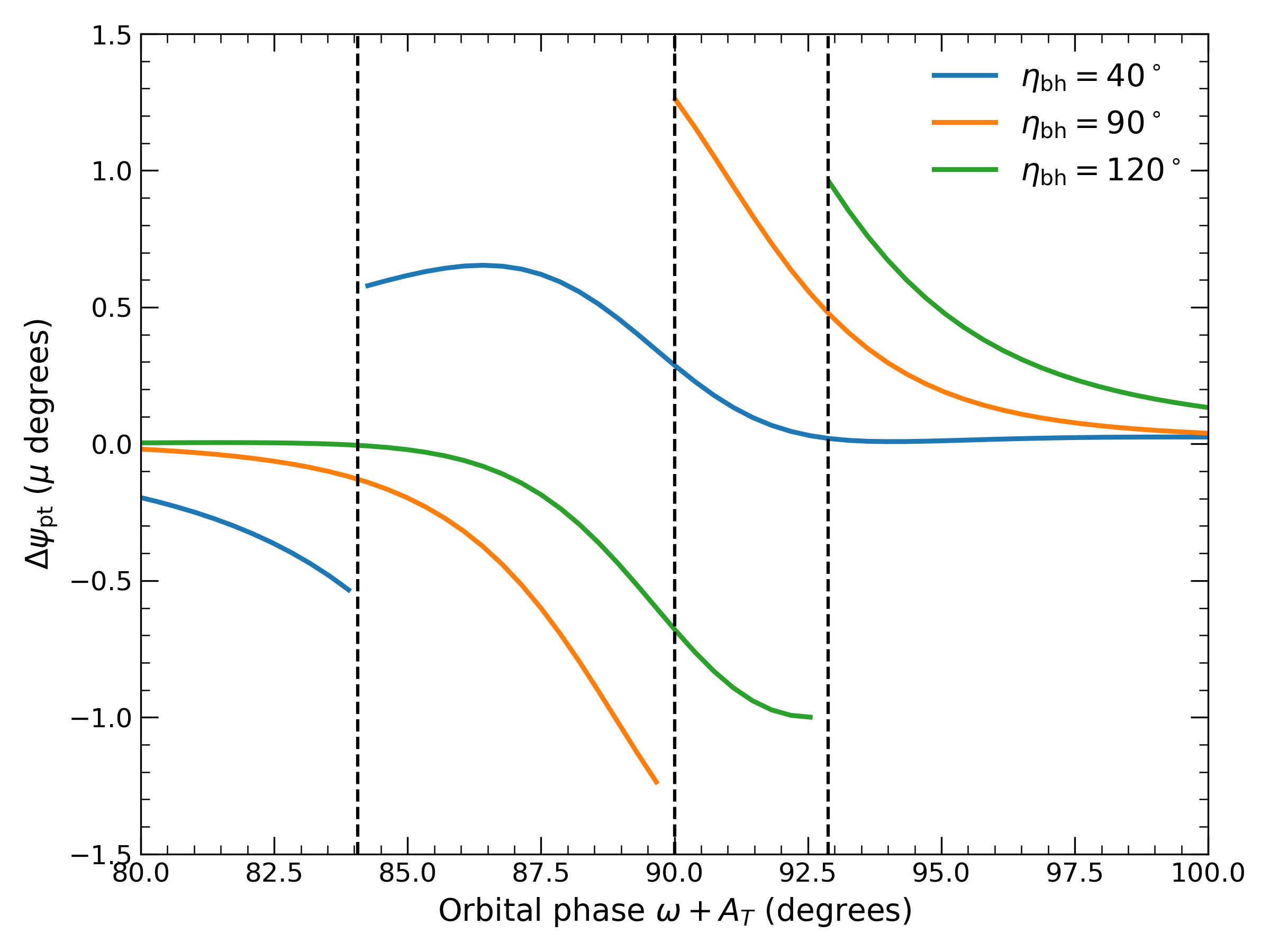}
    \caption{$\Delta\psi_{\rm pt}$ for different values of $\eta_{\rm bh}$.}
    \label{fig:diff_etabh_pt}
\end{subfigure}
\hfill
\begin{subfigure}{0.32\textwidth}
    \centering
    \includegraphics[width=\linewidth]{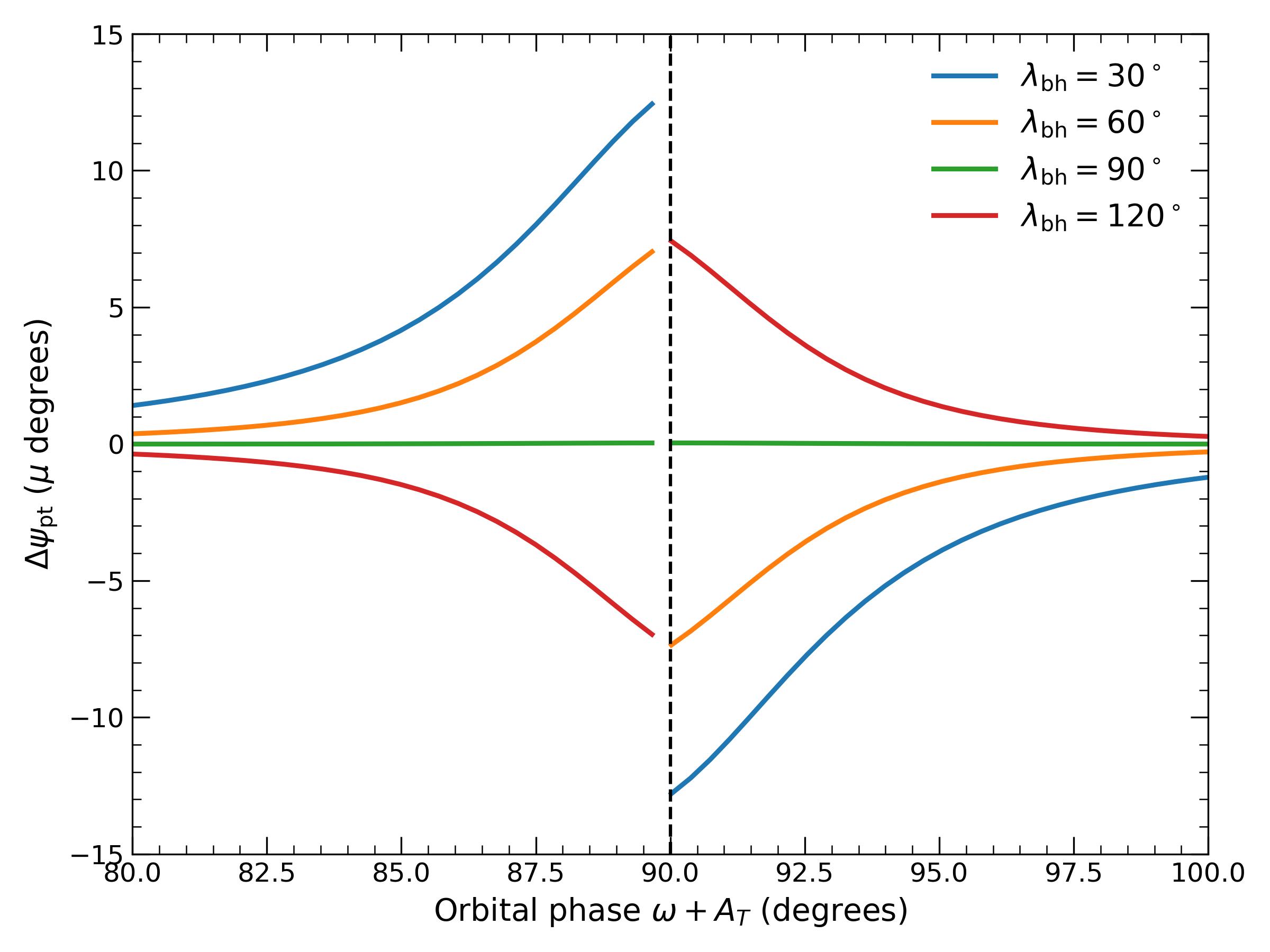}
    \caption{$\Delta\psi_{\rm pt}$ for different values of $\lambda_{\rm bh}$.}
    \label{fig:diff_lamdabh_pt}
\end{subfigure}

\caption{Dependence of the polarization transport rotation
$\Delta\psi_{\rm pt}$ on the companion spin parameter $\tilde{a}$
and the spin orientation angles $\eta_{\rm bh}$ and
$\lambda_{\rm bh}$ for the hypothetical pulsar--IMBH binary.}
\label{fig:diff_pt}
\end{figure*}

We next investigate the dependence of the polarization rotations
$\Delta\psi_{\rm FD,geo}$ and $\Delta\psi_{\rm pt}$ on the companion
spin parameter $\tilde{a}$ and the orientation of its spin axis,
specified by the angles $\eta_{\rm bh}$ and $\lambda_{\rm bh}$.
Figures~\ref{fig:diff_fdgeo} and \ref{fig:diff_pt} show the
corresponding variations for different values of these parameters.

Figures~\ref{fig:diff_spin_fdgeo} and \ref{fig:diff_spin_pt} show
the dependence on the companion spin parameter $\tilde{a}$. Both
$\Delta\psi_{\rm FD,geo}$ and $\Delta\psi_{\rm pt}$ increase in
magnitude with increasing spin,
indicating a stronger spin-induced contribution to the polarization
rotation.

The dependence on the spin orientation angle $\eta_{\rm bh}$ is
shown in Figs.~\ref{fig:diff_etabh_fd} and \ref{fig:diff_etabh_pt}.
Changing $\eta_{\rm bh}$ changes the orientation of the companion's
spin axis relative to the observer. Consequently, both polarization
rotations change, and the orbital phase at which the discontinuity
occurs also shifts in accordance with
Eq.~\ref{eq:discontinuity_condition}.

Finally, Figs.~\ref{fig:diff_lamdabh_fd} and
\ref{fig:diff_lamdabh_pt} show the dependence on the spin orientation
angle $\lambda_{\rm bh}$. This angle changes the orientation of the
spin axis relative to the orbital geometry without changing its
orientation relative to the observer in the same way as
$\eta_{\rm bh}$. The magnitudes of both polarization rotations
therefore vary with $\lambda_{\rm bh}$, while the orbital phase of
the discontinuity remains fixed.

One of the most important predictions of the rotating vector model
(RVM) is the characteristic polarization position-angle curve, commonly
referred to as the polarization S-curve. It describes the variation of
the polarization position angle $\psi$ of the observed radiation as a
function of the rotational phase $\Phi_m$ of the pulsar. Since the
pulsar emission beam has a finite width, the observer receives
radiation over a range of rotational phases $\Phi_m$, which determines
the width of the observed pulse profile \citep{DB26}. If
$\widehat{F}_{\rm sky}(\Phi_m)$ denotes the polarization vector of the
observed light ray at the rotational phase
$\Phi_m$, the corresponding polarization position angle is given by
\begin{equation}
\label{PPA}
\psi
=
\operatorname{atan2}
\left(
\widehat{F}_{{\rm sky},y}(\Phi_m),
\widehat{F}_{{\rm sky},x}(\Phi_m)
\right),
\end{equation}
where $\widehat{F}_{{\rm sky},x}$ and
$\widehat{F}_{{\rm sky},y}$ are the $x$- and $y$-components,
respectively, of the polarization vector in the sky plane.
The gravitational bending of the light ray changes its direction of
propagation and consequently modifies the polarization direction
observed on the sky. The rotation of the polarization vector discussed
in Section~\ref{definitions} therefore produces a corresponding
modification of the intrinsic polarization S-curve. The magnitude of
this modification depends on the orbital position of the pulsar, since
the gravitational effect of the companion varies along the orbit.
\begin{figure*}[htbp]
\centering
\begin{subfigure}{0.48\textwidth}
    \centering
    \includegraphics[width=\linewidth]{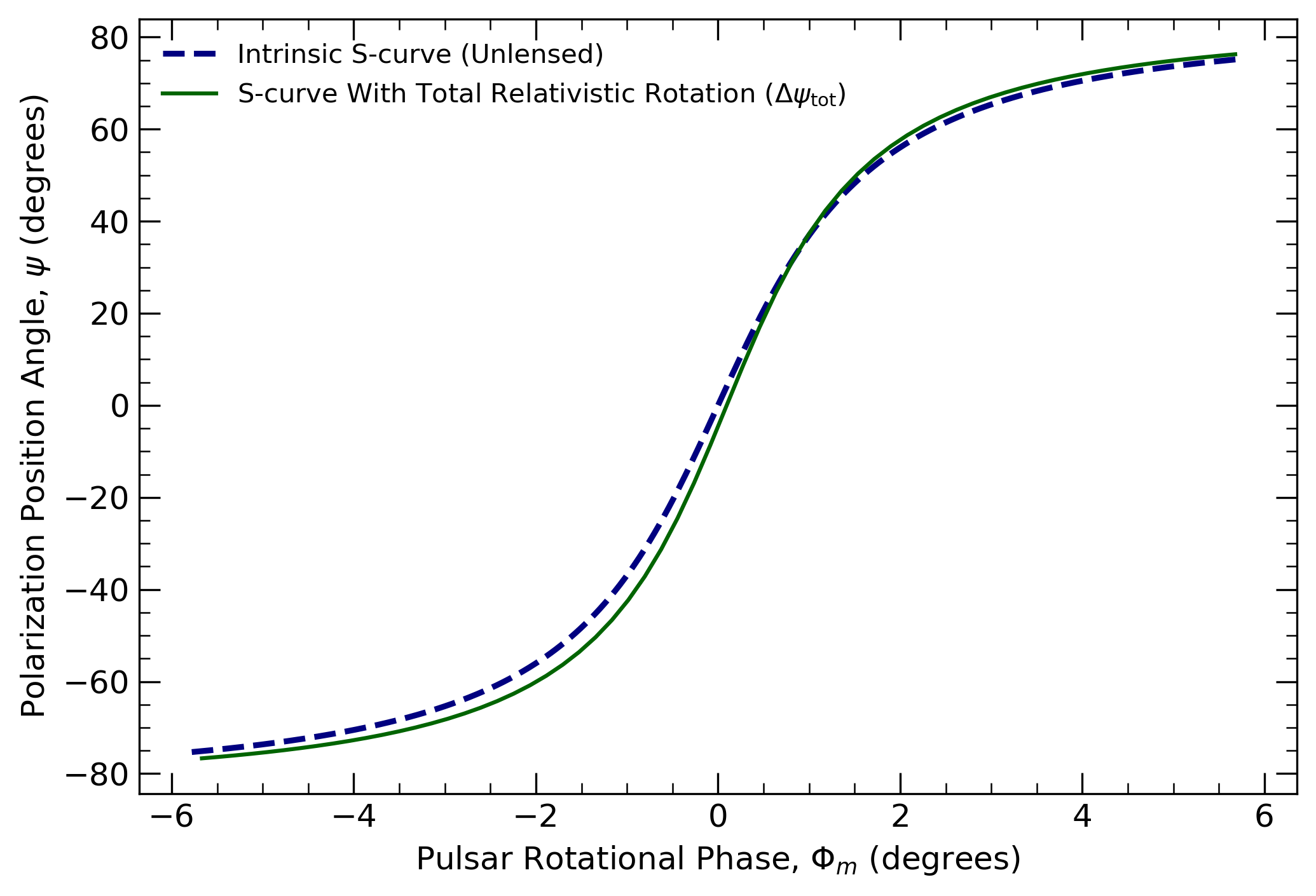}
    \caption{Orbital phase $A_T+\omega=89^\circ$.}
    \label{fig:scurve_89}
\end{subfigure}
\hfill
\begin{subfigure}{0.48\textwidth}
    \centering
    \includegraphics[width=\linewidth]{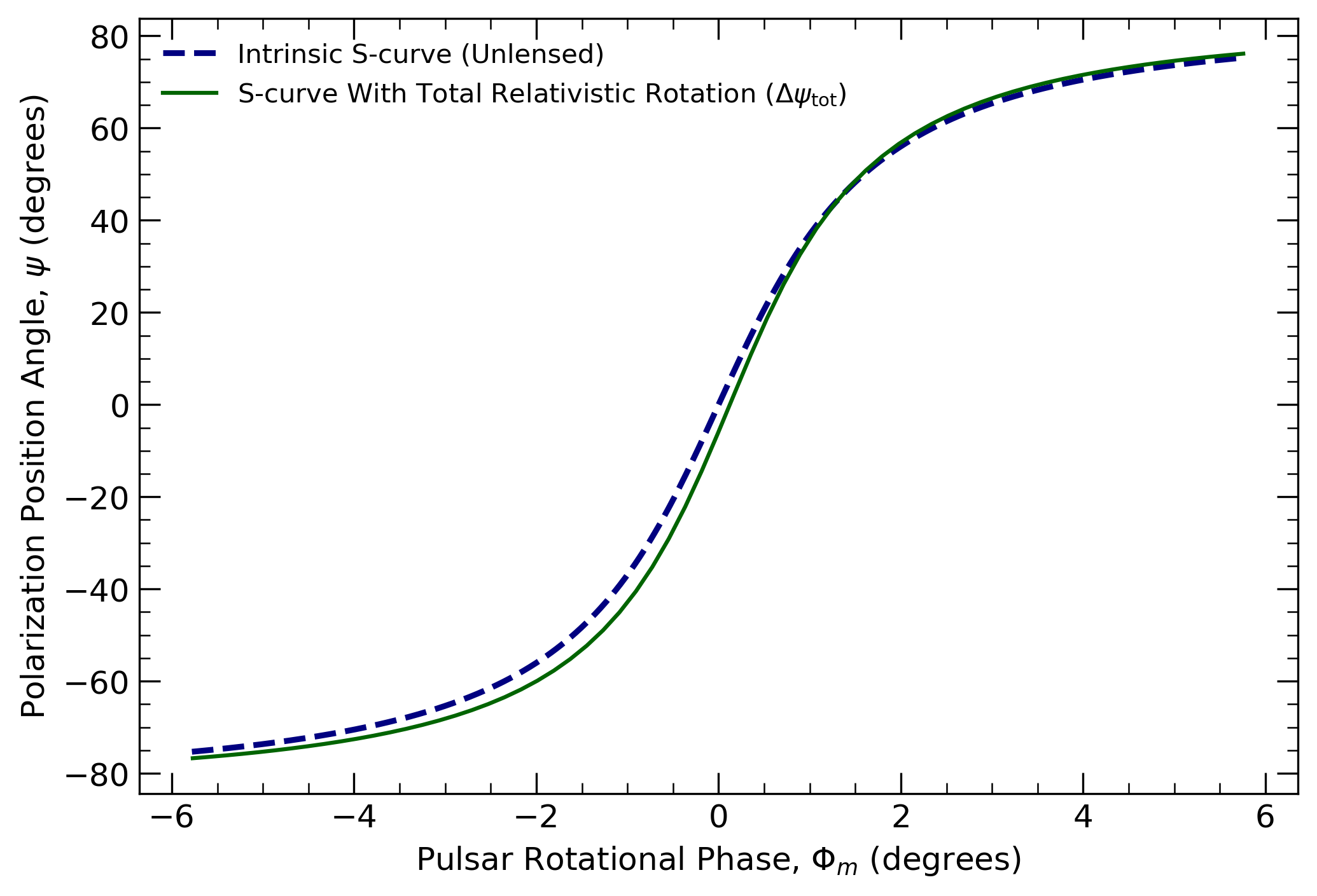}
    \caption{Orbital phase $A_T+\omega=90^\circ$.}
    \label{fig:scurve_90}
\end{subfigure}

\vspace{0.2cm}

\begin{subfigure}{0.48\textwidth}
    \centering
    \includegraphics[width=\linewidth]{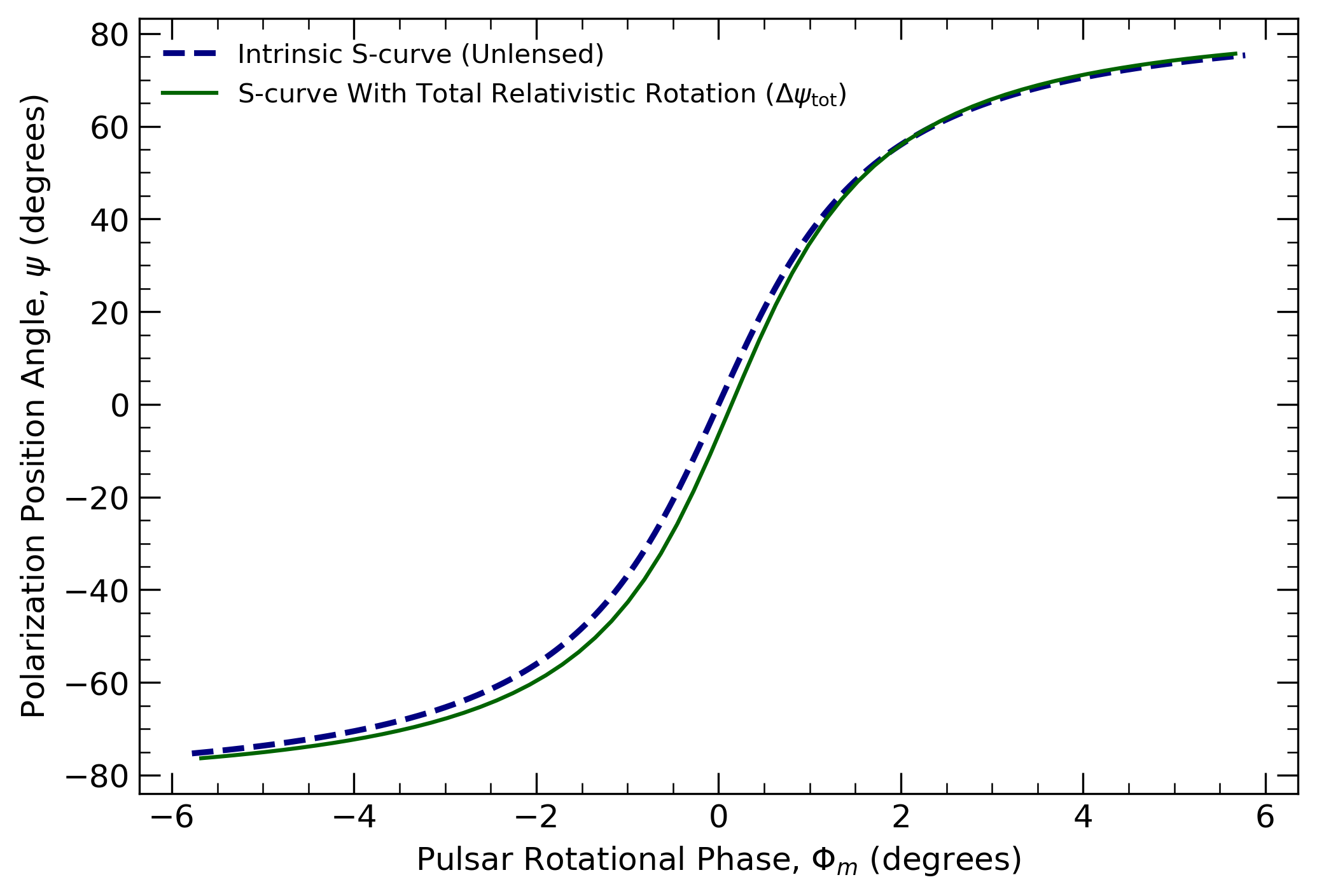}
    \caption{Orbital phase $A_T+\omega=91^\circ$.}
    \label{fig:scurve_91}
\end{subfigure}
\hfill
\begin{subfigure}{0.48\textwidth}
    \centering
    \includegraphics[width=\linewidth]{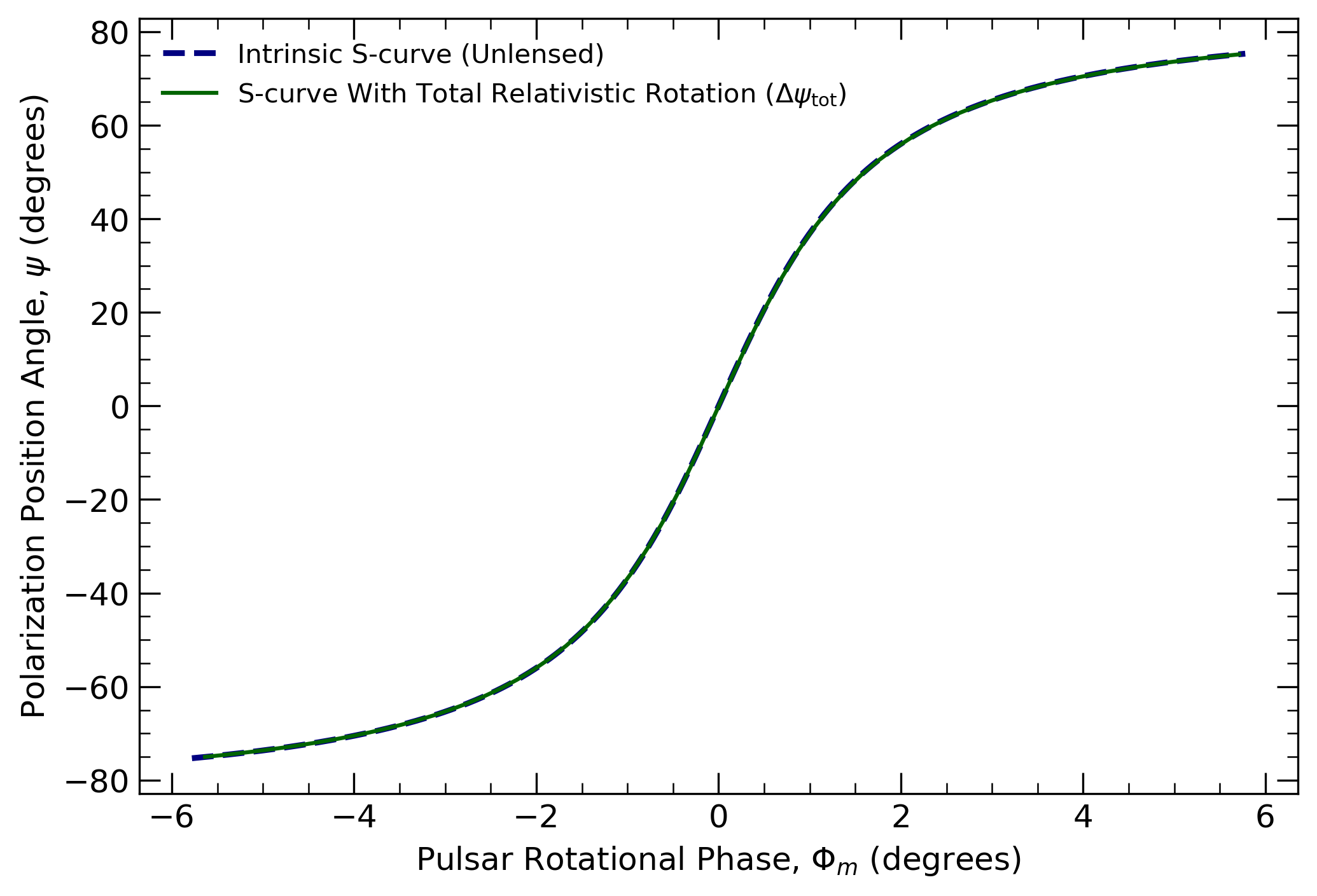}
    \caption{Orbital phase $A_T+\omega=200^\circ$.}
    \label{fig:scurve_200}
\end{subfigure}

\caption{Variations of the polarization position angle $\psi$ with the
pulsar rotational phase $\Phi_m$ at different orbital phases of the
hypothetical pulsar--IMBH binary. The orbital
phases $A_T+\omega=89^\circ$, $90^\circ$, and $91^\circ$ correspond to
the region close to superior conjunction, where the gravitational
effect of the companion is strongest. The result at
$A_T+\omega=200^\circ$ is also shown for comparison, where the
gravitational effect is considerably weaker.}
\label{S-curve}
\end{figure*}

In Fig.~\ref{S-curve}, we show the polarization position angle as a
function of the pulsar rotational phase at the orbital phases
$A_T+\omega=89^\circ$, $90^\circ$, $91^\circ$, and $200^\circ$. The
polarization S-curve is significantly modified at the orbital phases
$A_T+\omega=89^\circ$, $90^\circ$, and $91^\circ$, where the pulsar is
close to superior conjunction and the gravitational effect on the
propagating light rays is strongest. The modification of the S-curve is
largest close to $A_T+\omega=90^\circ$, corresponding to the region
where the gravitational bending is expected to be maximum.

In contrast, at the orbital phase $A_T+\omega=200^\circ$, the
gravitational effect of the companion is much weaker. Consequently, the
corresponding change in the polarization S-curve becomes almost
insignificant. These results show that the distortion of the observed
polarization S-curve depends strongly on the orbital phase and is
largest when the light rays propagate through the strongest region of
the gravitational field of the companion.

\subsection{A Test Case of a Pulsar--Supermassive Black Hole Close Binary}
\label{subsec:supermassive_bh}

We next consider a hypothetical pulsar--supermassive black hole (SMBH)
close binary. In such a system, strong-field general relativistic (GR)
effects can significantly modify the viewing geometry of the pulsar.
The orbital and spin evolution can also be sufficiently strong that
spin-orbit evolution cannot be ignored, even over a single orbital
period. Since no such binary has been discovered so far, we adopt the
baseline parameters in Table~\ref{tab:PSRintBH}, except that we set the
companion mass to $M_c=10^6\,{\rm M_{\odot}}$, the orbital period to
$P_b=5.5$ days, and the inclination to $i=70^\circ$. For these
parameters, the maximum and minimum separations are
$833.83\,GM_c/c^2$ and $401.5\,GM_c/c^2$, respectively. These
separations allow us to use the formalism described in
Section~\ref{subsec:orbitandspinevolution} to model the spin-orbit
evolution of the binary.

We choose a relatively low inclination angle because the Einstein
radius of the massive companion can be much larger than the pulsar
beam, leading to substantial beam and pulse-profile distortion
\citep{DB25}. In \citep{DB25}, we showed that this distortion remains
significant for inclination angles up to $80^\circ$. Since severe beam
distortion would complicate the analysis of the linear polarization
geometry, we set $i=70^\circ$ in the present study.

We first evaluate the changes in the system parameters over one
complete orbit. We adopt the baseline values at $t=0$ from
Table~\ref{tab:PSRintBH}, with $M_c=10^6\,{\rm M_{\odot}}$,
$P_b=5.5$ days, and $i=\lambda_{\rm bh}=70^\circ$. The evolutionary
framework described in Section~\ref{subsec:orbitandspinevolution}
predicts appreciable changes in some parameters over a single orbital
period. The longitude of periastron $\omega$ changes by $2.01^\circ$
and $1.95^\circ$ for companion spin parameters $\tilde{a}=0$ and
$0.5$, respectively. The pulsar spin-orientation angle $\eta_p$
decreases by $0.207^\circ$ and $0.204^\circ$, while $\lambda_p$
decreases by $0.66^\circ$ and $0.65^\circ$, respectively. In
contrast, the secular changes in the orbital elements are nearly
insensitive to the companion spin. Over one complete orbit, the orbital
period changes by $\Delta P_b=-2.37\times10^{-5}$~s, the eccentricity
by $\Delta e=-1.23\times10^{-11}$, and the inclination by
$\Delta i=7.95\times10^{-12\circ}$.

More rigorous formalisms have been developed for tight and highly
relativistic binary systems. In particular, \citep{Dsing2014},
\citep{KJLi2019}, and \citep{lwls22} investigated pulsars orbiting
massive and supermassive black holes by including spin-spin and
spin-curvature interactions through the Mathisson--Papapetrou--Dixon
(MPD) equations in Kerr spacetime. These studies considered fast
millisecond pulsars with $P_s=1$~ms at separations of
$10\text{--}100\,GM_c/c^2$, where strong-field GR effects become
important. In particular, \citep{Dsing2014} found that out-of-plane
orbital motion remains negligible when the mean orbital separation is
larger than approximately $40\,GM_c/c^2$. This is consistent with the
very small change in orbital inclination found in our calculation,
despite the use of a different theoretical framework.

\begin{figure}
    \centering
    \includegraphics[width=\linewidth]{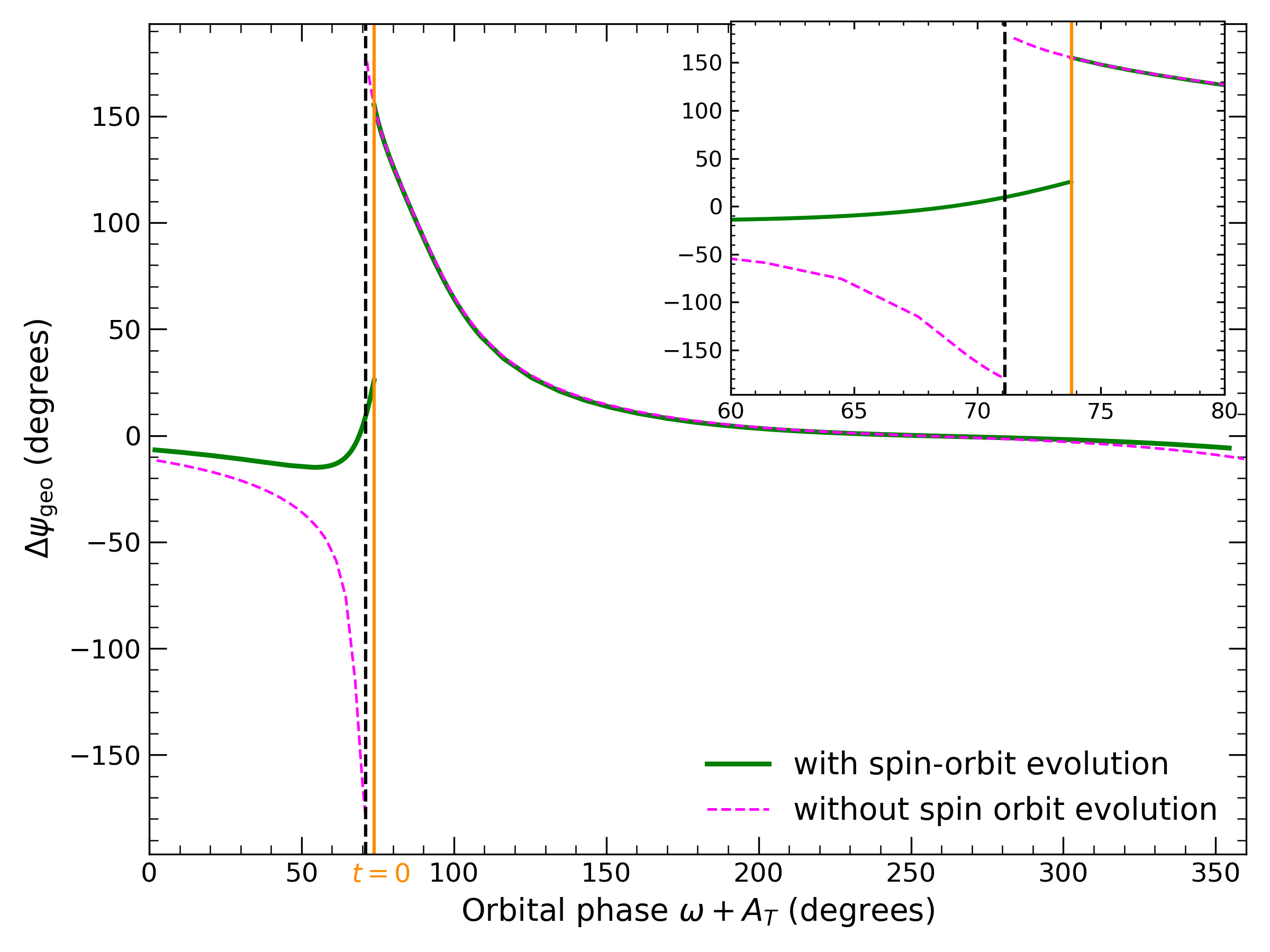}
    \caption{Geometric polarization shift, $\Delta\psi_{\rm geo}$, as a function of orbital phase $A_T+\omega$ for the hypothetical pulsar--SMBH binary. The dashed magenta curve shows the result without spin-orbit evolution, and the solid green curve shows the result with spin-orbit evolution. The vertical yellow line marks the initial orbital phase, while the black dashed line marks the apparent discontinuity in the case without spin-orbit evolution.}
    \label{fig:psi_geo_smbh}
\end{figure}

We first consider a Schwarzschild companion and then examine the
effect of companion spin. Figure~\ref{fig:psi_geo_smbh} shows the
geometric polarization shift, $\Delta\psi_{\rm geo}$, as a function
of orbital phase $A_T+\omega$. The dashed magenta curve represents the
case in which spin-orbit evolution is neglected, while the solid green
curve includes spin-orbit evolution. We use the initial longitude of
periastron $\omega=73.804^\circ$ from Table~\ref{tab:PSRintBH}; thus,
the initial orbital phase at $t=0$ is $A_T+\omega=73.804^\circ$,
indicated by the vertical yellow line. When spin-orbit evolution is
neglected, an apparent discontinuity occurs at
$A_T+\omega=71.12^\circ$, indicated by the black dashed line.

\begin{figure*}[htbp]
\centering
\begin{subfigure}{0.48\textwidth}
    \centering
    \includegraphics[width=\linewidth]{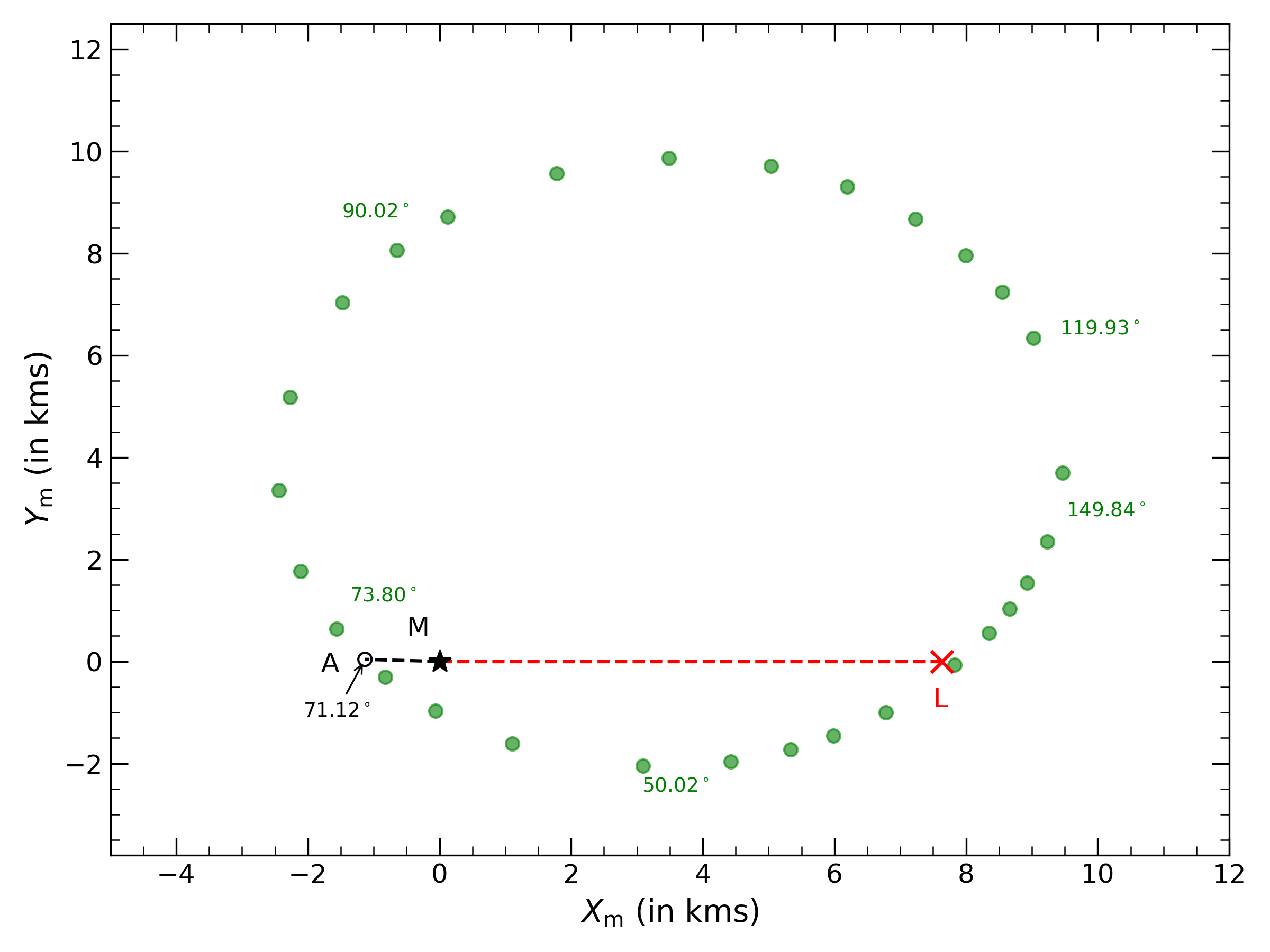}
    \caption{Emission points on the $X_mY_m$ plane from which the gravitationally bent radiation reaches the observer when spin-orbit evolution is neglected.}
    \label{fig:beamplot_smbh_no_so}
\end{subfigure}
\hfill
\begin{subfigure}{0.48\textwidth}
    \centering
    \includegraphics[width=\linewidth]{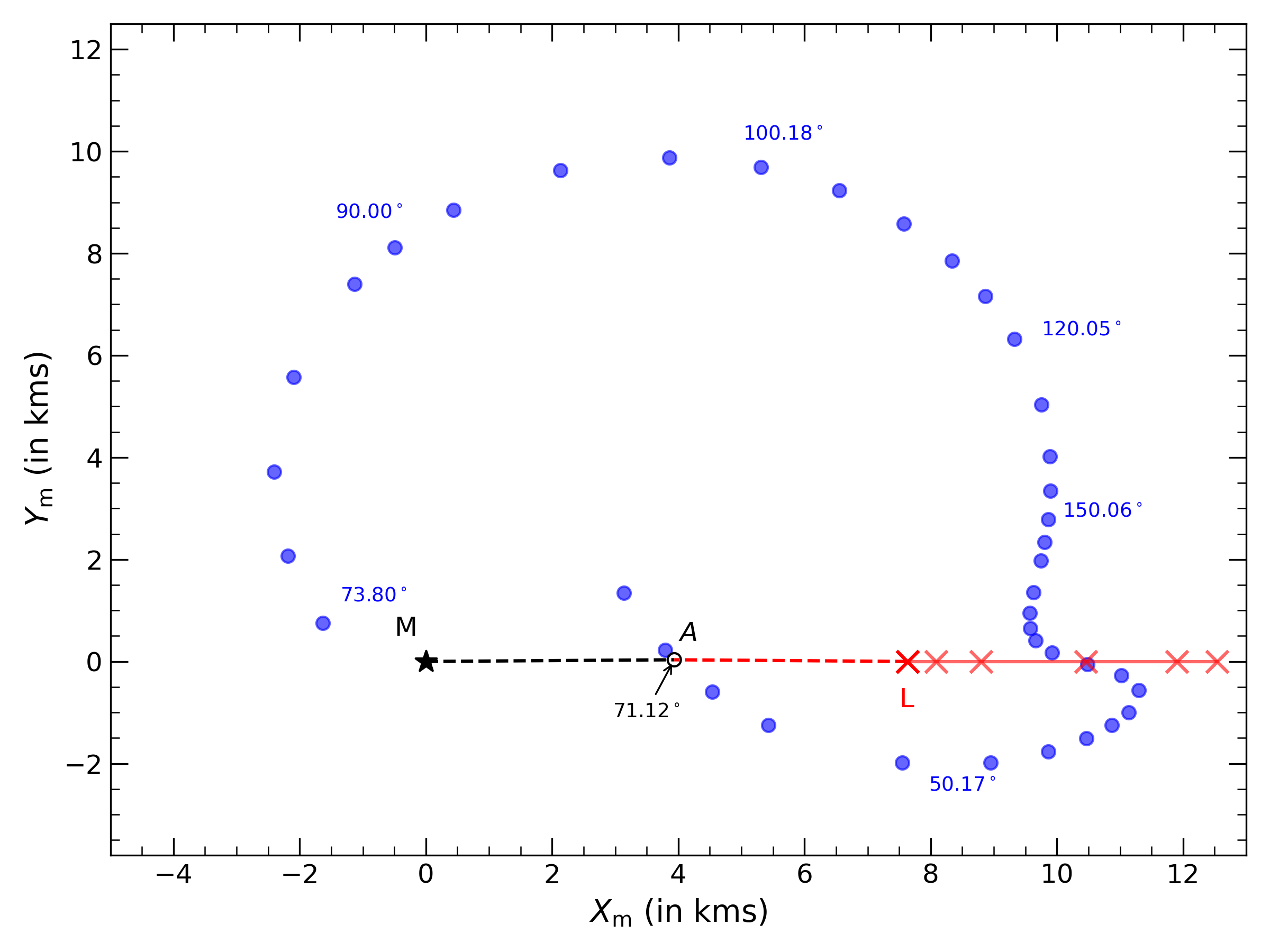}
    \caption{Same as the left panel, but including spin-orbit evolution. In this case, the LoS point $L$ moves along the $X_m$ axis with time rather than remaining fixed.}
    \label{fig:beamplot_smbh_with_so}
\end{subfigure}
\caption{Geometrical representation of the emission points on the $X_mY_m$ plane from which gravitationally bent light rays reach the observer at different orbital phases, with the pulsar rotational phase set to zero. The magnetic-axis intersection is marked by $M$, and the LoS intersection is marked by $L$.}
\label{fig:beamplot}
\end{figure*}

To understand this behavior of $\Delta\psi_{\rm geo}$, we show in
Figs.~\ref{fig:beamplot_smbh_no_so} and
\ref{fig:beamplot_smbh_with_so} the Cartesian coordinates $(x_m,y_m)$
in the $m$-frame (Section~\ref{subsec:analytical_polarization_m_frame})
of the emission points from which the gravitationally bent rays reach
the observer at different orbital phases. Figure~\ref{fig:beamplot_smbh_no_so}
shows the case without spin-orbit evolution, while
Fig.~\ref{fig:beamplot_smbh_with_so} shows the case with spin-orbit
evolution.

In these figures, we show the $X_mY_m$ plane at the emission height
$h_{\nu{\rm em}}$. The magnetic axis is directed along the $Z_m$ axis
of the $m$-frame and intersects this plane at the origin, marked by
the black star labeled $M$. The pulsar lies below this plane along the
$Z_m$ axis, at a distance $h_{\nu{\rm em}}$ from the plane. The point
where the line of sight (LoS) intersects the $X_mY_m$ plane is marked
by the red cross and labeled $L$.

We compare the polarization shift at different orbital phases when
the pulsar is at rotational phase zero, so that the LoS vector, the
magnetic axis, and the pulsar spin axis lie in the same plane. The LoS
therefore lies on the $X_m$ axis. Without spin-orbit evolution, the
LoS position remains fixed and is represented by the red cross. When
spin-orbit evolution is included, the LoS position changes continuously
along the $X_m$ axis, as indicated by the red curve in
Fig.~\ref{fig:beamplot_smbh_with_so}. The emission points from which
the observer receives gravitationally bent radiation are shown in
green when the spin-orbit evolution is neglected and in blue when it is
included. Representative orbital phases are also marked.

In Fig.~\ref{fig:beamplot_smbh_no_so}, when spin-orbit evolution is
neglected, the emission point $A$ lies on the same line as $M$ and
$L$. The direction from $M$ to $L$ is opposite to the direction from
$M$ to $A$. The orbital phase corresponding to the point $A$ is given by
\begin{equation}
    \tan^{-1}\left(\frac{\tan\eta_p}{\cos i}\right).
\end{equation}
We find that the orbital phase corresponding to point $A$ is
$A_T+\omega=71.12^\circ$. At this phase, the absolute value of the
geometric polarization shift reaches $180^\circ$.

When spin-orbit evolution is included, point $A$ again lies on the
same line as $M$ and $L$ at $A_T+\omega=71.12^\circ$. However, the
directions from $M$ to $L$ and from $M$ to $A$ are now the same, so
$\Delta\psi_{\rm geo}$ becomes zero at this orbital phase.

The apparent discontinuity in Fig.~\ref{fig:psi_geo_smbh} when
spin-orbit evolution is neglected therefore occurs because the
absolute value of the geometric shift reaches $180^\circ$, the maximum
value allowed for $\Delta\psi_{\rm geo}$ by the adopted RVM
description. It does not represent a physical discontinuity in the
polarization vector, but instead arises from the adopted definition of
the polarization angle.

When spin-orbit evolution is included, $\Delta\psi_{\rm geo}$ starts
at nearly $150^\circ$ at $t=0$, corresponding to
$A_T+\omega=73.804^\circ$ (Fig.~\ref{fig:psi_geo_smbh}). Initially,
its variation closely follows the case without spin-orbit evolution,
but the two curves subsequently diverge. The curve that includes
spin-orbit evolution does not return to its initial value after one
complete orbit because the orbital configuration evolves during the
motion. In particular, the final orbital phase is approximately
$75^\circ$, rather than $73.804^\circ$, mainly because of periastron
advance.

\begin{figure*}[htbp]
\centering
\begin{subfigure}{0.48\textwidth}
    \centering
    \includegraphics[width=\linewidth]{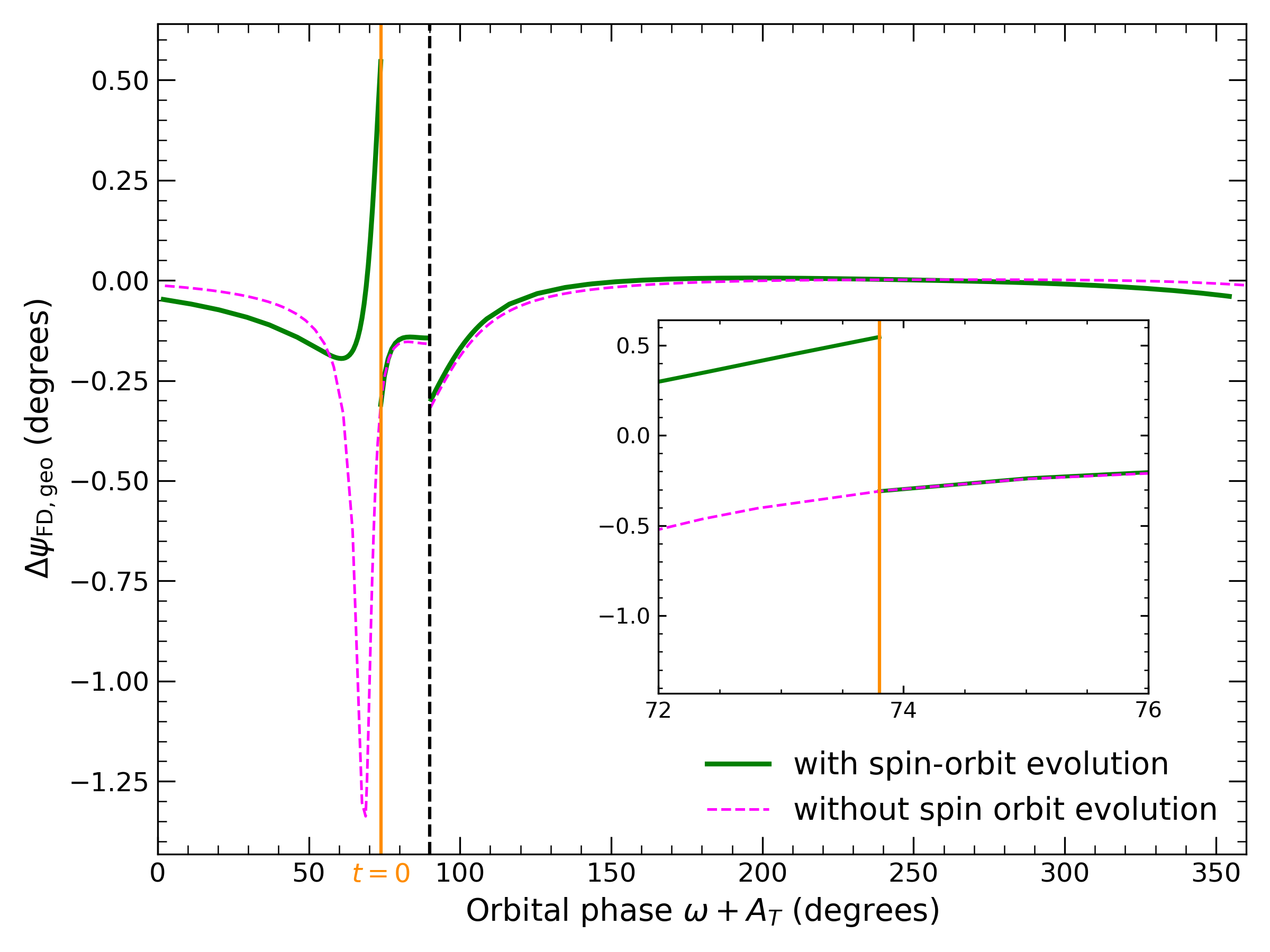}
    \caption{Absolute frame-dragging contribution to the geometric polarization shift, $\Delta\psi_{\rm FD,geo}$.}
    \label{fig:psi_fdgeo_smbh}
\end{subfigure}
\hfill
\begin{subfigure}{0.48\textwidth}
    \centering
    \includegraphics[width=\linewidth]{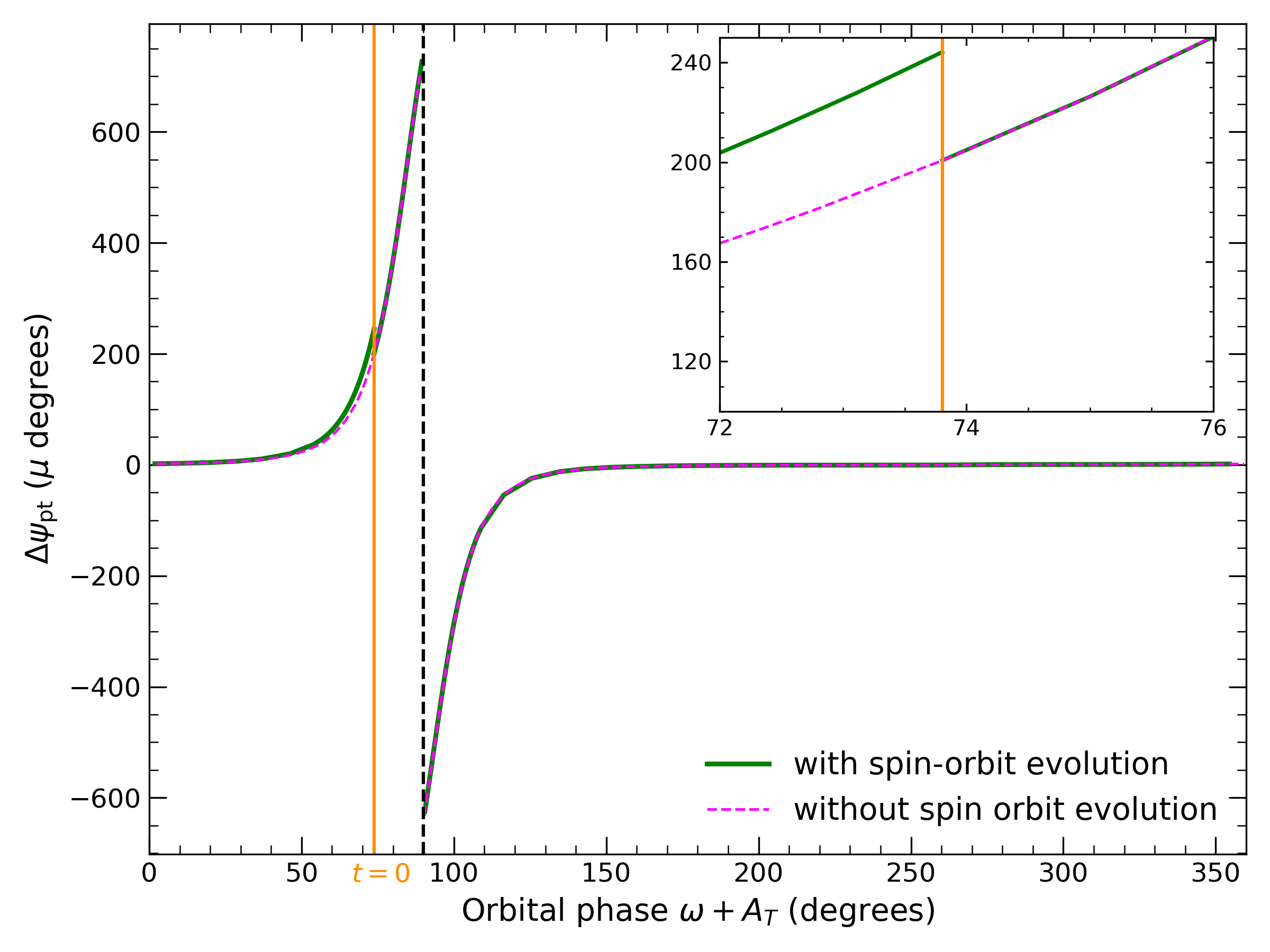}
    \caption{Polarization transport rotation, $\Delta\psi_{\rm pt}$.}
    \label{fig:psi_pt_smbh}
\end{subfigure}
\caption{Spin-dependent polarization changes for the hypothetical pulsar--SMBH binary with companion spin parameter $\tilde{a}=0.5$, shown as functions of orbital phase $A_T+\omega$. The left panel shows $\Delta\psi_{\rm FD,geo}$, and the right panel shows $\Delta\psi_{\rm pt}$.}
\label{fig:psi_rotation_smbh}
\end{figure*}

Next, we examine the effect of the SMBH spin on the polarization
direction. We set the dimensionless companion spin parameter to
$\tilde{a}=0.5$, as listed in Table~\ref{tab:PSRintBH}, and examine
the frame-dragging contribution to the geometric shift,
$\Delta\psi_{\rm FD,geo}$, and the polarization transport rotation,
$\Delta\psi_{\rm pt}$. The left panel of Fig.~\ref{fig:psi_rotation_smbh}
shows $\Delta\psi_{\rm FD,geo}$ as a function of orbital phase. Its
variation differs substantially from the case in which spin-orbit
evolution is considered, because the emission points on the beam from
which the observer receives radiation differ between the two cases,
as discussed above. The right panel shows $\Delta\psi_{\rm pt}$ as a
function of orbital phase. The magnitude of $\Delta\psi_{\rm pt}$ is
almost 500 times larger than in the intermediate-mass black hole case
considered above, although the shape of the curve remains unchanged.

\begin{figure*}[htbp]
\centering
\begin{subfigure}{0.48\textwidth}
    \centering
    \includegraphics[width=\linewidth]{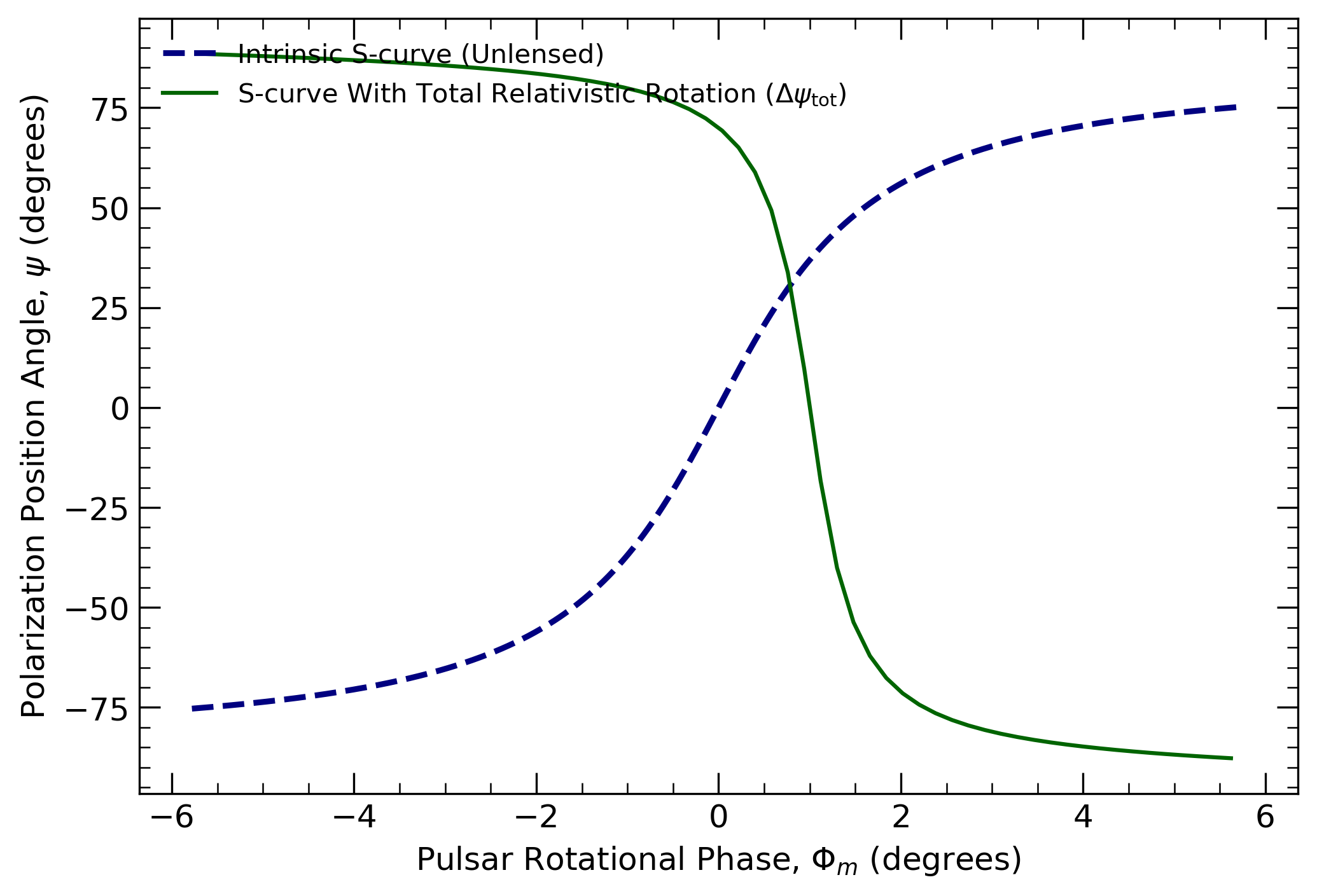}
    \caption{Orbital phase $A_T+\omega=85^\circ$.}
    \label{fig:scurve_smbh_85}
\end{subfigure}
\hfill
\begin{subfigure}{0.48\textwidth}
    \centering
    \includegraphics[width=\linewidth]{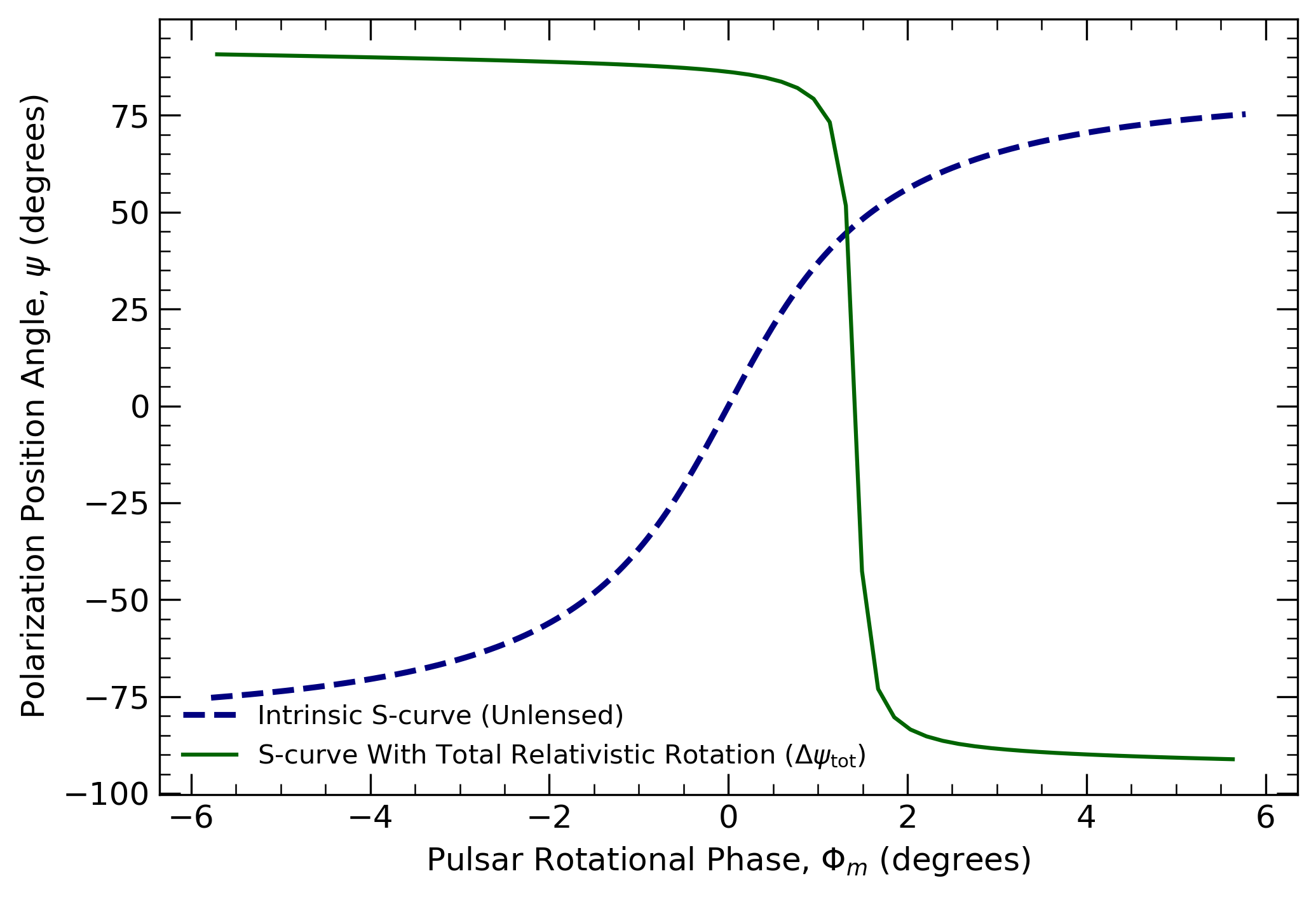}
    \caption{Orbital phase $A_T+\omega=90^\circ$.}
    \label{fig:scurve_smbh_90}
\end{subfigure}

\vspace{0.2cm}

\begin{subfigure}{0.48\textwidth}
    \centering
    \includegraphics[width=\linewidth]{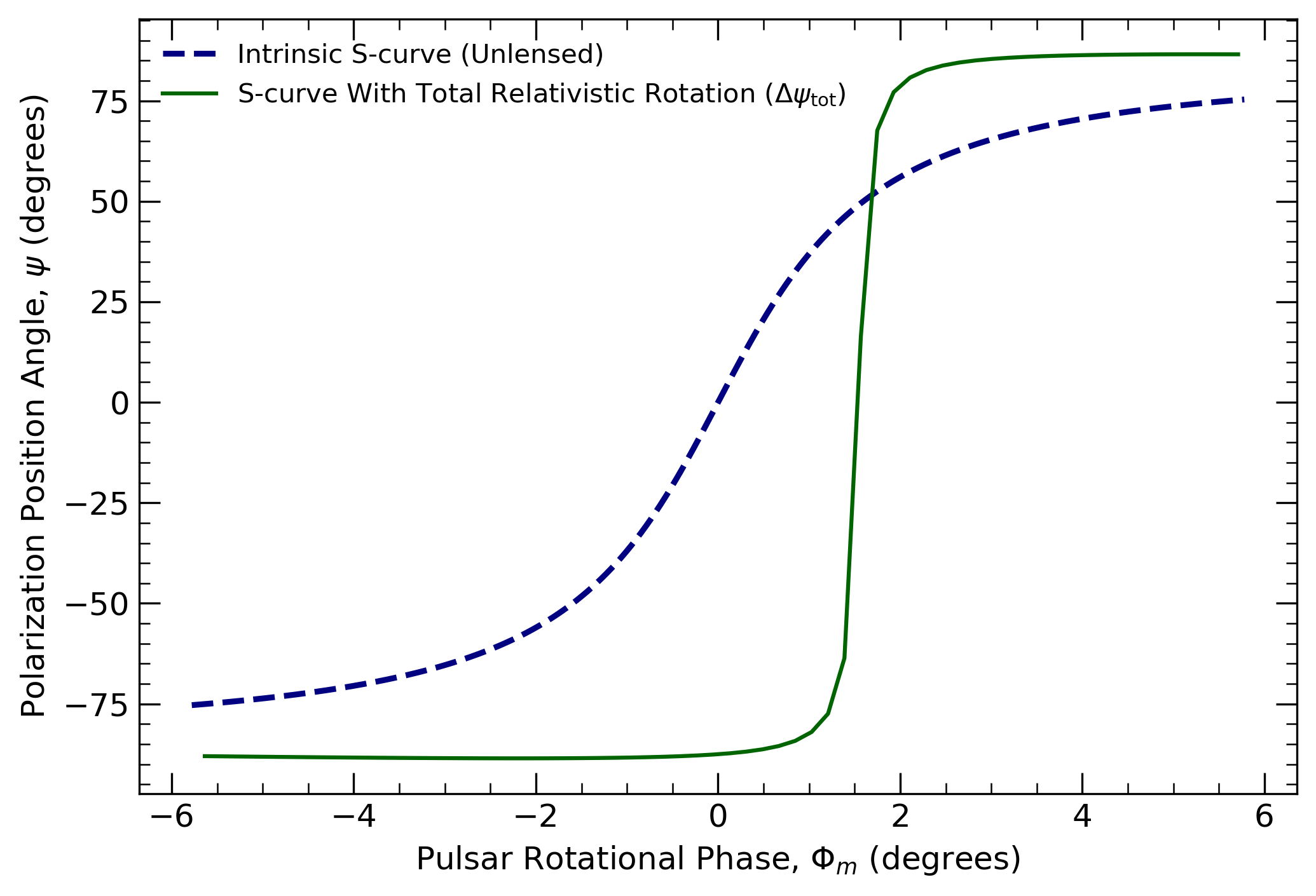}
    \caption{Orbital phase $A_T+\omega=92^\circ$.}
    \label{fig:scurve_smbh_92}
\end{subfigure}
\hfill
\begin{subfigure}{0.48\textwidth}
    \centering
    \includegraphics[width=\linewidth]{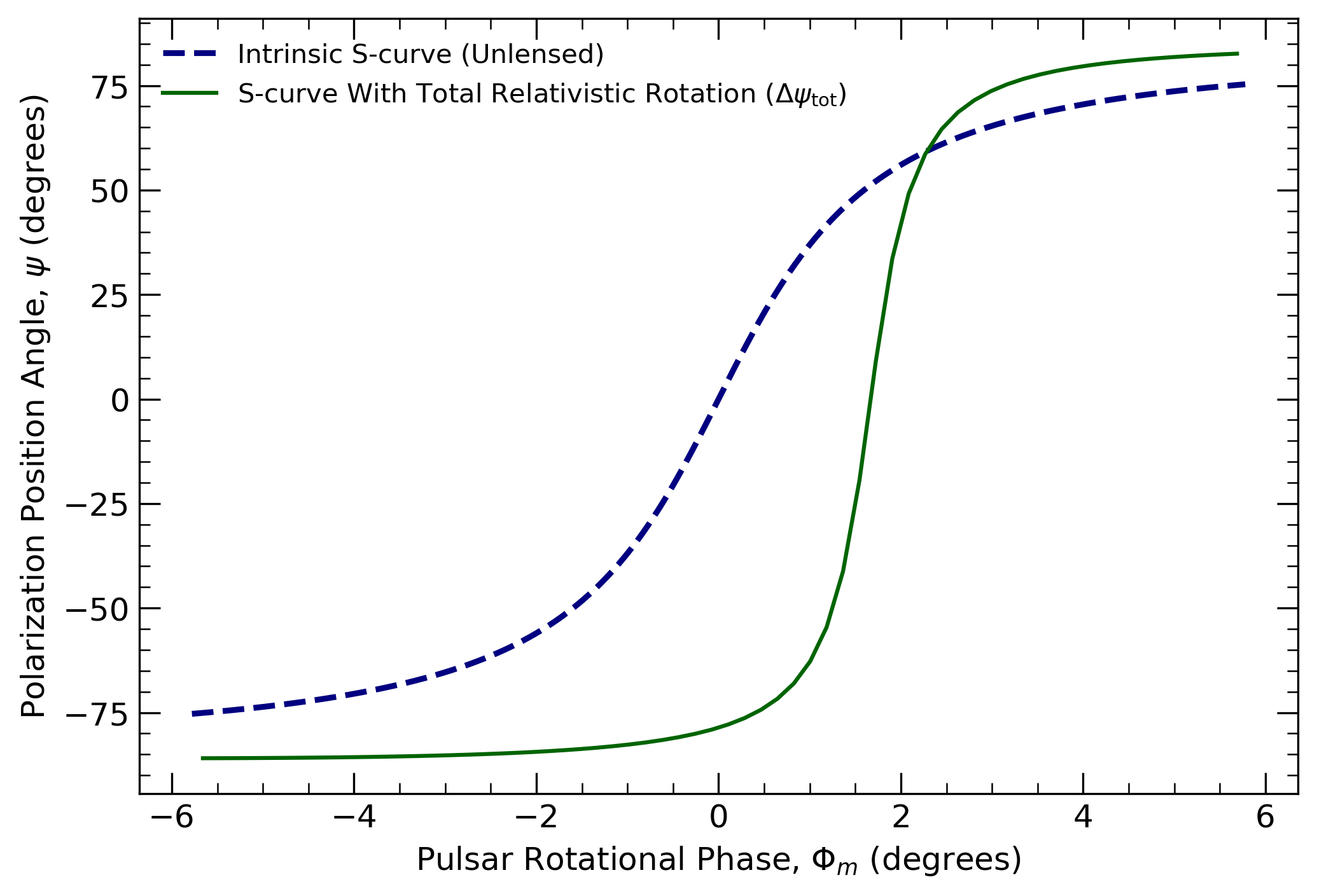}
    \caption{Orbital phase $A_T+\omega=95^\circ$.}
    \label{fig:scurve_smbh_95}
\end{subfigure}

\vspace{0.2cm}

\begin{subfigure}{0.48\textwidth}
    \centering
    \includegraphics[width=\linewidth]{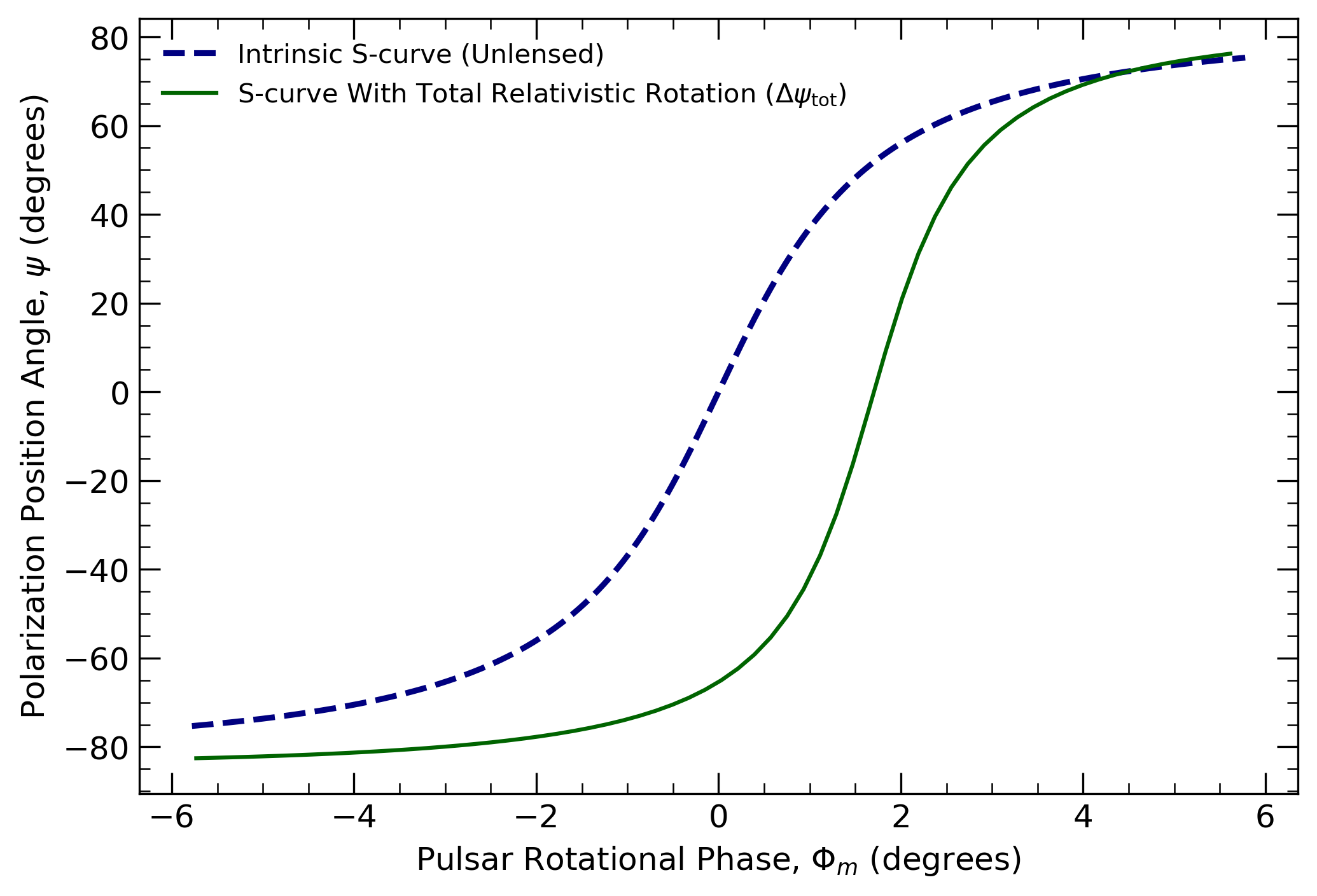}
    \caption{Orbital phase $A_T+\omega=100^\circ$.}
    \label{fig:scurve_smbh_100}
\end{subfigure}
\hfill
\begin{subfigure}{0.48\textwidth}
    \centering
    \includegraphics[width=\linewidth]{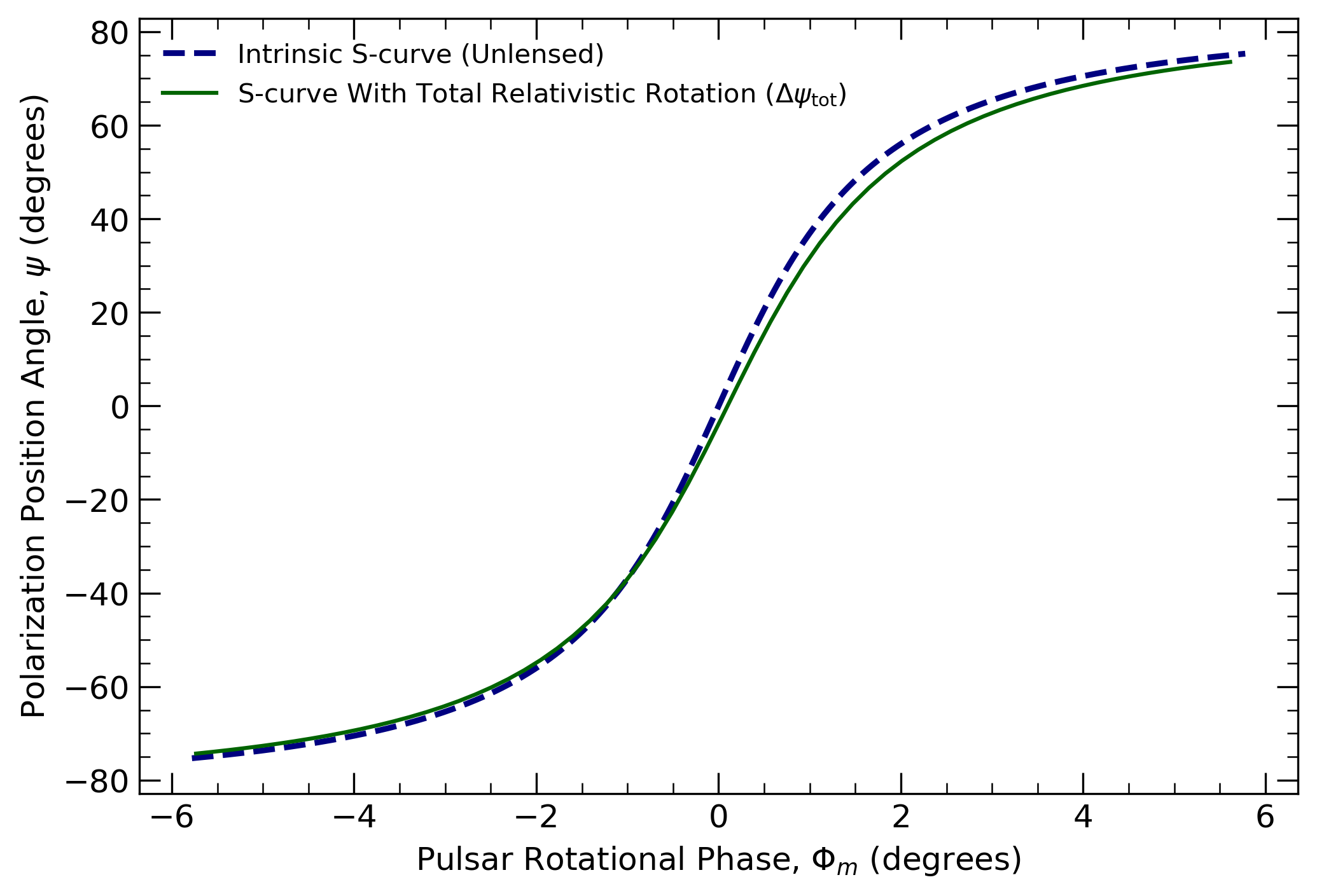}
    \caption{Orbital phase $A_T+\omega=200^\circ$.}
    \label{fig:scurve_smbh_200}
\end{subfigure}

\caption{Polarization S-curves at different orbital phases for the hypothetical pulsar--SMBH binary. The polarization position angle $\psi$ is plotted as a function of pulsar rotational phase. The largest changes occur near superior conjunction, where gravitational bending is strongest. At $A_T+\omega=85^\circ$ and $90^\circ$, the overall sense of the S-shaped variation is reversed compared with the corresponding unlensed curves, although the lensed curves are not exact reflections. The distortion gradually decreases at larger orbital phases as gravitational bending becomes weaker.}
\label{S-curve_smbh}
\end{figure*}

Next, we examine how the polarization S-curve changes with orbital
phase. The S-curve shows the polarization position angle $\psi$,
defined in Eq.~\eqref{PPA}, as a function of pulsar rotational phase.
Figure~\ref{S-curve_smbh} shows the S-curves at orbital phases
$A_T+\omega=85^\circ$, $90^\circ$, $92^\circ$, $95^\circ$,
$100^\circ$, and $200^\circ$.

At $A_T+\omega=85^\circ$ and $90^\circ$, where gravitational bending
is strong, the overall sense of the S-shaped variation is reversed
compared to the corresponding unlensed curves. The lensed curves are
not exact reflections of the unlensed S-curves. At
$A_T+\omega=92^\circ$, this reversal is no longer present, although
gravitational bending still produces significant distortion. As the
pulsar moves away from the region of strong bending, the distortion
gradually decreases, as shown in Figs.~\ref{fig:scurve_smbh_95} and
\ref{fig:scurve_smbh_100}. At $A_T+\omega=200^\circ$, gravitational
bending is almost insignificant. A small difference in the S-curve
remains, arising from the evolution of the emission geometry due to
pulsar spin precession.

This behavior can be understood from the motion of the LoS on the
pulsar beam plane. In the co-rotating $m$-frame, the LoS follows a
curved trajectory on the $X_mY_m$ plane as the pulsar rotates
\citep{DB26}. Gravitational bending displaces the apparent LoS
trajectory on the beam plane. We refer to the displacement along the
$X_m$ direction as the latitudinal displacement and that along the
$Y_m$ direction as the longitudinal displacement, since the pulsar
rotates along the $Y_m$ direction (see Fig.~\ref{polarization_geometry}).

The reversal in the overall S-shaped variation can be understood from
the latitudinal displacement of the LoS trajectory. At orbital phases
$A_T+\omega=85^\circ$ and $90^\circ$, the displacement is large enough
to move the lensed LoS trajectory to the opposite side of the magnetic
axis compared to the unlensed trajectory. Consequently, the overall
sense of the polarization S-curve is reversed. Because gravitational
bending also changes the detailed LoS trajectory on the beam plane,
the resulting curve is not an exact reflection of the unlensed S-curve.

Near $A_T+\omega=92^\circ$, the latitudinal displacement is no longer
large enough to move the LoS trajectory to the opposite side of the
magnetic axis. The transition between these configurations occurs when
the LoS trajectory crosses the magnetic axis. At this position, the
curvature acceleration vanishes and the polarization direction is
undefined in the present simplified model, as discussed in
Section~\ref{subsec:analytical_polarization_m_frame}. The overall
shape of the polarization curve therefore changes as the lensed LoS
trajectory moves from one side of the magnetic axis to the other.

A further feature is the shift of the distorted S-curve along the
pulsar rotational-phase axis. This shift is visible when bending is
strong, at $A_T+\omega=85^\circ$, $90^\circ$, $92^\circ$, $95^\circ$,
and $100^\circ$. It arises from the longitudinal displacement of the
LoS trajectory on the $X_mY_m$ plane. At $A_T+\omega=200^\circ$, where
gravity-induced bending is much weaker, the longitudinal displacement
and the resulting shift of the polarization S-curve are both very small.

\section{Conclusion}
\label{conclusion}

In this work, we investigated the effect of gravitational light bending on the polarization of radiation from a pulsar in a binary system with a black hole companion. We developed a formalism to calculate the null geodesic connecting the emission point to a distant observer and to track the polarization vector along the photon trajectory. The null geodesics are calculated using the formalism of \citep{KDL20,GSL20}, while the parallel transport of the polarization vector is treated using the null tetrad given in \citep{VA2025}. Since the formalism assumes that the motion of the companion can be neglected during photon propagation, we applied it to hypothetical pulsar--intermediate-mass black hole (IMBH) and pulsar--supermassive black hole (SMBH) binary systems.

For the IMBH system, we adopted orbital parameters close to those used in our previous work \citep{DBB23}. Although that study considered a stellar-mass black hole companion, the adopted parameters are comparable to those considered in other studies of pulsar--IMBH systems \citep{Devecchi:2007}. We considered companion masses of $10^5\,M_\odot$ and $10^6\,M_\odot$ for the IMBH and SMBH cases, respectively. For the SMBH system, we adopted an orbital period of $5.5$ days. This provides a compact and strongly relativistic configuration compared with systems having orbital periods of several years, as considered in studies motivated by the orbits of the S0 and S2 stars \citep{zs2017,liu2012}.

Our results show that gravitational light bending produces an apparent geometric polarization shift, $\Delta\psi_{\rm geo}$, that reaches a few degrees for the IMBH system and can reach $180^\circ$ for the SMBH system considered here. The spin of the companion introduces two additional contributions to the observed polarization direction: a geometric shift due to frame dragging, $\Delta\psi_{\rm FD,geo}$, and a polarization rotation due to parallel transport along the curved-spacetime null geodesic, $\Delta\psi_{\rm pt}$. To quantify the latter contribution, we construct a reference polarization vector by connecting the initial and final ray directions through a sequence of rotations. Therefore, $\Delta\psi_{\rm pt}$ depends on our chosen reference construction and should not be interpreted directly as the gravitational Faraday rotation. Separating the gravitomagnetic contribution from the dependence on the reference construction would require an appropriate reference triad along the null geodesic and an analysis of the rotation of the polarization vector relative to the propagation plane of the light ray \citep{Brodutch:2011}. Such an analysis is beyond the scope of this work. We therefore interpret $\Delta\psi_{\rm pt}$ as the polarization rotation associated with parallel transport, measured relative to our chosen reference construction.

For the systems considered here, $\Delta\psi_{\rm FD,geo}$ is of the order of a few millidegrees for the IMBH system and reaches the order of degrees for the SMBH system. In comparison, $\Delta\psi_{\rm pt}$ is of the order of microdegrees for the IMBH system and reaches the order of millidegrees for the SMBH system. We also included the evolution of the orbital geometry due to relativistic spin-orbit coupling. For the pulsar--IMBH system, the resulting changes are small over one orbital period. However, for the pulsar--SMBH system, the orbital geometry evolves significantly even over one orbit. Consequently, the polarization rotation angle varies significantly with orbital phase from one orbit to another.

Gravitational light bending also modifies the polarization position angle (PPA) profile. For the IMBH system, the characteristic S-shaped profile undergoes a relatively small but noticeable distortion. For the SMBH system, strong bending can reverse the overall sense of the S-shaped variation near the superior conjunction. The profile is also shifted in the pulsar rotational phase due to the longitudinal displacement of the light ray. These results demonstrate that relativistic photon propagation can modify both the shape and the phase dependence of the observed polarization profile.

From an observational perspective, the orbital-phase-dependent distortion of the PPA profile may provide a potential signature of relativistic photon propagation in a pulsar--BH binary. Modeling the polarization profile over the binary orbit could complement pulsar timing measurements and help constrain the geometry of the system \citep{kimpson2019}. However, the frame-dragging contribution to the geometric polarization shift and the polarization rotation associated with parallel transport are much smaller than the main geometric contribution for the systems considered here. This makes them more challenging to isolate observationally. Their detectability will depend on the properties of the binary system, the precision of the polarization measurements, and the ability to distinguish relativistic propagation effects from intrinsic and interstellar contributions to the observed signal.

Future studies incorporating realistic pulsar magnetosphere models together with relativistic photon propagation may provide a more complete interpretation of pulsar polarization profiles \citep{Oswald:2026, css21_need_to_rename_if_req, rfg22, cfr23, han21}. The capabilities of the Square Kilometre Array Observatory (SKAO) may also expand the scope of high-precision polarization studies. Such studies could help establish whether the predicted relativistic signatures can be distinguished from other effects that influence the observed polarization.

\begin{acknowledgments}
JD, DP and PT acknowledge financial support from the Istituto Nazionale di Astrofisica (INAF) 2023 Large Grant: “Gravitational Wave Detection using Pulsar Timing Arrays” (PI: D. Perrodin). AC and DP also acknowledge financial support under the National Recovery and Resilience Plan (NRRP), Mission 4, Component 2, Investment 1.1, Call for tender No. 104 published on 2.2.2022 by the Italian Ministry of University and Research (MUR), funded by the European Union
- NextGenerationEU - Project Title GUVIRP-Gravity
tests in the UltraViolet and InfraRed with Pulsar timing (PI: E. Barausse \& D. Perrodin).
\end{acknowledgments}
\appendix
\section{Asymptotic Tetrad and Polarization Vector at Infinity}
\label{Appendix}

In Section~\ref{subsec:polarization_transport}, we showed that the initial direction of the polarization vector and its final direction after parallel transport to infinity are related through Eq.~\ref{eq:condition1}. The transformation matrix $\tilde{E}^{\mu}{}_{\nu}$ is defined as \cite{Yang_2013}
\begin{equation}
\tilde{E}^{\mu}{}_{\nu} =
\begin{pmatrix}
e^\nu & 0 & 0 & -\Omega e^\psi \\
0 & e^{\mu_1} & 0 & 0 \\
0 & 0 & e^{\mu_2} & 0 \\
0 & 0 & 0 & e^\psi
\end{pmatrix},
\label{eq:LNRF_metric}
\end{equation}
where
\begin{subequations}
\label{eq:init1}
\begin{align}
e^\nu &= \sqrt{\frac{\rho^2\Delta}{\Sigma}}, \\
e^{\mu_1} &= \sqrt{\frac{\rho^2}{\Delta}}, \\
e^{\mu_2} &= \sqrt{\rho^2}, \\
e^\psi &= \sin\theta\sqrt{\frac{\Sigma}{\rho^2}}.
\end{align}
\end{subequations}

In the Boyer--Lindquist (BL) coordinate system, the null tetrad $(l^\mu,e_2^\mu,\tilde{e}_3^\mu,\tilde{n}^\mu)$ can be expressed as \cite{VA2025}
\begin{align}
l^{\mu}
={}& \frac{1}{\rho^2}
\Bigg(
-\left(\tilde{a}\frac{GM_c}{c^2}\right)
\left[
\left(\tilde{a}\frac{GM_c}{c^2}\right)\sin^2\theta
-D_\lambda
\right]
\nonumber\\
&\qquad
+\frac{r^2+\left(\tilde{a}\frac{GM_c}{c^2}\right)^2}{\Delta}
\left[
r^2+\left(\tilde{a}\frac{GM_c}{c^2}\right)^2
-\left(\tilde{a}\frac{GM_c}{c^2}\right)D_\lambda
\right],
\nonumber\\
&\qquad
\operatorname{sign}(p_r)\sqrt{\mathcal{R}},
\quad
\operatorname{sign}(p_\theta)\sqrt{\Theta},
\nonumber\\
&\qquad
-\left[
\left(\tilde{a}\frac{GM_c}{c^2}\right)
-\frac{D_\lambda}{\sin^2\theta}
\right]
\nonumber\\
&\qquad
+\frac{\left(\tilde{a}\frac{GM_c}{c^2}\right)}{\Delta}
\left[
r^2+\left(\tilde{a}\frac{GM_c}{c^2}\right)^2
-\left(\tilde{a}\frac{GM_c}{c^2}\right)D_\lambda
\right]
\Bigg),
\label{eq:l}
\end{align}

\begin{align}
e_2^{\mu}
={}& \frac{1}{\rho^2\Delta\sqrt{C}}
\Bigg(
\left(\tilde{a}\frac{GM_c}{c^2}\right)\cos\theta
\left[
r^2+\left(\tilde{a}\frac{GM_c}{c^2}\right)^2
\right]
\operatorname{sign}(p_r)\sqrt{\mathcal{R}}
\nonumber\\
&\qquad
-\left(\tilde{a}\frac{GM_c}{c^2}\right)
r\Delta\sin\theta\,
\operatorname{sign}(p_\theta)\sqrt{\Theta},
\nonumber\\
&\qquad
\left(\tilde{a}\frac{GM_c}{c^2}\right)
\Delta\cos\theta
\left[
r^2+\left(\tilde{a}\frac{GM_c}{c^2}\right)^2
-\left(\tilde{a}\frac{GM_c}{c^2}\right)D_\lambda
\right],
\nonumber\\
&\qquad
-r\Delta\sin\theta
\left[
\left(\tilde{a}\frac{GM_c}{c^2}\right)
-\frac{D_\lambda}{\sin^2\theta}
\right],
\nonumber\\
&\qquad
\left(\tilde{a}\frac{GM_c}{c^2}\right)^2
\cos\theta\,\operatorname{sign}(p_r)\sqrt{\mathcal{R}}
-r\Delta\csc\theta\,
\operatorname{sign}(p_\theta)\sqrt{\Theta}
\Bigg),
\label{eq:e2}
\end{align}

\begin{align}
\tilde{e}_3^{\mu}
={}& -\frac{1}{\rho^2\Delta\sqrt{C}}
\Bigg(
r\left[
r^2+\left(\tilde{a}\frac{GM_c}{c^2}\right)^2
\right]
\operatorname{sign}(p_r)\sqrt{\mathcal{R}}
\nonumber\\
&\qquad
+\left(\tilde{a}\frac{GM_c}{c^2}\right)^2
\sin\theta\cos\theta\,\Delta\,
\operatorname{sign}(p_\theta)\sqrt{\Theta},
\nonumber\\
&\qquad
r\Delta
\left[
r^2+\left(\tilde{a}\frac{GM_c}{c^2}\right)^2
-\left(\tilde{a}\frac{GM_c}{c^2}\right)D_\lambda
\right],
\nonumber\\
&\qquad
\left(\tilde{a}\frac{GM_c}{c^2}\right)
\Delta\sin\theta\cos\theta
\left[
\left(\tilde{a}\frac{GM_c}{c^2}\right)
-\frac{D_\lambda}{\sin^2\theta}
\right],
\nonumber\\
&\qquad
\left(\tilde{a}\frac{GM_c}{c^2}\right)
r\,\operatorname{sign}(p_r)\sqrt{\mathcal{R}}
\nonumber\\
&\qquad\qquad
+\left(\tilde{a}\frac{GM_c}{c^2}\right)
\Delta\cot\theta\,
\operatorname{sign}(p_\theta)\sqrt{\Theta}
\Bigg),
\label{eq:e3}
\end{align}

\begin{align}
\tilde{n}^{\mu}
={}& \frac{1}{2C}
\Bigg(
\left(\tilde{a}\frac{GM_c}{c^2}\right)
\left[
\left(\tilde{a}\frac{GM_c}{c^2}\right)
\sin^2\theta-D_\lambda
\right]
\nonumber\\
&\qquad
+\frac{
r^2+\left(\tilde{a}\frac{GM_c}{c^2}\right)^2
}{\Delta}
\left[
r^2+\left(\tilde{a}\frac{GM_c}{c^2}\right)^2
\right.
\nonumber\\
&\qquad\qquad
\left.
-\left(\tilde{a}\frac{GM_c}{c^2}\right)D_\lambda
\right],
\nonumber\\
&\qquad
\operatorname{sign}(p_r)\sqrt{\mathcal{R}},
\quad
-\operatorname{sign}(p_\theta)\sqrt{\Theta},
\nonumber\\
&\qquad
\left[
\left(\tilde{a}\frac{GM_c}{c^2}\right)
-\frac{D_\lambda}{\sin^2\theta}
\right]
+\frac{\left(\tilde{a}\frac{GM_c}{c^2}\right)}{\Delta}
\left[
r^2
\right.
\nonumber\\
&\qquad\qquad
\left.
+\left(\tilde{a}\frac{GM_c}{c^2}\right)^2
-\left(\tilde{a}\frac{GM_c}{c^2}\right)D_\lambda
\right]
\Bigg),
\label{eq:n}
\end{align}
where $C$ is a constant of motion of the null geodesic, defined as
\begin{align}
\label{eq:const_c}
C = D_q + \left(D_\lambda-\tilde{a}\frac{GM_c}{c^2}\right)^2.
\end{align}

However, $\tilde{e}_3^\mu$ and $\tilde{n}^\mu$ are not parallel transported along the null geodesic. The corresponding parallel-transported tetrad vectors are defined as \cite{VA2025}
\begin{subequations}
\label{eq:nullrot}
\begin{align}
e_{3}^{\mu} &= \tilde{e}_{3}^{\mu} + \tilde{\Phi}\, l^\mu, \\
k^\mu &= \tilde{n}^\mu + \tilde{\Phi}\, \tilde{e}_{3}^{\mu}
+ \frac{\tilde{\Phi}^{2}}{2}l^\mu, \\
\tilde{\Phi} &= \frac{1}{\sqrt{C}}J(r,\theta),
\end{align}
\end{subequations}

The function $J(r,\theta)$ is given by \cite{VA2025}
\begin{multline}
\label{eq:J}
J(r,\theta)
=
\operatorname{sign}(p_r)
\fint_r
\frac{r^2}
{\operatorname{sign}(p_r)\sqrt{\mathcal{R}(r)}}\,dr
\\
+
\tilde{a}\frac{GM_c}{c^2}
\fint_\theta
\frac{\cos^2\theta}
{\operatorname{sign}(p_\theta)\sqrt{\Theta(\theta)}}\,d\theta.
\end{multline}

The tetrad $(l^\mu,e_2^\mu,e_3^\mu,k^\mu)$ therefore represents the parallel-transported null tetrad along the null geodesic.

In Eq.~\ref{eq:condition1}, the quantities at the source,
$\left(l^\mu\tilde{E}^{\nu}{}_{\mu}\right)_s$,
$\left(e_2^\mu\tilde{E}^{\nu}{}_{\mu}\right)_s$, and
$\left(e_3^\mu\tilde{E}^{\nu}{}_{\mu}\right)_s$,
can be evaluated by tracing the position of the pulsar in the BH frame, i.e., at $r=r_{\rm bh,s}$ and $\theta=\theta_{\rm bh,s}$. The integration constants in $J(r,\theta)$ are chosen such that
\begin{equation}
J(r_{\rm bh,s},\theta_{\rm bh,s})=0
\end{equation}
at the position of the source. Consequently, $\tilde{\Phi}=0$ at the source, and hence
\begin{equation}
e_3^\mu=\tilde{e}_3^\mu
\qquad \text{and} \qquad
k^\mu=\tilde{n}^\mu.
\end{equation}

To evaluate the corresponding quantities at infinity,
$\left(l^\mu\tilde{E}^{\nu}{}_{\mu}\right)_f$,
$\left(e_2^\mu\tilde{E}^{\nu}{}_{\mu}\right)_f$, and
$\left(e_3^\mu\tilde{E}^{\nu}{}_{\mu}\right)_f$,
we use the algebraic form of $J(r,\theta)$ proposed by \citep{KDL20,GSL20}. Following their work, we decompose $J(r,\theta)$ as
\begin{equation}
\label{eq:split_J}
J(r,\theta)
=
\operatorname{sign}(p_r)
\frac{\sqrt{\mathcal{R}}}{r-r_3}
+
\mathcal{J}(r,\theta),
\end{equation}
where $r_3$ is the third root of the radial potential $\mathcal{R}$ defined in Eq.~\ref{R_rparticle}, with the four roots satisfying
\begin{equation}
r_1<r_2<r_3<r_4.
\end{equation}
The quantity $\mathcal{J}(r,\theta)$ is given by
\begin{equation}
\begin{aligned}
\mathcal{J}(r,\theta)
={}&
-\operatorname{sign}(p_r)_f
\Biggl[
\frac{r_1r_4+r_2r_3}
{\sqrt{(r_3-r_1)(r_4-r_2)}}
F\!\left(\sin^{-1}\!X(r),K\right)
\\
&\quad
+\sqrt{(r_3-r_1)(r_4-r_2)}
E\!\left(\sin^{-1}\!X(r),K\right)
\Biggr]
\\
&+
q\,\operatorname{sign}(p_r)_f
\Biggl[
\frac{\sqrt{\mathcal{R}(r_{\rm initial})}}
{r_{\rm initial}-r_3}
\\
&\quad
-\frac{r_1r_4+r_2r_3}
{\sqrt{(r_3-r_1)(r_4-r_2)}}
F\!\left(\sin^{-1}\!X(r_{\rm initial}),K\right)
\\
&\quad
-\sqrt{(r_3-r_1)(r_4-r_2)}
E\!\left(\sin^{-1}\!X(r_{\rm initial}),K\right)
\Biggr]
\\
&+
\tilde{a}\frac{GM_c}{c^2}
\fint_{\theta}
\frac{\cos^2\theta}
{\operatorname{sign}(p_\theta)\sqrt{\Theta(\theta)}}\,d\theta .
\end{aligned}
\label{J_expression}
\end{equation}
Here, $F$ and $E$ denote the incomplete elliptic integrals of the
first and second kinds, respectively, and $r_{\rm initial}$ denotes the
initial radial position of the photon. In the present case,
$r_{\rm initial}=r_{\rm bh,s}$. The quantities $K$ and $X(r)$
are defined as
\begin{equation}
K=
\frac{(r_3-r_2)(r_4-r_1)}
{(r_3-r_1)(r_4-r_2)},
\qquad
X(r)=
\sqrt{
\frac{(r-r_4)(r_3-r_1)}
{(r-r_3)(r_4-r_1)}
}.
\label{k_X}
\end{equation}
The parameter $q$ takes the value $+1$ when the light ray initially
moves towards the black hole before reaching the observer, and
$-1$ when it initially moves away from the black hole and reaches the
observer directly. The quantity $\mathcal{J}(r,\theta)$ is finite in
the region $r_4\leq r\leq\infty$. In the present case, the observer is
located at $r=\infty$ and $\theta=\theta_{\rm bh,f}$, and therefore
$\mathcal{J}(r,\theta)$ is evaluated in the limit $r\rightarrow\infty$
at $\theta=\theta_{\rm bh,f}$ \citep{KDL20,GSL20}.

Using Eqs.~\ref{eq:LNRF_metric}, \ref{eq:l}, \ref{eq:e2} and taking the limit $r\rightarrow\infty$, we obtain
\begin{align}
\label{eq:L_mu}
\left(
l^\mu\tilde{E}^{\nu}{}_{\mu}
\right)_f
=
\Bigg(
1,\,
\operatorname{sign}(p_r)_f,\,
0,\,
0
\Bigg),
\end{align}
and,
\begin{align}
\label{eq:e2_mu}
\left(
e_2^\mu\tilde{E}^{\nu}{}_{\mu}
\right)_f
={}&
\frac{1}{\sqrt{C}}
\Bigg(
\operatorname{sign}(p_r)_f
\left(\tilde{a}\frac{GM_c}{c^2}\right)
\cos\theta_{\rm bh,f},
\nonumber\\
&\qquad
\left(\tilde{a}\frac{GM_c}{c^2}\right)
\cos\theta_{\rm bh,f},
\nonumber\\
&\qquad
-\sin\theta_{\rm bh,f}
\left[
\left(\tilde{a}\frac{GM_c}{c^2}\right)
-\frac{D_\lambda}{\sin^2\theta_{\rm bh,f}}
\right],
\nonumber\\
&\qquad
-\operatorname{sign}(p_\theta)_f\sqrt{\Theta(\theta_{\rm bh,f})}
\Bigg).
\end{align}

Similarly, using Eqs.~\ref{eq:LNRF_metric}, \ref{eq:e3}, \ref{eq:nullrot} and \ref{eq:split_J}, we obtain
\begin{align}
\label{eq:e3_mu}
\left(
e_3^\mu\tilde{E}^{\nu}{}_{\mu}
\right)_f
={}&
\frac{1}{\sqrt{C}}
\Bigg(
\mathcal{J}(\infty,\theta_{\rm bh,f})
+
\operatorname{sign}(p_r)_f r_3,
\nonumber\\
&\qquad
\operatorname{sign}(p_r)_f
\left[
\mathcal{J}(\infty,\theta_{\rm bh,f})
+
\operatorname{sign}(p_r)_f r_3
\right],
\nonumber\\
&\qquad
\operatorname{sign}(p_r)_f
\operatorname{sign}(p_\theta)_f
\sqrt{\Theta(\theta_{\rm bh,f})},
\nonumber\\
&\qquad
\operatorname{sign}(p_r)_f
\sin\theta_{\rm bh,f}
\left[
\frac{D_\lambda}{\sin^2\theta_{\rm bh,f}}
-\tilde{a}\frac{GM_c}{c^2}
\right]
\Bigg).
\end{align}

Here, $\operatorname{sign}(p_r)_f$ denotes the sign of the radial momentum of the light ray at infinity. Since the light ray reaches the observer at infinity, it is outgoing, and therefore
\begin{equation}
\operatorname{sign}(p_r)_f=1.
\end{equation}
The sign of the polar momentum at infinity is given by
\begin{align}
\label{eq:m}
\operatorname{sign}(p_\theta)_f
=
(-1)^m
\operatorname{sign}(p_\theta)_s.
\end{align}

Here, $m$ denotes the number of turning points in the polar motion of the light ray along its trajectory from the pulsar to the observer. Each turning point changes the sign of the polar momentum $p_\theta$. Therefore, the sign of the polar momentum at infinity is related to its initial value at the source by Eq. \ref{eq:m}.
For the pulsar--BH binary configurations considered in this work, we consider trajectories with either $m=0$ or $m=1$. When the light ray is emitted from the pulsar in a direction initially away from the black hole, the trajectory does not encounter a turning point in its polar motion, and hence $m=0$. On the other hand, when the light ray is initially emitted towards the black hole, its trajectory undergoes one polar turning point before reaching the observer, corresponding to $m=1$. Consequently, the sign of the polar momentum remains unchanged for $m=0$, whereas it changes sign for $m=1$. We do not consider trajectories with $m>1$, corresponding to multiple turning points in the polar motion, as such trajectories are not relevant to the direct light-ray trajectories considered in the present study.
Using Eqs.~\ref{eq:L_mu}, \ref{eq:e2_mu}, and \ref{eq:e3_mu}, we solve Eq.~\ref{eq:condition1} to determine the final direction of the polarization vector, $\widehat{f}_{(\rm bh,f)\,\mu}$.

At infinity, the radiation propagates in an asymptotically flat spacetime, and the polarization vector must remain transverse to the final propagation direction. Therefore,
\begin{equation}
\widehat{f}_{(\rm bh,f)\,\nu}
\left(
l^\mu\tilde{E}^{\nu}{}_{\mu}
\right)_f
=0.
\end{equation}
From Eq.~\ref{eq:L_mu}, the final propagation direction is given by
\begin{equation}
\left(
l^\mu\tilde{E}^{\nu}{}_{\mu}
\right)_f
=
(1,1,0,0),
\end{equation}
where we have used $\operatorname{sign}(p_r)_f=1$. Consequently, the transversality condition requires the temporal and
radial components of $\widehat{f}_{(\rm bh,f)\,\mu}$ to be equal in
magnitude and opposite in sign, i.e., 
$\widehat{f}_{(\rm bh,f)\,t}=-\widehat{f}_{(\rm bh,f)\,r}$.
However, the polarization vector is defined up to an arbitrary component
along the propagation direction. Therefore, using the gauge freedom of
the polarization vector, we can add a term proportional to the final
propagation direction,
\begin{equation}
\widehat{f}_{(\rm bh,f)\,\mu}
\rightarrow
\widehat{f}_{(\rm bh,f)\,\mu}
+
\kappa
\left(E^\nu{}_{\mu}l_\nu\right)_f,
\end{equation}
without changing the physical polarization state, where $\kappa$ is an
arbitrary gauge parameter and $l_\mu$ denotes the covariant component of
$l^\mu$ \citep{VA2025}. Here, $E^\nu{}_{\mu}$ is the inverse
transformation matrix of $\tilde{E}^\nu{}_{\mu}$, which transforms
vectors from the Boyer--Lindquist (BL) coordinate system to the local
Cartesian BH frame. For the final propagation direction, we find
\begin{equation}
\left(E^\nu{}_{\mu}l_\nu\right)_f
=
(-1,1,0,0).
\end{equation}
Since the polar and azimuthal components of
$\left(E^\nu{}_{\mu}l_\nu\right)_f$ vanish, the gauge transformation
changes only the temporal and radial components of the polarization
vector. We can therefore choose the gauge parameter as
$\kappa=\widehat{f}_{(\rm bh,f)\,t}$, such that the temporal and radial
components of the transformed polarization vector both vanish. In other words, the gauge
freedom allows us to set the temporal and radial components of
$\widehat{f}_{(\rm bh,f)\,\mu}$ to zero. The remaining polar and
azimuthal components then completely determine the physical direction of
the polarization vector observed at infinity.

For completeness, we also give the expression for
$\left(k^\mu \tilde{E}^\nu{}_\mu\right)_f$:

\begin{multline}
\left(k^\mu \tilde{E}^\nu{}_\mu\right)_f
=
\Bigg(
\frac{1}{2C}
\Big[
\left(\mathcal{J}(\infty,\theta_{\rm bh,f})
+\operatorname{sign}(p_r)_f r_3\right)^2
\\
+\left(\tilde{a}\frac{GM_c}{c^2}\right)^2
\cos^2\theta_{\rm bh,f}+C
\Big],
\\
\frac{\operatorname{sign}(p_r)_f}{2C}
\Big[
\left(\mathcal{J}(\infty,\theta_{\rm bh,f})
+\operatorname{sign}(p_r)_f r_3\right)^2
\\
+\left(\tilde{a}\frac{GM_c}{c^2}\right)^2
\cos^2\theta_{\rm bh,f}-C
\Big],
\\
-\operatorname{sign}(p_r)_f
\frac{\tilde{a}\frac{GM_c}{c^2}
\sin\theta_{\rm bh,f}\cos\theta_{\rm bh,f}}{C}
\left(
\tilde{a}\frac{GM_c}{c^2}
-\frac{D_\lambda}{\sin^2\theta_{\rm bh,f}}
\right)
\\
+\operatorname{sign}(p_r)_f
\mathcal{J}(\infty,\theta_{\rm bh,f})
\operatorname{sign}(p_\theta)_f
\frac{\sqrt{\Theta(\theta_{\rm bh,f})}}{C}
+\operatorname{sign}(p_\theta)_f
\frac{r_3\sqrt{\Theta(\theta_{\rm bh,f})}}{C},
\\
-\frac{\sin\theta_{\rm bh,f}}{C}
\tilde{a}\frac{GM_c}{c^2}\cot\theta_{\rm bh,f}\,
\operatorname{sign}(p_r)_f
\operatorname{sign}(p_\theta)_f
\sqrt{\Theta(\theta_{\rm bh,f})}
\\
-\frac{\sin\theta_{\rm bh,f}}{C}
\left(
\tilde{a}\frac{GM_c}{c^2}
-\frac{D_\lambda}{\sin^2\theta_{\rm bh,f}}
\right)
\operatorname{sign}(p_r)_f
\mathcal{J}(\infty,\theta_{\rm bh,f})
\\
-r_3\frac{\sin\theta_{\rm bh,f}}{C}
\left(
\tilde{a}\frac{GM_c}{c^2}
-\frac{D_\lambda}{\sin^2\theta_{\rm bh,f}}
\right)
+\tilde{a}\left(\frac{GM_c}{c^2}\right)^2
\frac{\sin\theta_{\rm bh,f}}{C}
\Bigg).
\label{nE_expression}
\end{multline}
\vspace{1em}
\nocite{*}

\bibliography{apssamp}

@ARTICLE{VA2025,
       author = {{Frolov}, Valeri P. and {Koek}, Alex},
        title = "{Spinoptics in the Kerr spacetime: Polarized wave scattering}",
      journal = {\prd},
         year = 2025,
        month = may,
       volume = {111},
       number = {10},
          eid = {104081},
        pages = {104081},
          doi = {10.1103/trh5-4sgq},
archivePrefix = {arXiv},
       eprint = {2503.19208},
 primaryClass = {gr-qc},
       adsurl = {https://ui.adsabs.harvard.edu/abs/2025PhRvD.111j4081F}
}

@article{DB25,
    author = "Debnath, Jyotijwal and Bagchi, Manjari",
    title = "{On the effect of the light-bending phenomenon for a pulsar in a binary with a Kerr black hole}",
    eprint = "2506.05783",
    archivePrefix = "arXiv",
    primaryClass = "gr-qc",
    doi = "10.1093/mnras/staf974",
    journal = "Mon. Not. Roy. Astron. Soc.",
    volume = "541",
    number = "2",
    pages = "821--846",
    year = "2025"
}

@article{rc69,
    author = {Radhakrishnan, V. and Cooke, D.~J.},
    title = {Magnetic Poles and the Polarization Structure of Pulsar Radiation},
    journal = {Astrophysics Letters},
    year = {1969},
    month = {January},
    volume = {3},
    pages = {225},
}

@article{CM2010,
    author = {Chung, C. T. Y. and Melatos, A.},
    title = {Stokes tomography of radio pulsar magnetospheres – I. Linear polarization},
    journal = {Monthly Notices of the Royal Astronomical Society},
    volume = {411},
    number = {4},
    pages = {2471-2529},
    year = {2011},
    month = {03},
    issn = {0035-8711},
    doi = {10.1111/j.1365-2966.2010.17858.x},
    url = {https://doi.org/10.1111/j.1365-2966.2010.17858.x},
    eprint = {https://academic.oup.com/mnras/article-pdf/411/4/2471/18485628/mnras0411-2471.pdf},
}

@article{WWH2014,
    author = {Wang, P. F. and Wang, C. and Han, J. L.},
    title = {Polarized curvature radiation in pulsar magnetosphere},
    journal = {Monthly Notices of the Royal Astronomical Society},
    volume = {441},
    number = {3},
    pages = {1943-1953},
    year = {2014},
    month = {07},
    issn = {0035-8711},
    doi = {10.1093/mnras/stu690},
    url = {https://doi.org/10.1093/mnras/stu690},
    eprint = {https://academic.oup.com/mnras/article-pdf/441/3/1943/3891372/stu690.pdf},
}

@ARTICLE{BCW1991,
       author = {{Blaskiewicz}, M. and {Cordes}, J.~M. and {Wasserman}, I.},
        title = "{A Relativistic Model of Pulsar Polarization}",
      journal = {\apj},
         year = 1991,
        month = apr,
       volume = {370},
        pages = {643},
          doi = {10.1086/169850},
       adsurl = {https://ui.adsabs.harvard.edu/abs/1991ApJ...370..643B}
}

@book{jackson98,
    author = "Jackson, John David",
    title = "{Classical Electrodynamics}",
    isbn = "978-0-471-30932-1",
    publisher = "Wiley",
    year = "1998"
}

@book{Lorimer2004,
       author    = {Lorimer, D.~R. and Kramer, M.},
       title     = {Handbook of Pulsar Astronomy},
       year      = {2004},
       volume    = {4},
       publisher = {Cambridge University Press},
       address   = {Cambridge},
       series    = {Cambridge Observing Handbooks for Research Astronomers},
       adsurl    = {https://ui.adsabs.harvard.edu/abs/2004hpa..book.....L}
}

@article{GJ69,
    author = {Goldreich, Peter and Julian, William H.},
    title = {Pulsar Electrodynamics},
    journal = {The Astrophysical Journal},
    year = {1969},
    month = {August},
    volume = {157},
    pages = {869-880},
    doi = {10.1086/150119},
    adsurl = {https://ui.adsabs.harvard.edu/abs/1969ApJ...157..869G}
}

@ARTICLE{Deutsch1955,
       author = {{Deutsch}, Arnim J.},
        title = "{The electromagnetic field of an idealized star in rigid rotation in vacuo}",
      journal = {Annales d'Astrophysique},
         year = 1955,
        month = jan,
       volume = {18},
        pages = {1},
       adsurl = {https://ui.adsabs.harvard.edu/abs/1955AnAp...18....1D}
}

@article{KG1998,
    author = {Kijak, Jaroslaw and Gil, Janusz},
    title = {Radio emission regions in pulsars},
    journal = {Monthly Notices of the Royal Astronomical Society},
    year = {1998},
    month = {September},
    volume = {299},
    number = {3},
    pages = {855-861},
    doi = {10.1046/j.1365-8711.1998.01832.x},
    adsurl = {https://ui.adsabs.harvard.edu/abs/1998MNRAS.299..855K}
}

@article{KG2003,
    author = {Kijak, J. and Gil, J.},
    title = {Radio emission altitude in pulsars},
    journal = {Astronomy \& Astrophysics},
    year = {2003},
    month = {January},
    volume = {397},
    pages = {969-972},
    doi = {10.1051/0004-6361:20021583},
    adsurl = {https://ui.adsabs.harvard.edu/abs/2003A&A...397..969K}
}

@article{Manchester1970,
    author        = "Manchester, R. N.",
    title         = "{Polarization of Pulsar Radiation}",
    journal       = "Nature",
    volume        = "228",
    pages         = "264--266",
    year          = "1970",
    doi           = "10.1038/228264a0"
}

@article{Sturrock1971,
    author = {Sturrock, P. A.},
    title = {A Model of Pulsars},
    journal = {The Astrophysical Journal},
    year = {1971},
    month = {March},
    volume = {164},
    pages = {529},
    doi = {10.1086/150865},
    adsurl = {https://ui.adsabs.harvard.edu/abs/1971ApJ...164..529S}
}

@article{Arons1983,
    author = {Arons, J.},
    title = {Pair creation above pulsar polar caps: Geometrical structure and energetics of slot gaps},
    journal = {The Astrophysical Journal},
    year = {1983},
    month = {March},
    volume = {266},
    pages = {215-241},
    doi = {10.1086/160771},
    adsurl = {https://ui.adsabs.harvard.edu/abs/1983ApJ...266..215A}
}

@article{Hibschman2001,
    author = {Hibschman, J. A. and Arons, J.},
    title = {Pair-production multiplicities in rotation-powered pulsars},
    eprint = {astro-ph/0107209},
    archivePrefix = {arXiv},
    doi = {10.1086/323069},
    journal = {The Astrophysical Journal},
    volume = {560},
    pages = {871-884},
    year = {2001}
}

@article{Arendt2002,
    author = {Arendt, Paul N., Jr. and Eilek, Jean A.},
    title = {Pair Creation in the Pulsar Magnetosphere},
    journal = {The Astrophysical Journal},
    year = {2002},
    month = {December},
    volume = {581},
    number = {1},
    pages = {451-469},
    doi = {10.1086/344133},
    archivePrefix = {arXiv},
    eprint = {astro-ph/0207638},
    primaryClass = {astro-ph},
    adsurl = {https://ui.adsabs.harvard.edu/abs/2002ApJ...581..451A}
}

@article{gil1984,
    author = {{Gil}, J. and {Gronkowski}, P. and {Rudnicki}, W.},
    title = "{Geometry of the emission region of PSR 0950+08}",
    journal = {Astronomy and Astrophysics},
    year = {1984},
    month = {March},
    volume = {132},
    pages = {312-316},
    adsurl = {https://ui.adsabs.harvard.edu/abs/1984A&A...132..312G}
}

@article{Komesaroff1970,
    author = {M. M. Komesaroff},
    title = {Possible Mechanism for the Pulsar Radio Emission},
    journal = {Nature},
    year = {1970},
    month = {February},
    volume = {225},
    number = {5233},
    pages = {612--614},
    doi = {10.1038/225612a0},
    adsurl = {https://ui.adsabs.harvard.edu/abs/1970Natur.225..612K}
}

@ARTICLE{Rankin83,
       author = {{Rankin}, J.~M.},
        title = "{Toward an empirical theory of pulsar emission. I. Morphological taxonomy}",
      journal = {The Astrophysical Journal},
         year = 1983,
        month = nov,
       volume = {274},
        pages = {333-358},
          doi = {10.1086/161450}
}

@article{lm88,
       author = {{Lyne}, A.~G. and {Manchester}, R.~N.},
        title = "{The shape of pulsar radio beams.}",
      journal = {Monthly Notices of the Royal Astronomical Society},
         year = 1988,
        month = oct,
       volume = {234},
        pages = {477-508},
          doi = {10.1093/mnras/234.3.477}       
}

@ARTICLE{gks93,
       author = {{Gil}, J.~A. and {Kijak}, J. and {Seiradakis}, J.~H.},
        title = "{On the two-dimensional structure of pulsar beams.}",
      journal = {Astronomy \& Astrophysics},
         year = 1993,
        month = may,
       volume = {272},
        pages = {268-276}
}

@ARTICLE{GM2001,
    author = {{Gangadhara}, R.~T. and {Gupta}, Y.},
     title = "{Understanding the Radio Emission Geometry of PSR B0329+54}",
   journal = {The Astrophysical Journal},
      year = 2001,
     month = jul,
    volume = {555},
    number = {1},
     pages = {31-39},
       doi = {10.1086/321439},
}

@article{lpp94,
  author    = {Lyne, A.G. and Pritchard, R.S. and Graham-Smith, F.},
  title     = {Pulsar polarimetry and the geometry of the pulsar magnetosphere},
  journal   = {Monthly Notices of the Royal Astronomical Society},
  year      = {1994},
  volume    = {268},
  pages     = {845--854}
}

@article{ppr05,
  author    = {Pétri, J. and Pétri, J. and Romani, R.W.},
  title     = {Pulsar current closure and acceleration in general relativistic magnetospheres},
  journal   = {The Astrophysical Journal},
  year      = {2005},
  volume    = {632},
  number    = {2},
  pages     = {1152--1165}
}

@article{fcl11,
  author    = {Frail, D.A. and Clausen-Brown, E. and Lommen, A.N.},
  title     = {Pulsar Emission Geometry from Wide-band Polarization Profiles},
  journal   = {The Astrophysical Journal},
  year      = {2011},
  volume    = {739},
  number    = {2},
  pages     = {93--102}
}

@article{css21_need_to_rename_if_req,
  author    = {Chen, A. and Spitkovsky, A.},
  title     = {Recent advances in understanding pulsar emission and magnetosphere physics},
  journal   = {Journal of Physics: Conference Series},
  year      = {2021},
  volume    = {1622},
  pages     = {012043}
}

@article{rfg22,
  author    = {Ramesh, G. and Fong, W.-f. and Ghosh, P.},
  title     = {High-energy Pulsar Emission: Models and Observations},
  journal   = {Journal of High Energy Astrophysics},
  year      = {2022},
  volume    = {33},
  pages     = {42--54}
}

@article{cfr23,
  author    = {Chen, X. and Feng, H. and Rea, N.},
  title     = {Observational Insights into Pulsar Magnetospheres},
  journal   = {The Astrophysical Journal},
  year      = {2023},
  volume    = {951},
  number    = {1},
  pages     = {87}
}

@article{han21,
  author    = {Han, J.L. and Wang, Y. and Lu, Y.},
  title     = {Magnetospheric Models of Pulsars and their Emission Mechanisms},
  journal   = {The Astrophysical Journal},
  year      = {2021},
  volume    = {908},
  number    = {2},
  pages     = {123}
}

@article{pan21,
  author    = {Pan, L. and Li, X. and Xie, Y.},
  title     = {Pulsar Timing and Geometry: Current Status and Future Prospects},
  journal   = {Monthly Notices of the Royal Astronomical Society},
  year      = {2021},
  volume    = {505},
  number    = {3},
  pages     = {3792--3805}
}

@article{hpb14,
  author    = {Harding, A.K. and Petrukhin, A. and Baring, M.G.},
  title     = {Electromagnetic Emission from Pulsar Magnetospheres},
  journal   = {The Astrophysical Journal},
  year      = {2014},
  volume    = {790},
  number    = {2},
  pages     = {128}
}

@ARTICLE{sch90,
       author = {{Schneider}, J.},
        title = "{Gravitational beam deviation effects in the timing formula of binary pulsars.}",
      journal = {\aap},
         year = 1990,
        month = jun,
       volume = {232},
        pages = {62}       
}

@article{dk1995,
    author = "Doroshenko, Oleg V and Kopeikin, Sergei M and (DK95)",
    title = "{Relativistic effect of gravitational deflection of light in binary pulsars}",
    journal = {MNRAS},
    volume = "274",
    number={4},
    pages = "1029--1038",
    year = "1995"    
}

@article{Rafikov2005,
    author = "Rafikov, Roman R. and Lai, Dong and (RL06)",
    title = "{Effects of pulsar rotation on timing measurements of the double pulsar system J0737-3039}",
    eprint = "astro-ph/0503461",
    archivePrefix = "arXiv",
    doi = "10.1086/500346",
    journal = {ApJ},
    volume = "641",
    pages = "438--446",
    year = "2006"    
}

@article{ksm21,
       author = {{Kramer}, M. and {Stairs}, I.~H. and {Manchester}, R.~N. and {Wex}, N. and {Deller}, A.~T. and {Coles}, W.~A. and {Ali}, M. and {Burgay}, M. and {Camilo}, F. and {Cognard}, I. and {Damour}, T. and {Desvignes}, G. and {Ferdman}, R.~D. and {Freire}, P.~C.~C. and {Grondin}, S. and {Guillemot}, L. and {Hobbs}, G.~B. and {Janssen}, G. and {Karuppusamy}, R. and {Lorimer}, D.~R. and {Lyne}, A.~G. and {McKee}, J.~W. and {McLaughlin}, M. and {M{\"u}nch}, L.~E. and {Perera}, B.~B.~P. and {Pol}, N. and {Possenti}, A. and {Sarkissian}, J. and {Stappers}, B.~W. and {Theureau}, G.},
        title = "{Strong-Field Gravity Tests with the Double Pulsar}",
      journal = {Physical Review X},
           year = 2021,
        month = oct,
       volume = {11},
       number = {4},
          eid = {041050},
        pages = {041050},
          doi = {10.1103/PhysRevX.11.041050}
}

@book{chandrasekhar84,
  author    = {Chandrasekhar, S.},
  title     = {The Mathematical Theory of Black Holes},
  year      = {1983},
  publisher = {Oxford University Press},
  address   = {New York},
  note      = {The International Series of Monographs on Physics. Series editors: R.~J.~Elliot, J.~A.~Krumhunsl, and D.~H.~Wilkinson}
}

@ARTICLE{pro16,
       author = {{Paganini}, Claudio F. and {Ruba}, Blazej and {Oancea}, Marius A.},
        title = "{Characterization of Null Geodesics on Kerr Spacetimes}",
      journal = {arXiv e-prints},
         year = 2016,
        month = nov,
          eid = {arXiv:1611.06927},
        pages = {arXiv:1611.06927},
          doi = {10.48550/arXiv.1611.06927}
}

@article{KDL20,
    author = "Kapec, Daniel and Lupsasca, Alexandru",
    title = "{Particle motion near high-spin black holes}",
    eprint = "1905.11406",
    archivePrefix = "arXiv",
    primaryClass = "hep-th",
    doi = "10.1088/1361-6382/ab519e",
    journal = "Class. Quant. Grav.",
    volume = "37",
    number = "1",
    pages = "015006",
    year = "2020"
}

@article{GSL20,
  title = {Null geodesics of the Kerr exterior},
  author = {Gralla, Samuel E. and Lupsasca, Alexandru},
  journal = {Phys. Rev. D},
  volume = {101},
  issue = {4},
  pages = {044032},
  numpages = {22},
  year = {2020},
  month = {02},
  publisher = {American Physical Society},
  doi = {10.1103/PhysRevD.101.044032},
  url = {https://link.aps.org/doi/10.1103/PhysRevD.101.044032}
}

@ARTICLE{BE22,
       author = {{Ben-Salem}, Bilel and {Hackmann}, Eva},
        title = "{Propagation time delay and frame dragging effects of lightlike geodesics in the timing of a pulsar orbiting SgrA*}",
      journal = {\mnras},
         year = 2022,
        month = oct,
       volume = {516},
       number = {2},
        pages = {1768-1780},
          doi = {10.1093/mnras/stac2337},
archivePrefix = {arXiv},
       eprint = {2203.10931},
 primaryClass = {gr-qc},
       adsurl = {https://ui.adsabs.harvard.edu/abs/2022MNRAS.516.1768B}
}

@ARTICLE{DBB23,
       author = {{Debnath}, Jyotijwal and {Bagchi}, Manjari and {Basu}, Avishek and (Paper-I)},
        title = "{A study of the light bending phenomenon under full general relativity for a pulsar in a binary with a Schwarzschild black hole}",
      journal = {\mnras},
         year = 2023,
        month = oct,
       volume = {524},
       number = {4},
        pages = {5411-5431},
          doi = {10.1093/mnras/stad2147},
archivePrefix = {arXiv},
       eprint = {2306.11431},
 primaryClass = {gr-qc},
       adsurl = {https://ui.adsabs.harvard.edu/abs/2023MNRAS.524.5411D}
}

@ARTICLE{BarkerOconnell1975,
       author = {{Barker}, B.~M. and {O'Connell}, R.~F.},
        title = "{Gravitational two-body problem with arbitrary masses, spins, and quadrupole moments}",
      journal = {\prd},
         year = 1975,
        month = jul,
       volume = {12},
       number = {2},
        pages = {329-335},
          doi = {10.1103/PhysRevD.12.329},
       adsurl = {https://ui.adsabs.harvard.edu/abs/1975PhRvD..12..329B}
}

@ARTICLE{damourschafer1988,
       author = {{Damour}, T. and {Schafer}, G.},
        title = "{Higher-order relativistic periastron advances and binary pulsars.}",
      journal = {Nuovo Cimento B Serie},
         year = 1988,
        month = jan,
       volume = {101B},
       number = {2},
        pages = {127-176},
          doi = {10.1007/BF02828697},
       adsurl = {https://ui.adsabs.harvard.edu/abs/1988NCimB.101..127D}
}

@ARTICLE{peters1964,
       author = {{Peters}, P.~C.},
        title = "{Gravitational Radiation and the Motion of Two Point Masses}",
      journal = {Physical Review},
         year = 1964,
        month = nov,
       volume = {136},
       number = {4B},
        pages = {1224-1232},
          doi = {10.1103/PhysRev.136.B1224},
       adsurl = {https://ui.adsabs.harvard.edu/abs/1964PhRv..136.1224P}
}

@article{bagchi2017,
    author = "Bagchi, Manjari",
    editor = "Blaschke, David and Ayriyan, Alexander and Friesen, Alexandra and Grigorian, Hovik",
    title = "{Prospects of Constraining the Dense Matter Equation of State from Timing Analysis of Pulsars in Double Neutron Star Binaries: The Cases of PSR J0737 - 3039A and PSR J1757 - 1854}",
    eprint = "1712.02632",
    archivePrefix = "arXiv",
    primaryClass = "astro-ph.HE",
    doi = "10.3390/universe4020036",
    journal = "Universe",
    volume = "4",
    number = "2",
    pages = "36",
    year = "2018"
}

@ARTICLE{bagchi2010,
       author = {{Bagchi}, Manjari},
        title = "{Rotational parameters of strange stars in comparison with neutron stars}",
      journal = {\na},
         year = 2010,
        month = jan,
       volume = {15},
       number = {1},
        pages = {126-134},
          doi = {10.1016/j.newast.2009.07.003},
archivePrefix = {arXiv},
       eprint = {0805.2721},
 primaryClass = {astro-ph},
       adsurl = {https://ui.adsabs.harvard.edu/abs/2010NewA...15..126B}
}

@ARTICLE{Dsing2014,
       author = {{Singh}, Dinesh and {Wu}, Kinwah and {Sarty}, Gordon E.},
        title = "{Fast spinning pulsars as probes of massive black holes' gravity}",
      journal = {\mnras},
         year = 2014,
        month = jun,
       volume = {441},
       number = {1},
        pages = {800-808},
          doi = {10.1093/mnras/stu614},
archivePrefix = {arXiv},
       eprint = {1403.7171},
 primaryClass = {astro-ph.HE},
       adsurl = {https://ui.adsabs.harvard.edu/abs/2014MNRAS.441..800S}
}

@ARTICLE{kimpson2019,
       author = {{Kimpson}, Tom and {Wu}, Kinwah and {Zane}, Silvia},
        title = "{Pulsar timing in extreme mass ratio binaries: a general relativistic approach}",
      journal = {\mnras},
         year = 2019,
        month = jun,
       volume = {486},
       number = {1},
        pages = {360-377},
          doi = {10.1093/mnras/stz845},
archivePrefix = {arXiv},
       eprint = {1903.08258},
 primaryClass = {astro-ph.HE},
       adsurl = {https://ui.adsabs.harvard.edu/abs/2019MNRAS.486..360K}
}

@ARTICLE{barker1982,
       author = {{Barker}, B.~M. and {Byrd}, G.~G. and {Oconnell}, R.~F.},
        title = "{Spin nutation in binary systems due to general relativistic and quadrupole effects}",
      journal = {\apj},
         year = 1982,
        month = feb,
       volume = {253},
        pages = {309-311},
          doi = {10.1086/159635},
       adsurl = {https://ui.adsabs.harvard.edu/abs/1982ApJ...253..309B}
}

@ARTICLE{Yang_2013,
       author = {{Yang}, Xiaolin and {Wang}, Jiancheng},
        title = "{YNOGK: A New Public Code for Calculating Null Geodesics in the Kerr Spacetime}",
      journal = {The Astrophysical Journal Supplement Series},
         year = 2013,
        month = 07,
       volume = {207},
       number = {1},
          eid = {6},
        pages = {6},
          doi = {10.1088/0067-0049/207/1/6}
}

@ARTICLE{RL06,
       author = {{Rafikov}, Roman R. and {Lai}, Dong},
        title = "{Effects of gravitational lensing and companion motion on the binary pulsar timing}",
      journal = {\prd},
         year = 2006,
        month = mar,
       volume = {73},
       number = {6},
          eid = {063003},
        pages = {063003},
          doi = {10.1103/PhysRevD.73.063003},
archivePrefix = {arXiv},
       eprint = {astro-ph/0512417},
 primaryClass = {astro-ph},
       adsurl = {https://ui.adsabs.harvard.edu/abs/2006PhRvD..73f3003R}
}

@ARTICLE{KJLi2019,
       author = {{Li}, Kaye Jiale and {Wu}, Kinwah and {Singh}, Dinesh},
        title = "{Spin dynamics of a millisecond pulsar orbiting closely around a massive black hole}",
      journal = {\mnras},
         year = 2019,
        month = may,
       volume = {485},
       number = {1},
        pages = {1053-1066},
          doi = {10.1093/mnras/stz389},
archivePrefix = {arXiv},
       eprint = {1902.03146},
 primaryClass = {astro-ph.HE},
       adsurl = {https://ui.adsabs.harvard.edu/abs/2019MNRAS.485.1053L}
}

@ARTICLE{majar2009,
       author = {{Maj{\'a}r}, J{\'a}nos},
        title = "{Spin-spin interaction in the spin-precession equations}",
      journal = {\prd},
         year = 2009,
        month = nov,
       volume = {80},
       number = {10},
          eid = {104028},
        pages = {104028},
          doi = {10.1103/PhysRevD.80.104028},
       adsurl = {https://ui.adsabs.harvard.edu/abs/2009PhRvD..80j4028M}
}

@ARTICLE{majar2012,
       author = {{Maj{\'a}r}, J{\'a}nos and {Mik{\'o}czi}, Bal{\'a}zs},
        title = "{Second order spin effects in the spin precession of compact binaries}",
      journal = {\prd},
         year = 2012,
        month = sep,
       volume = {86},
       number = {6},
          eid = {064028},
        pages = {064028},
          doi = {10.1103/PhysRevD.86.064028},
       adsurl = {https://ui.adsabs.harvard.edu/abs/2012PhRvD..86f4028M}
}

@ARTICLE{liu2012,
       author = {{Liu}, K. and {Wex}, N. and {Kramer}, M. and {Cordes}, J.~M. and {Lazio}, T.~J.~W.},
        title = "{Prospects for Probing the Spacetime of Sgr A* with Pulsars}",
      journal = {\apj},
         year = 2012,
        month = mar,
       volume = {747},
       number = {1},
          eid = {1},
        pages = {1},
          doi = {10.1088/0004-637X/747/1/1},
archivePrefix = {arXiv},
       eprint = {1112.2151},
 primaryClass = {astro-ph.HE},
       adsurl = {https://ui.adsabs.harvard.edu/abs/2012ApJ...747....1L}
}

@ARTICLE{lwls22,
       author = {{Li}, Kaye Jiale and {Wu}, Kinwah and {Leung}, Po Kin and {Singh}, Dinesh},
        title = "{Relativistic scattering of a fast spinning neutron star by a massive black hole}",
      journal = {\mnras},
         year = 2022,
        month = apr,
       volume = {511},
       number = {3},
        pages = {3602-3617},
          doi = {10.1093/mnras/stab2925},
archivePrefix = {arXiv},
       eprint = {2110.03494},
 primaryClass = {astro-ph.HE},
       adsurl = {https://ui.adsabs.harvard.edu/abs/2022MNRAS.511.3602L}
}

@book{Maggiore:2007ulw,
    author = "Maggiore, Michele",
    title = "{Gravitational Waves. Vol. 1: Theory and Experiments}",
    doi = "10.1093/acprof:oso/9780198570745.001.0001",
    isbn = "978-0-19-171766-6, 978-0-19-852074-0",
    publisher = "Oxford University Press",
    year = "2007"
}

@ARTICLE{km02,
       author = {{Kopeikin}, Sergei and {Mashhoon}, Bahram},
        title = "{Gravitomagnetic effects in the propagation of electromagnetic waves in variable gravitational fields of arbitrary-moving and spinning bodies}",
      journal = {\prd},
         year = 2002,
        month = mar,
       volume = {65},
       number = {6},
          eid = {064025},
        pages = {064025},
          doi = {10.1103/PhysRevD.65.064025},
archivePrefix = {arXiv},
       eprint = {gr-qc/0110101},
 primaryClass = {gr-qc},
       adsurl = {https://ui.adsabs.harvard.edu/abs/2002PhRvD..65f4025K}
}

@ARTICLE{zs2017,
       author = {{Zhang}, Fupeng and {Saha}, Prasenjit},
        title = "{Probing the Spinning of the Massive Black Hole in the Galactic Center via Pulsar Timing: A Full Relativistic Treatment}",
      journal = {\apj},
         year = 2017,
        month = nov,
       volume = {849},
       number = {1},
          eid = {33},
        pages = {33},
          doi = {10.3847/1538-4357/aa8f47},
archivePrefix = {arXiv},
       eprint = {1709.08341},
 primaryClass = {astro-ph.GA},
       adsurl = {https://ui.adsabs.harvard.edu/abs/2017ApJ...849...33Z}
}

@ARTICLE{hkc22,
       author = {{Hu}, H. and {Kramer}, M. and {Champion}, D.~J. and {Wex}, N. and {Parthasarathy}, A. and {Pennucci}, T.~T. and {Porayko}, N.~K. and {van Straten}, W. and {Venkatraman Krishnan}, V. and {Burgay}, M. and {Freire}, P.~C.~C. and {Manchester}, R.~N. and {Possenti}, A. and {Stairs}, I.~H. and {Bailes}, M. and {Buchner}, S. and {Cameron}, A.~D. and {Camilo}, F. and {Serylak}, M.},
        title = "{Gravitational signal propagation in the double pulsar studied with the MeerKAT telescope}",
      journal = {\aap},
         year = 2022,
        month = nov,
       volume = {667},
          eid = {A149},
        pages = {A149},
          doi = {10.1051/0004-6361/202244825},
archivePrefix = {arXiv},
       eprint = {2209.11798},
 primaryClass = {astro-ph.HE}
}

@ARTICLE{FCA2005,
       author = {{Jenet}, Fredrick A. and {Creighton}, Teviet and {Lommen}, Andrea},
        title = "{Pulsar Timing and the Detection of Black Hole Binary Systems in Globular Clusters}",
      journal = {\apjl},
         year = 2005,
        month = jul,
       volume = {627},
       number = {2},
        pages = {L125-L128},
          doi = {10.1086/431949},
archivePrefix = {arXiv},
       eprint = {astro-ph/0505585},
 primaryClass = {astro-ph},
       adsurl = {https://ui.adsabs.harvard.edu/abs/2005ApJ...627L.125J}
}

@unpublished{DB26,
    author        = "Debnath, Jyotijwal and Bagchi, Manjari",
    title         = "{On coordinate frames relevant for pulsar physics}",
    eprint        = "2608.18501",
    archivePrefix = "arXiv",
    primaryClass  = "astro-ph.HE",
    month         = "8",
    year          = "2026",
    note          = "arXiv:2608.18501 [astro-ph.HE]"
}

@article{Ishihara1987,
    author = "Ishihara, Hideki and Takahashi, Masaaki and Tomimatsu, Akira",
    title = "{GRAVITATIONAL FARADAY ROTATION INDUCED BY KERR BLACK HOLE}",
    reportNumber = "DPNU-87-17",
    doi = "10.1103/PhysRevD.38.472",
    journal = "Phys. Rev. D",
    volume = "38",
    pages = "472",
    year = "1988"
}

@ARTICLE{Corner1980,
       author = {{Connors}, P.~A. and {Piran}, T. and {Stark}, R.~F.},
        title = "{Polarization features of X-ray radiation emitted near black holes.}",
      journal = {\apj},
         year = 1980,
        month = jan,
       volume = {235},
        pages = {224-244},
          doi = {10.1086/157627},
       adsurl = {https://ui.adsabs.harvard.edu/abs/1980ApJ...235..224C}
}

@article{Chen:2025,
    author = "Chen, Xian and V{\'a}zquez-Aceves, Ver{\'o}nica and Chen, Siyuan and Lee, Kejia and Guo, Yanjun and Liu, Kuo",
    title = "{Detecting intermediate-mass black holes using miniature pulsar timing arrays in globular clusters}",
    eprint = "2507.08201",
    archivePrefix = "arXiv",
    primaryClass = "astro-ph.HE",
    doi = "10.1103/nb5d-qzm8",
    journal = "Phys. Rev. Res.",
    volume = "7",
    number = "3",
    pages = "033300",
    year = "2025"
}

@ARTICLE{Stairs2003,
       author = {{Stairs}, Ingrid H.},
        title = "{Testing General Relativity with Pulsar Timing}",
      journal = {Living Reviews in Relativity},
         year = 2003,
        month = sep,
       volume = {6},
       number = {1},
          eid = {5},
        pages = {5},
          doi = {10.12942/lrr-2003-5},
archivePrefix = {arXiv},
       eprint = {astro-ph/0307536},
 primaryClass = {astro-ph}      
}

@ARTICLE{agazie24,
       author = {Agazie, G. and others},
        title = "{Comparing Recent Pulsar Timing Array Results on the Nanohertz Stochastic Gravitational-wave Background}",
      journal = {\apj},
         year = 2024,
        month = may,
       volume = {966},
       number = {1},
          eid = {105},
        pages = {105},
          doi = {10.3847/1538-4357/ad36be},
archivePrefix = {arXiv},
       eprint = {2309.00693},
 primaryClass = {astro-ph.HE},
       adsurl = {https://ui.adsabs.harvard.edu/abs/2024ApJ...966..105A}
}

@article{Krishnakumar:2021,
    author = "Krishnakumar, M. A. and others",
    title = "{High precision measurements of interstellar dispersion measure with the upgraded GMRT}",
    eprint = "2101.05334",
    archivePrefix = "arXiv",
    primaryClass = "astro-ph.HE",
    doi = "10.1051/0004-6361/202140340",
    journal = "Astron. Astrophys.",
    volume = "651",
    pages = "A5",
    year = "2021"
}

@article{Seymour:2018,
    author = "Seymour, Brian C. and Yagi, Kent",
    title = "{Testing General Relativity with Black Hole-Pulsar Binaries}",
    eprint = "1808.00080",
    archivePrefix = "arXiv",
    primaryClass = "gr-qc",
    doi = "10.1103/PhysRevD.98.124007",
    journal = "Phys. Rev. D",
    volume = "98",
    number = "12",
    pages = "124007",
    year = "2018"
}

@article{DellaMonica:2025,
    author = "Della Monica, Riccardo and de Martino, Ivan",
    title = "{Pulsar timing in the Galactic Center}",
    eprint = "2501.03912",
    archivePrefix = "arXiv",
    primaryClass = "gr-qc",
    doi = "10.1103/9zwc-gk79",
    journal = "Phys. Rev. D",
    volume = "114",
    number = "2",
    pages = "024048",
    year = "2026"
}

@article{Liu:2014,
    author = "Liu, K. and Eatough, R. P. and Wex, N. and Kramer, M.",
    title = "{Pulsar{\textendash}black hole binaries: prospects for new gravity tests with future radio telescopes}",
    eprint = "1409.3882",
    archivePrefix = "arXiv",
    primaryClass = "astro-ph.GA",
    doi = "10.1093/mnras/stu1913",
    journal = "Mon. Not. Roy. Astron. Soc.",
    volume = "445",
    number = "3",
    pages = "3115--3132",
    year = "2014"
}

@article{Walker:1970,
    author = "Walker, M. and Penrose, R.",
    title = "{On quadratic first integrals of the geodesic equations for type [22] spacetimes}",
    doi = "10.1007/BF01649445",
    journal = "Commun. Math. Phys.",
    volume = "18",
    pages = "265--274",
    year = "1970"
}

@article{Frolov:2020,
    author = "Frolov, Valeri P.",
    title = "{Maxwell equations in a curved spacetime: Spin optics approximation}",
    eprint = "2007.03743",
    archivePrefix = "arXiv",
    primaryClass = "gr-qc",
    doi = "10.1103/PhysRevD.102.084013",
    journal = "Phys. Rev. D",
    volume = "102",
    number = "8",
    pages = "084013",
    year = "2020"
}

@article{Dovciak:2008,
    author = "Dovciak, M. and Muleri, F. and Goosmann, R. W. and Karas, V. and Matt, G.",
    title = "{Thermal disc emission from a rotating black hole: X-ray polarization signatures}",
    eprint = "0809.0418",
    archivePrefix = "arXiv",
    primaryClass = "astro-ph",
    doi = "10.1111/j.1365-2966.2008.13872.x",
    journal = "Mon. Not. Roy. Astron. Soc.",
    volume = "391",
    pages = "32",
    year = "2008"
}

@article{kerr63,
  title = {Gravitational Field of a Spinning Mass as an Example of Algebraically Special Metrics},
  author = {Kerr, Roy P.},
  journal = {Phys. Rev. Lett.},
  volume = {11},
  issue = {5},
  pages = {237--238},
  numpages = {0},
  year = {1963},
  month = {Sep},
  publisher = {American Physical Society},
  doi = {10.1103/PhysRevLett.11.237},
  url = {https://link.aps.org/doi/10.1103/PhysRevLett.11.237}
}

@article{Devecchi:2007,
    author = "Devecchi, B. and Colpi, M. and Mapelli, M. and Possenti, A.",
    title = "{Millisecond pulsars around intermediate-mass black holes in globular clusters}",
    eprint = "0706.1656",
    archivePrefix = "arXiv",
    primaryClass = "astro-ph",
    doi = "10.1111/j.1365-2966.2007.12160.x",
    journal = "Mon. Not. Roy. Astron. Soc.",
    volume = "380",
    pages = "691",
    year = "2007"
}

@article{NouriZonoz1999,
  title = {Gravitoelectromagnetic approach to the gravitational Faraday rotation in stationary spacetimes},
  author = {Nouri-Zonoz, Mohammad},
  journal = {Phys. Rev. D},
  volume = {60},
  issue = {2},
  pages = {024013},
  numpages = {5},
  year = {1999},
  month = {Jun},
  publisher = {American Physical Society},
  doi = {10.1103/PhysRevD.60.024013},
  url = {https://link.aps.org/doi/10.1103/PhysRevD.60.024013}
}

@article{Sereno2005,
    author = {Sereno, M.},
    title = {Detecting gravitomagnetism with rotation of polarization by a gravitational lens},
    journal = {Monthly Notices of the Royal Astronomical Society},
    volume = {356},
    number = {1},
    pages = {381-385},
    year = {2005},
    month = {01},
    issn = {0035-8711},
    doi = {10.1111/j.1365-2966.2004.08456.x},
    url = {https://doi.org/10.1111/j.1365-2966.2004.08456.x},
    eprint = {https://academic.oup.com/mnras/article-pdf/356/1/381/3601429/356-1-381.pdf},
}

@unpublished{Oswald:2026,
    author        = "Oswald, L. S. and others",
    title         = "{Understanding Pulsar Magnetospheres with the SKAO}",
    eprint        = "2607.05432",
    archivePrefix = "arXiv",
    primaryClass  = "astro-ph.HE",
    reportNumber  = "AASKAII/Oswald01",
    month         = jul,
    year          = "2026",
    note          = "arXiv:2607.05432 [astro-ph.HE]"
}

@article{Brodutch:2011,
    author = "Brodutch, Aharon and Terno, Daniel R.",
    title = "{Polarization rotation, reference frames and Mach's principle}",
    eprint = "1107.1274",
    archivePrefix = "arXiv",
    primaryClass = "gr-qc",
    doi = "10.1103/PhysRevD.84.121501",
    journal = "Phys. Rev. D",
    volume = "84",
    pages = "121501",
    year = "2011"
}

\end{document}